\documentclass[namedreferences]{solarphysics}
\usepackage[hyperref,optionalrh]{spr-sola-addons} % For Solar Physics 
\usepackage{rotating,graphicx}        % For eps figures, newer & more powerfull
\usepackage[export]{adjustbox}%
\usepackage{amssymb}        % useful mathematical symbols
\usepackage{color}           % For color text: \color command
\usepackage{breakurl}        % For breaking URLs easily trough lines
\usepackage{placeins}
\newcommand{\etal}{{\it et al.}}

\chardef\us=`\_

\begin{document}

\begin{article}
\begin{opening}

\title{Formation of Isolated Radio Type II Bursts at Low Frequencies}

\author[addressref={aff1},corref,email={silpoh@utu.fi}]{\inits{S.}\fnm{Silja}~\lnm{Pohjolainen}}%\sep
\author[addressref={aff2},corref,email={natash@utu.fi}]{\inits{N.}\fnm{Nasrin}~\lnm{Talebpour Sheshvan}}

%\author{\inits{}\fnm{}~\lnm{}\orcid{}}
%\author{P.~\surname{Author-a}$^{1}$\sep
%        E.~\surname{Author-b}$^{1}$\sep
%        M.~\surname{Author-c}$^{2}$      
%       }

%   \institute{$^{1}$ First affiliation
%                     email: \url{e.mail-a} email: \url{e.mail-b}\\ 
%              $^{2}$ Second affiliation
%                     email: \url{e.mail-c} \\
%             }
\address[id=aff1]{Tuorla Observatory, Department of Physics and Astronomy, University of Turku, Turku, Finland}
\address[id=aff2]{Space Research Laboratory, Department of Physics and Astronomy, University of Turku, Turku, Finland }

\runningauthor{S. Pohjolainen, N. Talebpour Sheshvan}
\runningtitle{Formation of Isolated Radio Type II Bursts}

\begin{abstract}
  The first appearance of radio type II burst emission at decameter-hectometer (DH) waves typically occurs in connection, and often simultaneously, with other types of radio emissions. As type II bursts are signatures of propagating shock waves that are associated with flares and coronal mass ejections (CMEs), a rich variety of radio emissions can be expected. However, sometimes DH type II bursts appear in the dynamic spectra without other or earlier radio signatures. One explanation for them could be that the flare-CME launch happens on the far side of the Sun, and the emission is observed only when the source gets high enough in the solar atmosphere. In this study we have analysed 26 radio type II bursts that started at DH waves and were well-separated ('isolated') from other radio emission features. These bursts were identified from all DH type II bursts observed in 1998-2016, and for 12 events we had observations from at least two different viewing angles with the instruments onboard {\it Wind} and STEREO satellites. We found that only 30\% of the type II bursts had their source origin on the far side of the Sun, but also that no bursts originated from the central region of the Sun (longitudes E30 - W40). Almost all of the isolated DH type II bursts could be associated with a shock near the CME leading front, and only few were determined to be shocks near the CME flank regions. In this respect our result differs from earlier findings. Our analysis, which included inspection of various CME and radio emission characteristics, suggests that the isolated DH type II bursts could be a special subgroup within DH type II bursts, where the radio emission requires particular coronal conditions to form and to die out. 
\end{abstract}
\end{opening}
%-------------------------------------------------

\section{Introduction}
     \label{Introduction} 

The first discovery of solar radio type II bursts dates back to 1947, by \cite{Payne-Scott1947}. Type II bursts appear as continuous and slowly drifting bursts, from high to low frequencies, in the radio dynamic spectra \citep{Wild1950}. The frequency drift was then found to display the decreasing electron density in the solar atmosphere, and the discovery of harmonics was taken as evidence that there was a common source producing oscillations at a fundamental plasma frequency and its second harmonic, see for example the historic review by \cite{pick2008}. The disturbance producing type II emission was later identified as a MHD shock \citep{uchida60}.

Type II bursts can be categorized into metric (m), decameter hectometric (DH), and kilometric (km) bursts, based on the wavelengths they are observed. DH waves cover the frequency range of 30 MHz-300~kHz, but as radio emission below $\sim$25 MHz is blocked by the Earth atmosphere the whole wavelength range cannot be observed with ground-based instruments. The existing space instruments have instrumental upper limits at 16 MHz (STEREO) and 14~MHz ({\it Wind}), and so it is customary to define DH and interplanetary (IP) type II bursts as those that occur at frequencies below these space instrument upper limits.  

Previous studies have shown that DH and km type II bursts are produced by coronal mass ejection (CME) driven shocks \citep{Sheeley1985, Cane1987,Gopalswamy2001}. CMEs are large-scale plasma structures ejected from the solar chromosphere and corona, and these transients propagate out to the IP space. In shock theory, a shock can form as a bow shock at the leading front of the CME, as a driven shock near the CME front (even well-separated from the driver after launch), or as a driven (expansion) shock near the CME flanks, see for example the reviews by \cite{warmuth2007} and \cite{vrsnak2008}.

A bow shock is formed when the speed of a coronal transient exceeds the local magnetosonic speed, which in the solar corona is well represented by the Alfv\'en speed. Modelling the magnetic field structure of an active region produced a local Alfv\'en speed minimum at a distance of 1.2 - 1.8 solar radii R$_{\odot}$, and a maximum of 740 km s$^{-1}$ at a distance of 3.8  R$_{\odot}$ \citep{mann2003}. The model by \cite{warmuth2005} found a local speed minimum at the coronal base, 300~km s$^{-1}$, and a local maximum at a distance of 3.5 R$_{\odot}$, 1100 km s$^{-1}$. For example \cite{evans2008} have compared the Alfv\'en speed values obtained for different types of regions (open field regions, streamers, active regions) with the different models and techniques. Due to the high magnetosonic speed, most metric type II bursts do not continue into the outer corona. And as the magnetosonic speed decreases at distances larger than 3.5 - 3.8  R$_{\odot}$, DH type bursts are more likely to form there. This distance corresponds to a frequency of 8-12 MHz in most coronal density models, see for example the comparison between heights and frequencies in \cite{pohjolainen2007}.

The characteristics of CMEs associated with DH type II bursts have been well-studied and reported \citep{Gopalswamy2000, Gopalswamy2001, Gopalswamy2002, sharma2008, Pappa2010}, and also the differences between radio-loud and radio-quiet CMEs have been investigated \citep{Gopalswamy2010}. 
Basically, CMEs associated with DH type II bursts have been found to be faster and wider. In addition,  \cite{vasanth2013} found that in the DH type II burst-associated events, CME properties like speed, width, and shock speed were higher in limb events than in disk events, suggesting that projection effects play a part in the measurements.

Also DH type II bursts have been studied statistically, by their start and end frequencies, their duration, burst source location, and emission bandwidth. \cite{sharma2017} concluded that the average duration of bursts that started between 16 and 1 MHz was considerably less than those that started below 1 MHz. In their analysis of type II burst end frequencies, \cite{vasanth2015} concluded that the lower the emission end frequency, the higher the possibility of having a strong geomagnetic storm near Earth.
\cite{shanmugaraju2018} divided the type II bursts into groups by their end frequency: (A) higher than 1~MHz, (B) intermediate, between 1~MHz and 100~kHz, and (C) lower than 100~kHz. Their analysis indicated that there was a correlation between the observed CME height and the estimated type II burst height for groups B and C, but no correlation for group A. They suggested that the type II shock could occur at the flanks of the CME in group A, and at the CME leading front in groups B and C. The statistical results obtained by \cite{aguilar2005} showed that the average bandwidth-to-frequency ratio was $\sim$0.3 in the DH type II bursts observed with the {\it Wind}/WAVES radio instruments, irrespective of their spectral domain.
The ratio distribution, however, covered a large interval of 0.05 - 0.8, indicating that in individual bursts the emission bandwidth can be quite different.  

The DH type II burst appearances have also been used to classify events.
\cite{pohjolainen2013} selected wide-band and diffuse IP type II bursts, to find out where they originated. Most of the analysed bursts (and shocks) could be considered to be due to plasma emission, formed at or just above the CME leading edge. However, for some radio events the source was unclear and could have been due to other mechanisms, like \cite{bastian2007} had earlier suggested. 

DH type II bursts are typically observed in the radio dynamic spectra mixed with other strong emissions like type III bursts (fast electron beams), but they can also appear quite suddenly and without other close-in-time radio features. Some of these well-separated, isolated DH type II bursts can be easily explained by their source location: 
If the flare-CME launch happens on the far side of the Sun, radio emission would be observed only when the radio emission source gets high enough in the solar atmosphere, where the emission is no longer blocked by the solar disk or by the dense solar atmosphere, 
{\it i.e.}, where the emission frequency exceeds the local plasma frequency. Observation of accelerated electron beams would then depend on the direction of the magnetic field lines. For type II bursts, propagation direction and wideness of the shock front could also play a part.

Our study presents analysis of selected DH type II radio bursts that were observed well-separated from other radio emission features. With statistical analysis of the burst, flare, and CME characteristics we try to determine the reasons for them to appear isolated in the  radio dynamic spectra, and discuss their appearances and source origins. Some of the most  interesting events are analysed in more detail. 

\section{Data Analysis}

We have selected radio type II bursts that appear to start at a low frequency,
at decameter-hectometer (DH) wavelengths, and which can be identified as a
separate structure from any other emission feature visible in the radio
spectrum at that time. The type II bursts were searched using the Wind/Plasma
and Radio Waves (WAVES) Type II lists at
\url{https://solar-radio.gsfc.nasa.gov/wind/data_products.html}
and the Type II list at
\url{https://cdaw.gsfc.nasa.gov/CME_list/radio/waves_type2.html},
including the years 1998-2016.

Coronagraph images and CME leading front heights were obtained from the LASCO
CME Catalog, maintained at the CDAW Data Center. This catalogue contains data
from both the Solar and Heliospheric Observatory (SOHO) and Solar Terrestrial
Relations Observatory (STEREO) missions, with white-light coronagraph images
and EUV images at several different wavelengths. Solar flare data were
obtained from the NOAA's Satellite and Information Service (NESDIS) website.
Information on radio type II bursts at meter wavelengths were obtained from
the NOAA radio spectral listings (up to year 2011), and from individual
spectral plots from different radio observatories (mainly from the Radio
Solar Telescope Network, RSTN, at
\url{https://www.ngdc.noaa.gov/stp/space-weather/solar-data/solar-features/solar-radio/rstn-spectral/}). 

For the type II burst height estimates we used the 'hybrid' atmospheric
density model by \cite{vrsnak2004}. As the plasma frequency $f_p$ depends only
on the plasma density $n_e$ ($f_p$ = 9000 $\sqrt{n_e}$, when $f$ is in Hz and
$n_e$ is in cm$^{-3}$), the type II emission frequency can be converted to
source height if we know how the density changes in the corona and
interplanetary space, see for example \cite{pohjolainen2007}. Only fundamental
emission lanes were used for the calculations. As at DH wavelengths
fundamental emission should be stronger than the harmonic \citep{lengyel85},
we assumed fundamental emission in events that showed only one emission lane. 

\subsection{Selection Criteria and Visual Appearances}

Our selection criteria for the DH type II bursts was that the emission lane (or lanes, if both fundamental and harmonic lanes were visible) must appear below 16 or 14 MHz (instrumental upper limits), must be separated from any other radio emission feature, and must show a frequency drift in order to verify it really is a type II burst. With these criteria we excluded events with separate multiple emission lanes (fundamental-harmonic bursts accepted but mixed multiple bursts excluded), and also excluded lanes where the start of the type II burst was superposed or mixed with any other type of burst activity. For example intense type III bursts usually make it impossible to identify the actual type II start time and frequency. For data analysis purposes we also demanded that adequate CME data was available for each event. We found 26 such events from the years 1998-2016, see Table \ref{table1} in the Appendix. 

\begin{figure}[h]
  \centering
  \includegraphics[width=0.35\textwidth]{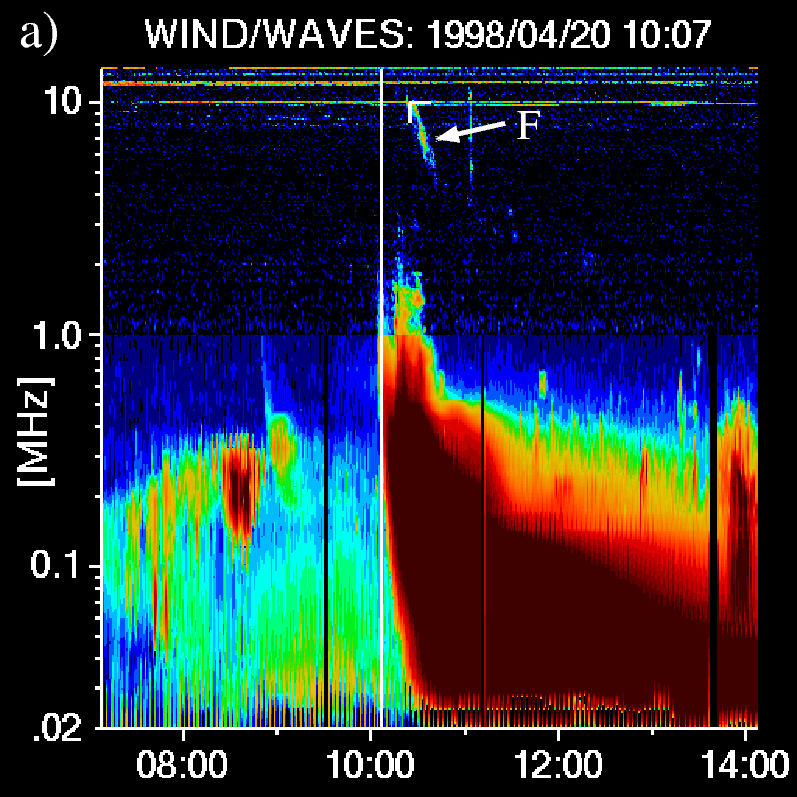}
  \includegraphics[width=0.35\textwidth]{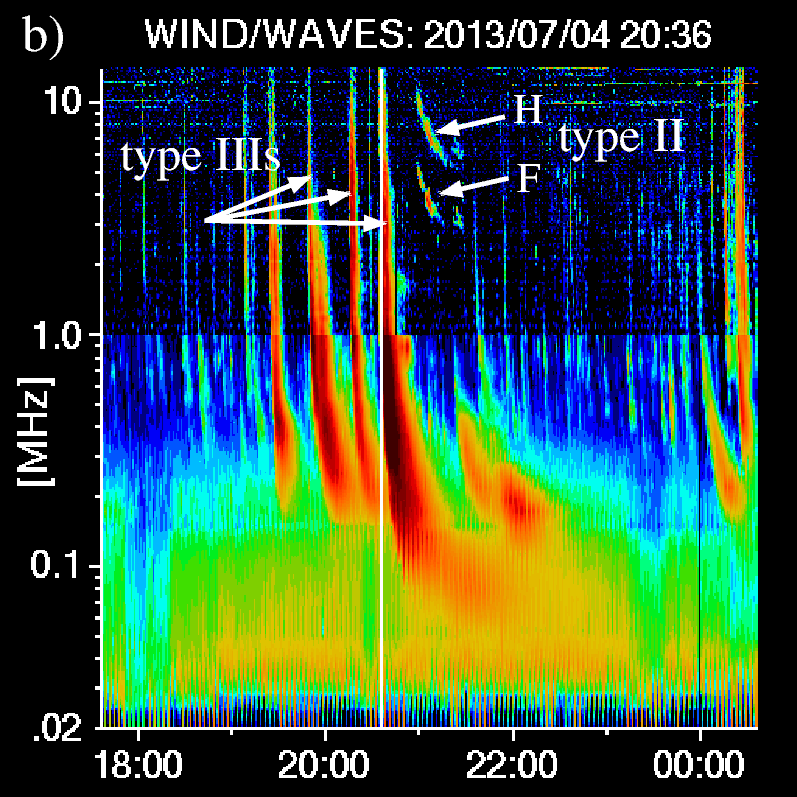}\\
  \includegraphics[width=0.35\textwidth]{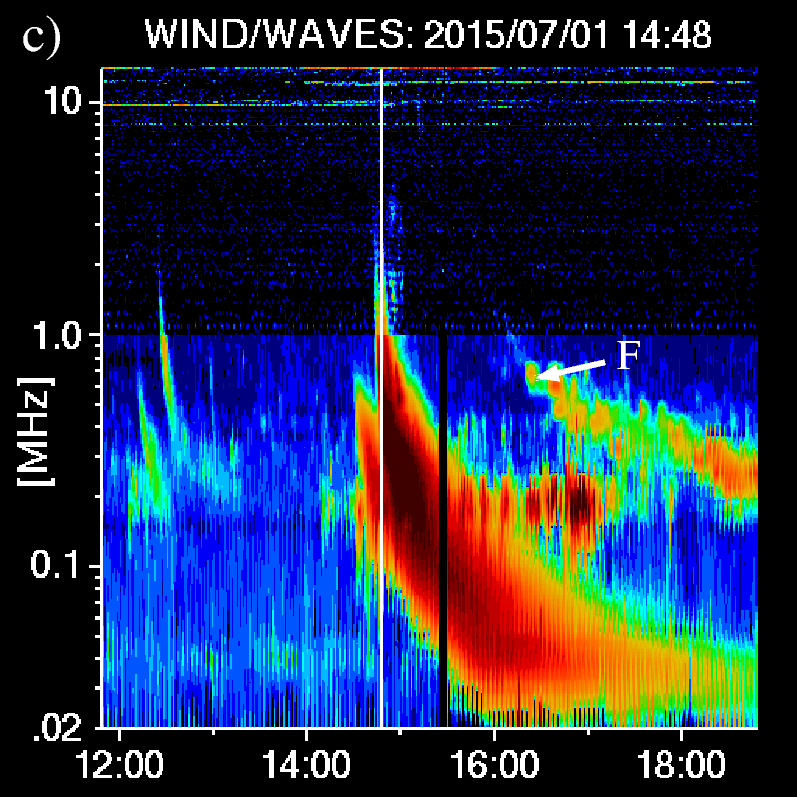}
  \includegraphics[width=0.35\textwidth]{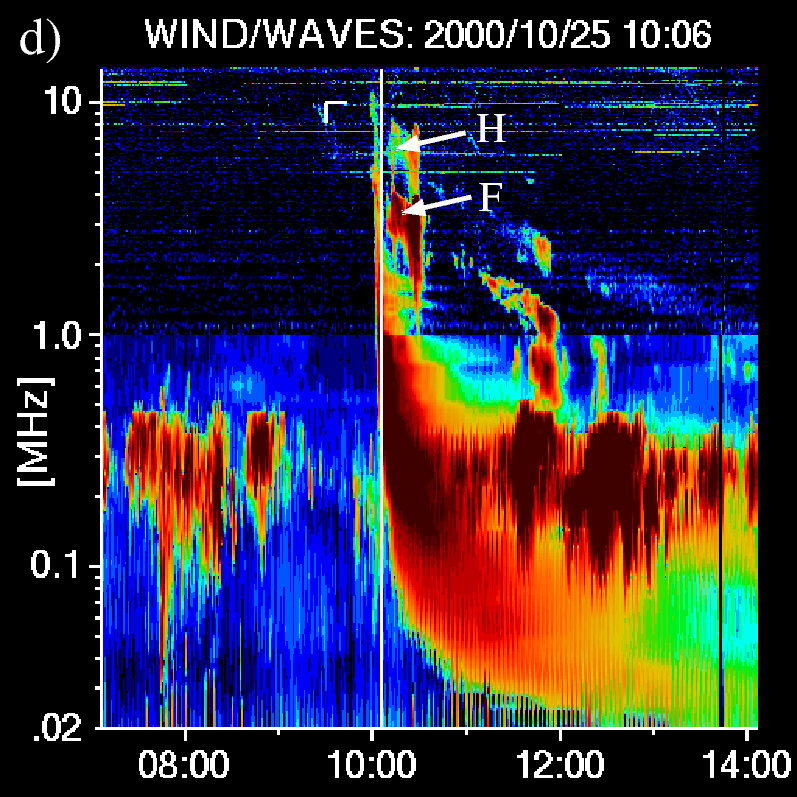}
  \caption{Examples of isolated type II bursts at DH wavelengths. 
    a) A short-duration type II burst with one single lane appears at 
    10:25 UT at 10~MHz start frequency on 20 April 1998.
    b) A short-duration type II burst with fundamental (F) and harmonic (H) emission lane pair 
    at 5 and 10 MHz appears at 20:57 UT on 4 July 2013. The spectrum also shows earlier 
    type III bursts (fast-drift bursts that are formed by propagating electron beams).  
    c) A long-duration type II burst with one single lane appears at 
    16:22 UT with a low-frequency start at 700 kHz on 1~July 2015.
    d) A strong type II burst with F and H emission lane pair appears at 4 and 8 MHz at 10:00 UT on 25 October 2000.
    }
\label{examples}%
\end{figure}

A visual inspection of the spectral events showed different appearances. Some of the bursts were very short (duration 5-8 min) and some were long-duration events that were visible for hours. Some showed narrow-band emission ($\Delta$f/f $\leq$ 0.1) and some were very wide-band ($\Delta$f/f $>$ 0.3). Some showed only one emission lane and some showed two emission lanes, where the second lane appeared at twice the frequency of the first. One lane events were interpreted to be due to fundamental plasma emission (F), and two-lane events due to fundamental and harmonic (F+H) plasma emission. Examples of the burst appearances are shown in Figure \ref{examples}. The type II burst duration, emission band, and F and F+H appearance are listed in Tables \ref{table1} and \ref{table2}, together with other burst characteristics and associated features. The radio spectra of all events are shown in the Appendix, in Figures  \ref{spectra19980420} - \ref{spectra20160815}.

%%%%%%%%%%%%%%

\subsection{Correlation Analysis}

The DH type II bursts were associated with 9 GOES M-class flares, 7 C-class flares, 3 eruptive prominences (EP), and 7 had their origin behind the limb, on the far side of the Sun, so that flare classification for them was not available. 

All the DH type II bursts were associated with a close-in-time CME. CME speeds were determined at the time of the DH type II burst appearance, using the second order fits to the CME leading front heights listed in the LASCO CME Catalog. The CME speeds ranged from 480 to 1800 km s$^{-1}$. Of the CMEs, 9 were accelerating, 14 were decelerating, and 3 had a constant speed.

\begin{figure}[ht!]
  \centering
   \includegraphics[width=1.0\textwidth]{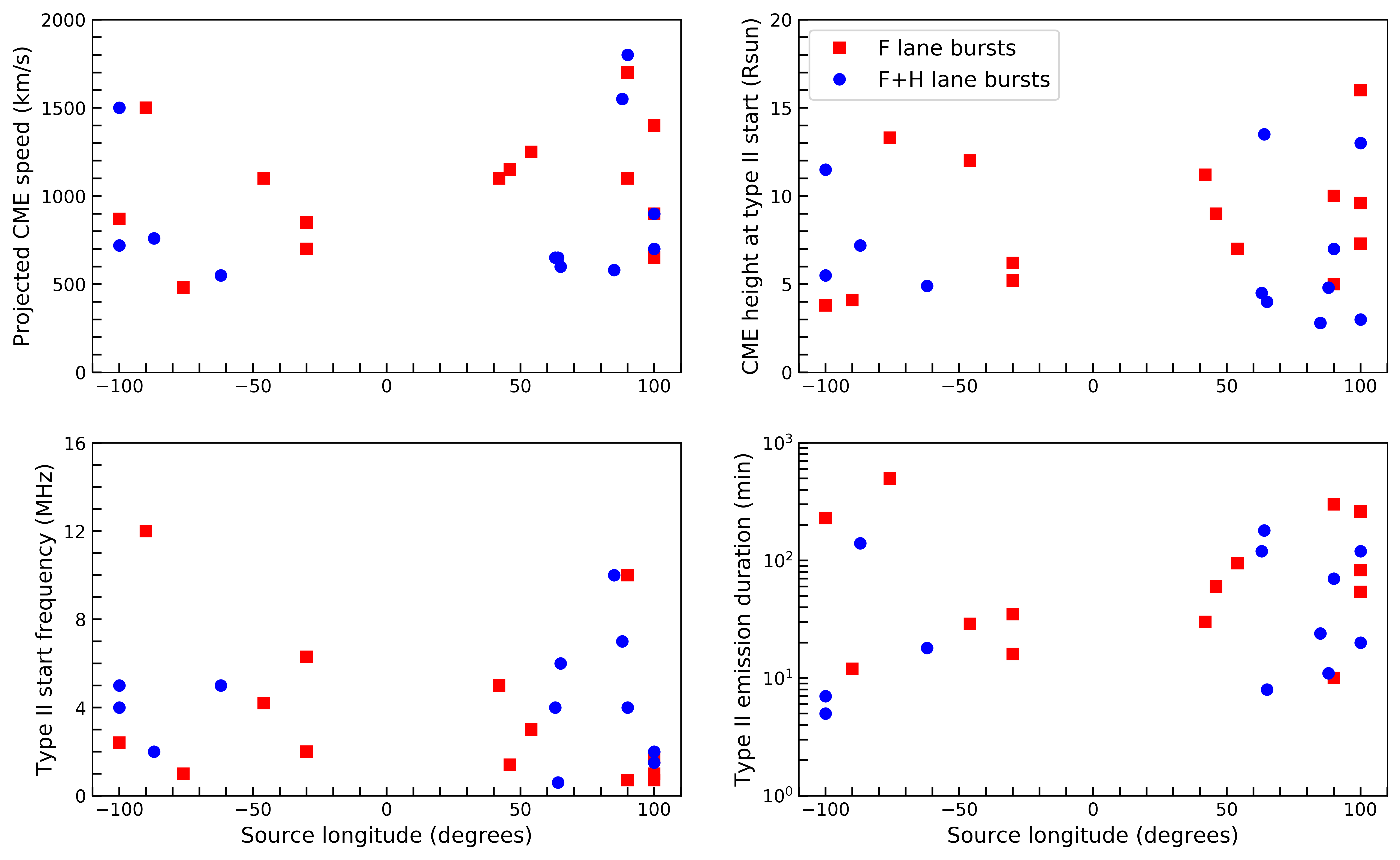}
   \caption{Flare/CME source longitude on the solar disk compared with 
      the projected CME speed ({\it top left}), the CME height at the 
      time of the type II burst start ({\it top right}),
      the type II burst start frequency ({\it bottom left}), and 
      the type II burst duration({\it bottom right}).
      Behind-the-limb events are given as -100 (behind the eastern limb)
      and +100 (behind the western limb).
      F and F+H lane bursts are plotted with different symbols.
   }
\label{longitude}%
\end{figure}

In Figure \ref{longitude} we compare the source longitude ({\it i.e.}, location of the active region from where the eruption originated) with projected CME speed, CME height at the type II start time, type II start frequency, and type II burst duration. Figure \ref{longitude} shows that near the disk center, inside E30-W42 longitudes there are no isolated DH type II bursts. Within the longitudes E50-E30 and W42-W50 all the isolated type II bursts are single (F) lane events.

The highest CME speeds are observed when the flare/CME source longitudes are $\gtrsim$\,90 degrees. It may be that the observed CME speeds for the on-the-disk events suffer from projection effects and look to be less than the real speeds, as was suggested by \cite{vasanth2013}. Comparison between the CME height at the time of the type II burst start and the source longitude gives a similar result as the comparison to the CME speed.

The type II start frequency does not seem to have a clear correlation with the source longitude, but the highest start frequencies are in bursts that that have their source origin near the solar limb.

The comparison between the type II emission duration and the source longitude shows that the type II events that originate near or behind the solar limb have the longest durations. Our sample of events shows quite an opposite tendency compared to that of \cite{mittal2017}, who found that the durations of the DH type II radio bursts located at the solar disc center are longer than the durations of the bursts located at the solar limb. However, their sample of 426 DH type II bursts and our sample of 26 bursts are not directly comparable, so ours could be a selection effect.
%%%%%%

\begin{figure}[h!]
  \centering
   \includegraphics[width=1.0\textwidth]{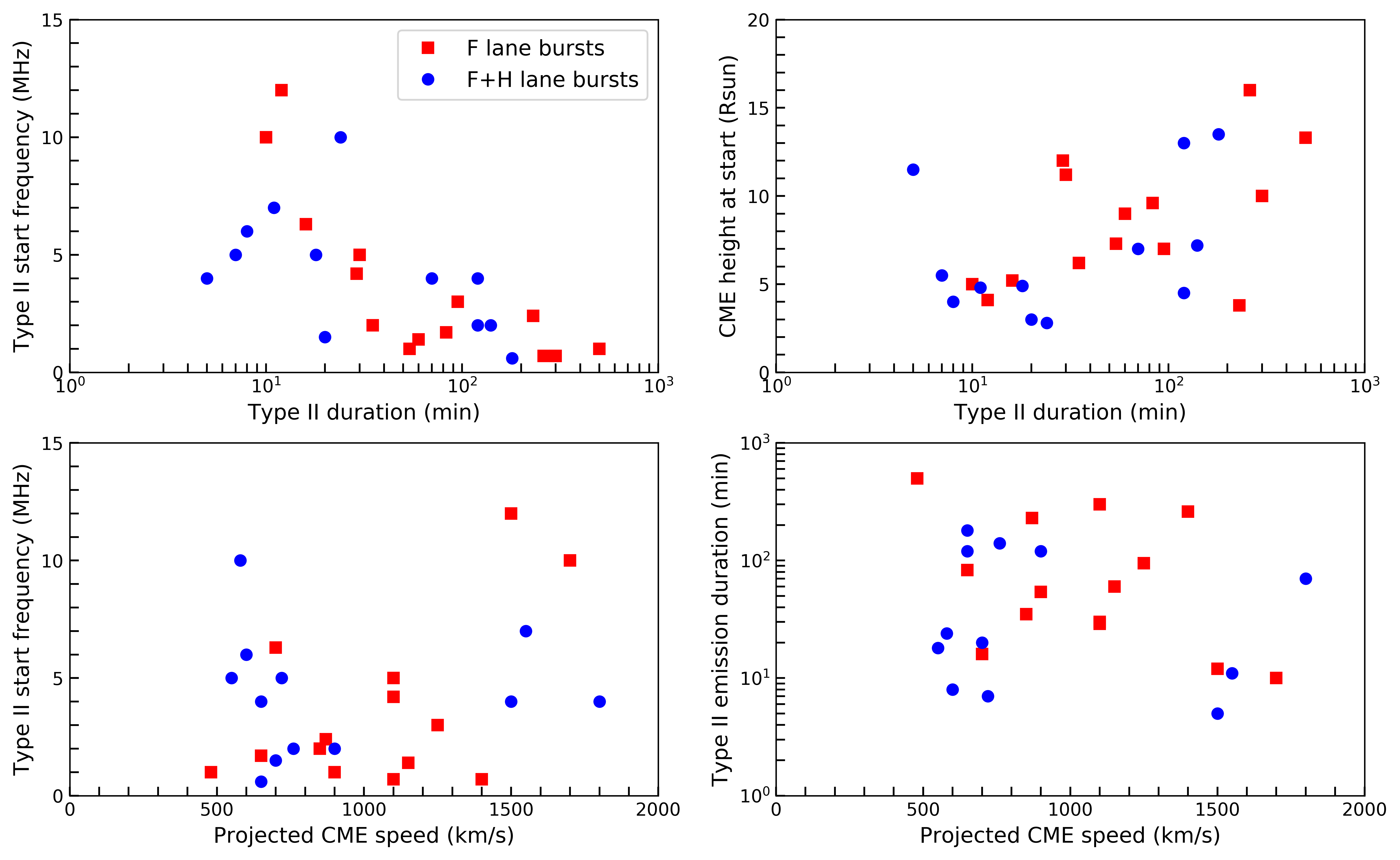}
 \caption{Type II burst duration compared with the type II burst start frequency ({\it top left}) and the CME height at the time of type II start ({\it top right}). Projected CME speed at the time of type II burst start compared with the type II burst start frequency ({\it bottom left}) and the type II burst duration ({\it bottom right}). }
\label{duration-speed}%
\end{figure}

Figure \ref{duration-speed} shows the type II burst start frequency compared to the type II burst duration. This plot is in accordance with the finding of \cite{sharma2017}, that bursts starting at lower frequencies have longer durations. We also compared the projected CME leading front height at the time of type II start with the type II burst duration, and this shows a tendency that the higher the CME is at the type II burst start, the longer the type II burst duration.

The DH type II burst start frequency does not look to have any clear correlation with the projected CME speed. When the projected CME speed is compared to the type II burst duration, we see that the highest speed CMEs do not produce long-duration DH type II bursts. 
%%%%%

\begin{figure}[h!]
   \centering
   \includegraphics[width=1.0\textwidth]{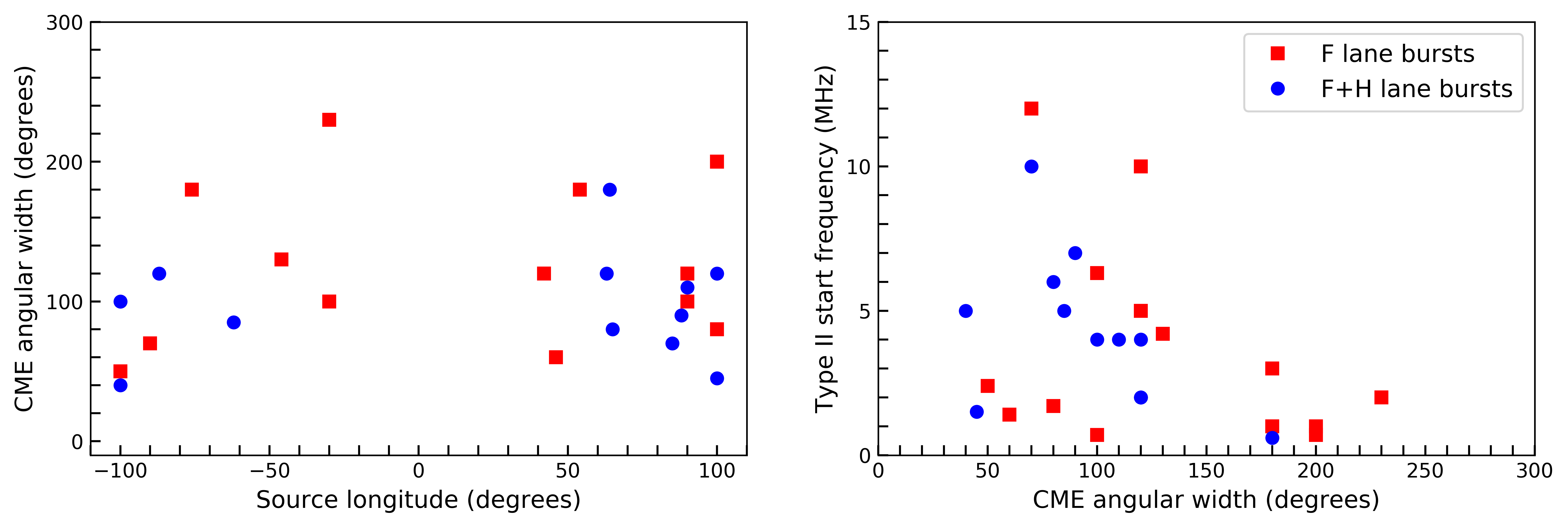}
   \caption{CME angular width near the time of type II start compared
     with the flare/CME source longitude ({\it left}) and the type II start frequency ({\it right}).
   }
\label{width}%
\end{figure}

We then compared the CME angular width with the source longitude and the type II burst start frequency, in Figure \ref{width}.
The CME Catalog lists the associated CMEs as full halos (10/26), partial halos (13/26), and with widths 94 - 101 degrees (3/26). We measured the CME widths near the time of the type II emission start, which gave more narrow widths that ranged from 40 to 230 degrees. The smallest CME angular widths were in the events that occurred at or behind the solar limb. The closer to the disk center the event originated, the larger the angular width. This could be a simple projection effect. 

A large CME angular width looks to correlate with a low type II start frequency ({\it i.e.}, high formation height). However, a narrow CME width does not look to imply a high type II formation height, as some of the very narrow CMEs were associated with bursts that had a very low start frequency. This may rule out the assumption that type II burst formation requires a certain angular width for the CME.  

We also compared the projected CME speed with the GOES soft X-ray maximum flux. Of the 26 events three were eruptive prominences without an observed X-ray flare and seven were far side events where GOES intensity could not be measured. The correlation plot (not shown here) suggested that the more intense the flare flux, the higher the CME speed. 

Furthermore, we checked if these events showed any metric type II emission, to see if a propagating shock existed already at lower coronal heights. Due to the atmospheric cut-off in ground-based radio observations (that are possible down to 30-25 MHz) and the instrumental limit of space observations (start frequency 16-14 MHz), type II emission lanes that start at meter waves cannot be followed directly to DH waves, because of the spectral gap in observations. 
Of the analysed events, 10 out of 26 were associated with earlier metric  type II emission, see Table \ref{table2}. It is obvious that far side events and eruptive prominences lack metric emission. An association to flare intensity and CME speed was also noted: in our data set metric type II bursts required an M-class GOES flare and/or a very high speed CME (1100-1800 km s$^{-1}$).

Our events were also compared with the solar energetic particle (SEP) event catalogue by \cite{paassilta2017} that covers the years in our list. A match was found in five events. In three events (20 April 1998, 25 October 2000, 17~July 2012) the SEP injection time, determined with the velocity dispersion analysis (VDA) method, matched well with the time of the DH type II burst appearance. In two events (22 August 2005, 1 July 2015) the SEP injection time was earlier, but could be explained with an earlier metric type II burst. Due to the small number of SEP events, no further correlation analysis was performed. We note, however, that considering the spectral structures of type II bursts may help to improve the SEP forecasting accuracy \citep{Iwai2020}.

%%%%%%%%%%%%%%%%%%%%%%%%%%%%%

\section{Events with Large Height Difference between the CME Leading Front
and the Type II Burst Source}
\label{Heightsep}

\begin{figure}[h!]
   \centering
   \includegraphics[width=0.6\textwidth]{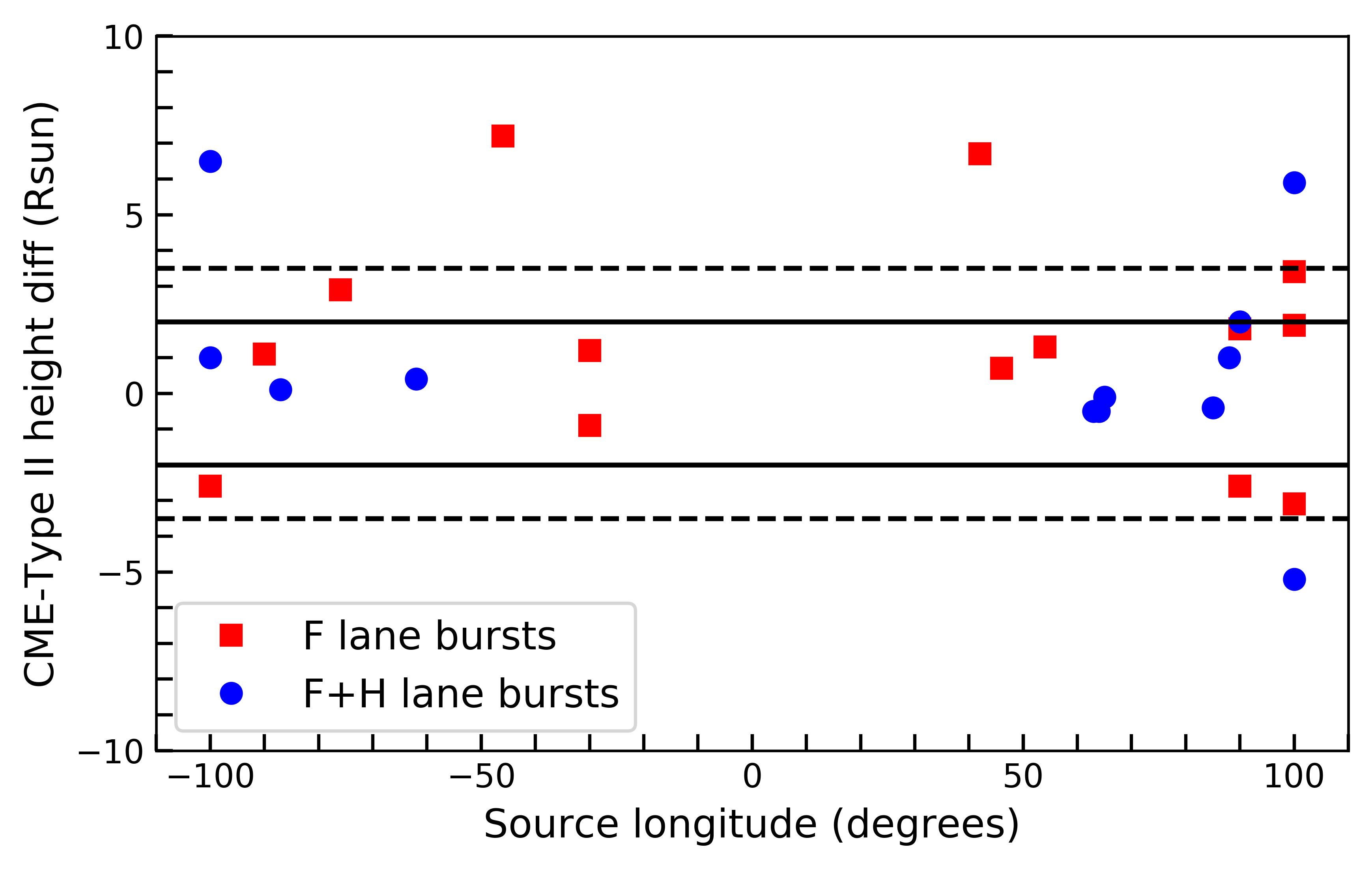}
   \caption{Flare/CME source longitude on the solar disk compared with the height difference between the CME leading front and the
     type II burst source at the time of the type II burst start (positive when the CME is higher than the type II burst). 
    Solid lines mark the $\pm$2~R$_{\odot}$ error margin for calculated heights (mainly from density models) and dashed lines indicate the values that we concluded to be within acceptable height error in individual events.
     }
\label{corr2}%
\end{figure}

To evaluate where our DH type II bursts could be formed, we calculated the height difference between the CME leading front and the type II burst for each event. These are listed in Table \ref{table1}. 
Figure \ref{corr2} shows the height difference versus source longitude, and the value is positive when the CME front is located higher than the type II burst source.

Errors in type II burst height estimation can be quite large, as the calculated heights depend on the chosen atmospheric density model and on the selected frequency within the emission lane taken for calculation (for example, center or leading edge of the lane, affected also by the wideness of the emission lane). In the frequency range of 10 - 1 MHz, the height difference that comes from using different (but generally accepted) density models is $\sim$2 R$_{\odot}$, see for example the height values in Table~1 in \cite{pohjolainen2007}.

Equally, the CME leading front height can be uncertain, especially if the front consists of a bright flux rope and a diffuse faint region ahead of it. The diffuse region has often been identified as the CME-driven shock front, see \cite{Schmidt2016A} and references therein. Usually, the shock is more visible toward the flank regions of the CME, and the shock's stand off distance from the CME's leading edge is not more than 1-2 R$_{\odot}$ in this height range \citep{Schmidt2016B}, and often much less \citep{Lee2017}. 

Of our 26 events, 16 could be considered to be a match between the type II burst height and the CME leading front height, within the possible $\pm$2 R$_{\odot}$ error. Similar heights suggest that the type II burst was most probably created by a CME bow shock.

Of the remaining ten events, five had an intermediate height difference of 2.6-3.4 R$_{\odot}$. These events showed relatively wide emission lanes. Wide-band emission is tricky in a sense that the frequency selected for height calculation (lower border, lane center, or some other point inside the lane) will have a large effect on the height. The spectral plots of these events show how the frequency drift of the type II burst follows that of the CME quite well, 
but due to the wideness of the emission lane the heights calculated from the lane center do not always match with the CME heights. Hence, some other frequency point within the emission lane could still give a good match. We also noted that in many of these events the type II height at the start of the burst was different from the CME leading front height, but later on the the heights became similar. Based on our analysis, we conclude that also the five type II bursts with 
intermediate height differences can reasonably well be associated with the CME leading fronts.

The height difference between the CME leading front and the type II shock was observed to be very large, 5.2-7.2 R$_{\odot}$, in five events. On 27 March 2012, the type II burst source was located 5.2 R$_{\odot}$ higher than the leading front of the CME. In the other four events, the type II burst sources were located 5.9--7.2 R$_{\odot}$ lower than leading front of the CME. We now examine the possible reasons for these large height differences.

%%%%%%%%%%%%%%%%

\subsection{Event on 27 March 2012}
\label{event27032012}
  
\begin{figure}[!h]
  \centering
  \includegraphics[width=0.35\textwidth]{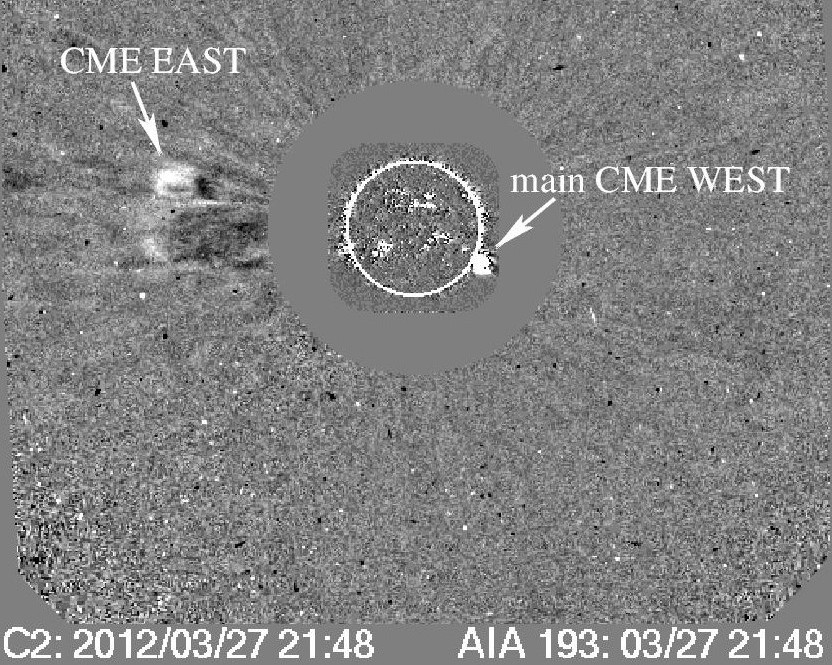}
  \includegraphics[width=0.35\textwidth]{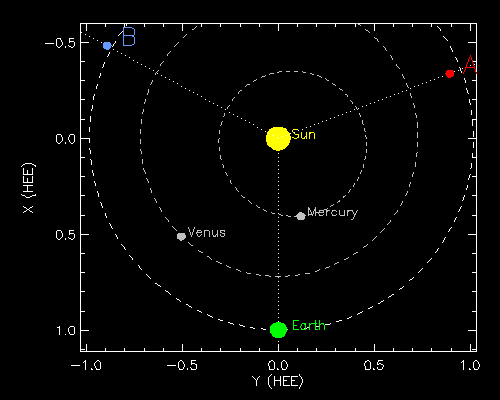}\\
  \includegraphics[width=0.35\textwidth]{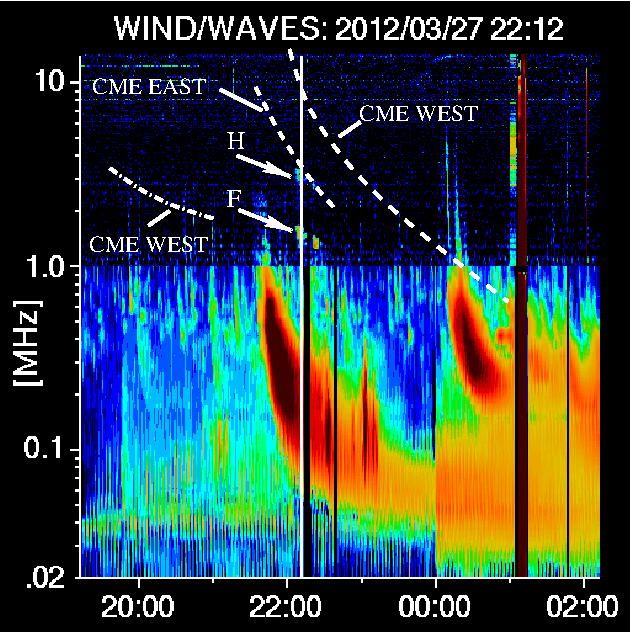}
  \includegraphics[width=0.35\textwidth]{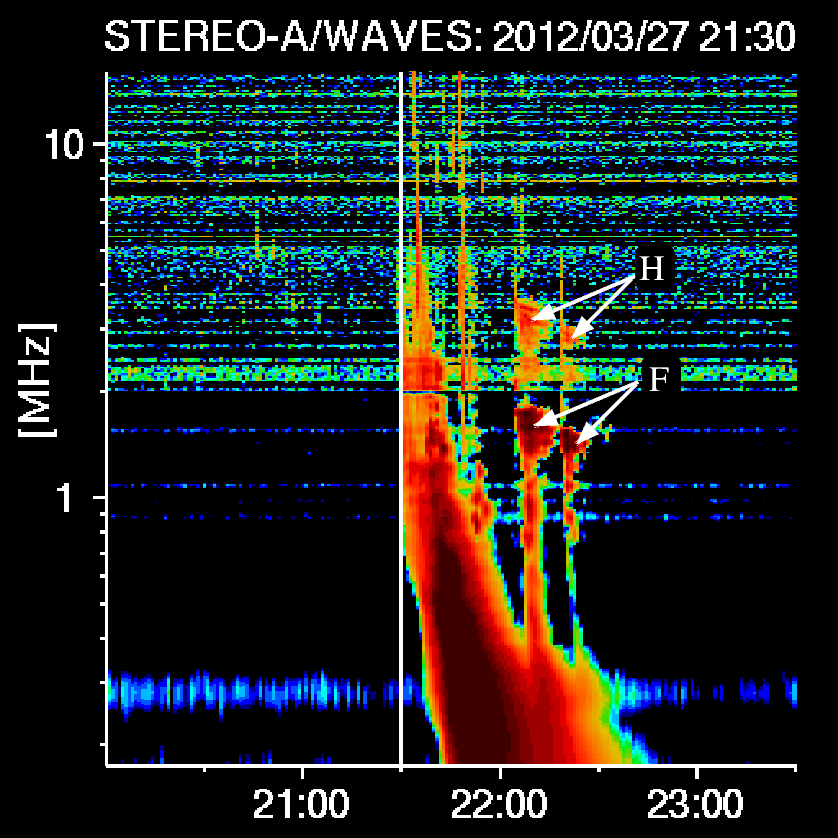}
  \caption{Event on 27 March 2012, source origin at SW90b (on the far side). The type II burst source is located 5.2 R$_{\odot}$ higher than the leading front of the main CME (CME WEST, dashed white line). There are earlier faint and narrow CMEs that propagate toward west (CME WEST, dash-dotted white line) and toward east (CME EAST, dashed white line). STEREO-A dynamic spectrum shows how radio type III emission lanes cross the features that were interpreted as type II emission. Some parts of these type III bursts are visible in the {\it Wind}/WAVES and STEREO-B/WAVES spectrum, but only at frequencies below 1 MHz. The positions of both STEREO spacecraft on 27 March 2012 are also shown. {\it Wind}/WAVES was located in L1, along the Sun-Earth line. }
  \label{newspectra20120327}%
\end{figure}

On 27 March 2012 the DH type II burst source was located 5.2 R$_{\odot}$ higher than the leading front of the main CME, which was propagating toward the west at speed 700 km s$^{-1}$ at the time when the type II burst appeared (dashed line marked CME WEST in Figure \ref{newspectra20120327}, here the CME heights have been converted to frequencies). The CME heights were also verified from the STEREO-A/COR1 observations, and they were not higher than what was observed
by SOHO/LASCO. The active region (AR) was most probably NOAA AR 11440, which had rotated to $\sim$W100, as it was last observed from Earth on 26 March at location S25W91.

On closer inspection, there were two earlier-launched faint and narrow CMEs which were catalogued as very poor events, and did not show up in the coronagraph images at the time of the type II burst appearance. The first one propagated toward the west (dash-dotted line marked CME WEST, the heights have been extrapolated onward from the last observed height of 4.5 R$_{\odot}$ at 19:00~UT), and the second one toward the east (marked CME EAST) in Figure~\ref{newspectra20120327}.
The heights of CME EAST are near the heights of the type II burst harmonic emission. However, as STEREO-B/WAVES did not record a type II burst at all, and only one type III burst was visible below 700 kHz, it does not look probable that the CME EAST could have created the burst recorded by {\it Wind}/WAVES and STEREO-A/WAVES.

STEREO-A/WAVES dynamic spectrum shows that the radio enhancements, which we interpreted as type II F+H lanes, were located along type III burst emission lanes that were not visible in the {\it Wind}/WAVES spectrum. The radio dynamic spectrum at 180-25 MHz from RSTN Sagamore Hill (Earth view, same as for {\it Wind}) does not show any metric type III or type II radio emission within the time range. 

We therefore conclude that the association of the type II burst to the westward propagating fast (main) CME is questionable. It is possible that the type II-like structures were not a type II burst, but flare-accelerated electrons propagating through the dense (earlier) CME structures, {\it i.e.}, they were type III burst enhancements when the beams travelled through local turbulent plasma. Fundamental-harmonic structures can occur also in type III bursts, as a significant number of type III bursts at meter and decameter waves have been found to show harmonics, see for example \cite{robinson98} and references
therein. 

%%%%%%%%%%%%%%%%

\subsection{Event on 11 December 2001}
\label{event11122001}
  
\begin{figure}[h!]
  \centering
  \includegraphics[width=0.35\textwidth]{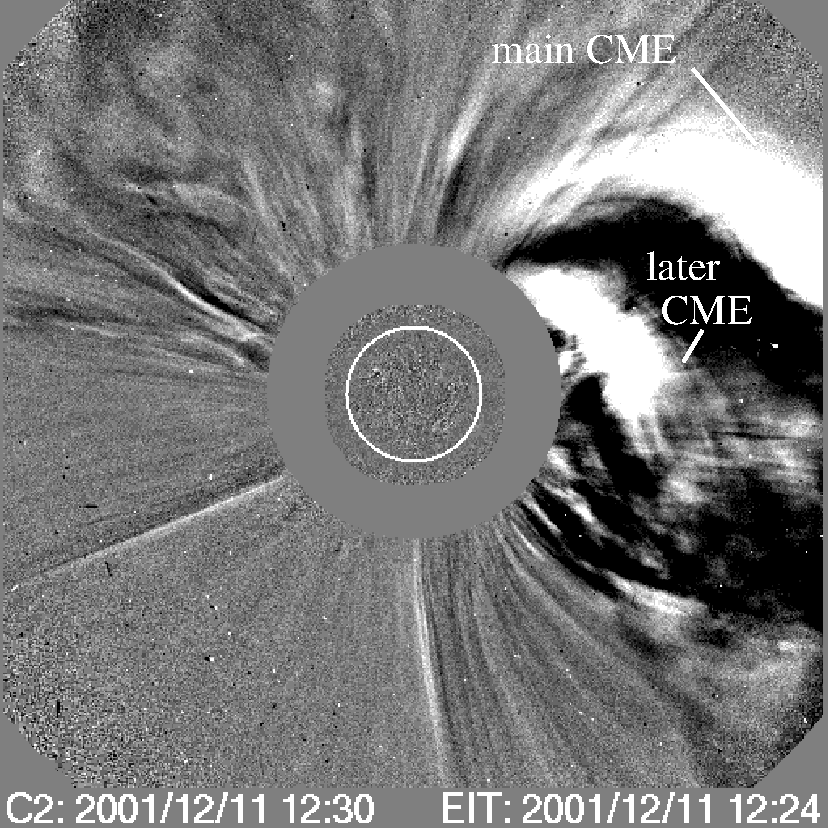}
  \includegraphics[width=0.35\textwidth]{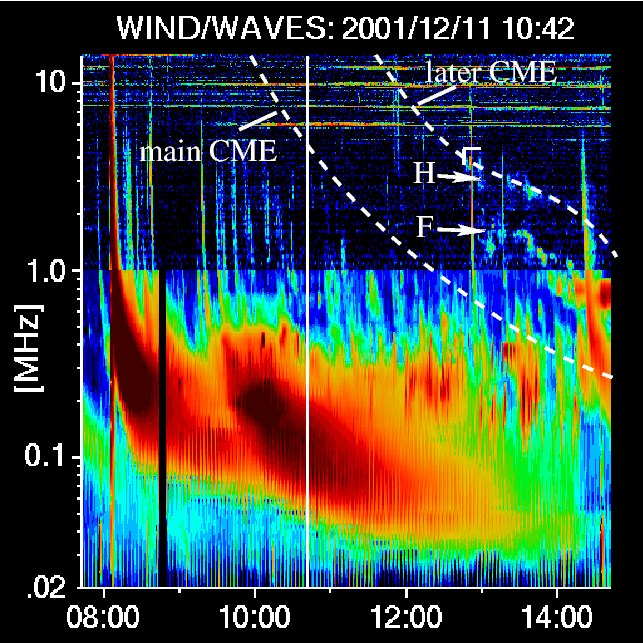}\\
  \includegraphics[width=0.6\textwidth]{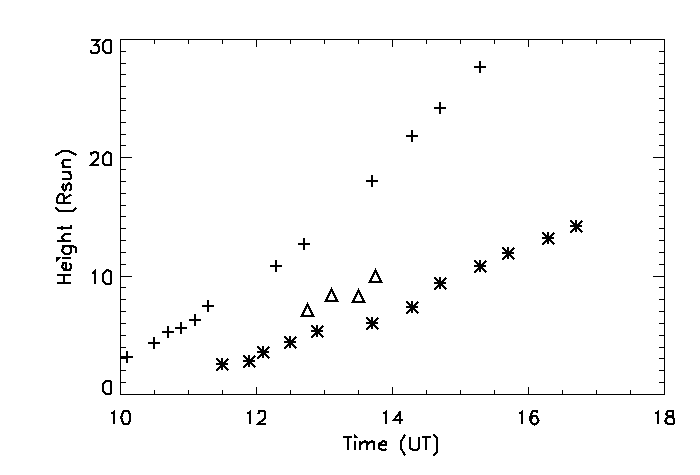}
\caption{Event on 11 December 2001, source location at SW90b (on the far side). The leading front of the main CME is located 5.9 R$_{\odot}$ higher than the type II burst source. However, there is a later CME propagating westward, in the wake of the main CME. The LASCO C2 difference image ({\it top left}) shows the northern edge of the main CME and the leading front of the later CME. The CME propagation heights, converted to frequencies, are indicated with the dashed white lines in the radio dynamic spectrum ({\it top right}). The height-time plot ({\it bottom}) shows the height evolution of these two CME fronts (crosses and stars) and the calculated type II burst heights from the fundamental emission lane (triangles). }
  \label{newspectra20011211}%
\end{figure}

The DH type II burst on 11 December 2001 showed relatively weak emission but had a clear start at 12:45 UT, with fundamental and harmonic emission lanes (Figure \ref{newspectra20011211}). After 14:00 UT there is uncertainty if and how the lane continues, narrow or wide, as we see enhanced emission at a much wider frequency range. However, some of these enhancements could be due to type III bursts.

The main CME was located 5.9 R$_{\odot}$ higher than the type II burst source at 12:45 UT, but a second bright structure had appeared inside the main CME at 11:30 UT. It propagated to the same westward direction, at speed 500 km s$^{-1}$. The height evolution of the main CME and the later, inner CME structure are shown in Figure \ref{newspectra20011211}, indicated with white dashed lines in the dynamic spectrum. The type II burst shows a curved structure, and a similar change is observed in the height evolution of the later CME structure (height-time plot in Figure~\ref{newspectra20011211}). The height difference between this later CME front and the type II burst is $\sim$2~R$_{\odot}$, which falls inside the error margin in height estimation.

We therefore conclude that the DH type II burst was created by the later CME structure, and that the curved type II structures were due to density changes along the CME propagation path, caused by the earlier (main) CME.

%%%%%%%%%%%%%%%%%%%%%%%%%

\begin{figure}[!h]
  \centering
    \includegraphics[width=0.35\textwidth]{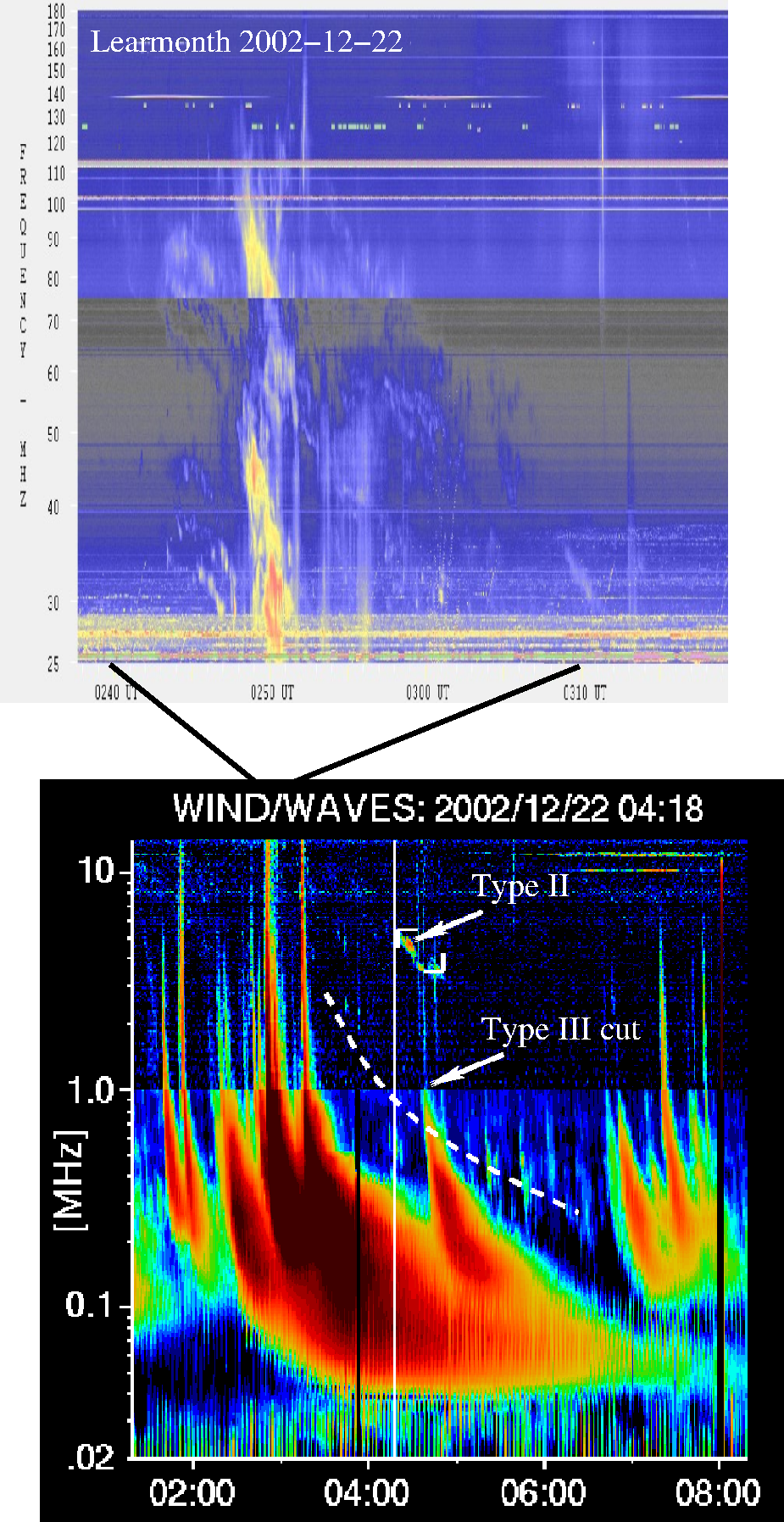}
    \hspace{10mm}
    \includegraphics[width=0.357\textwidth]{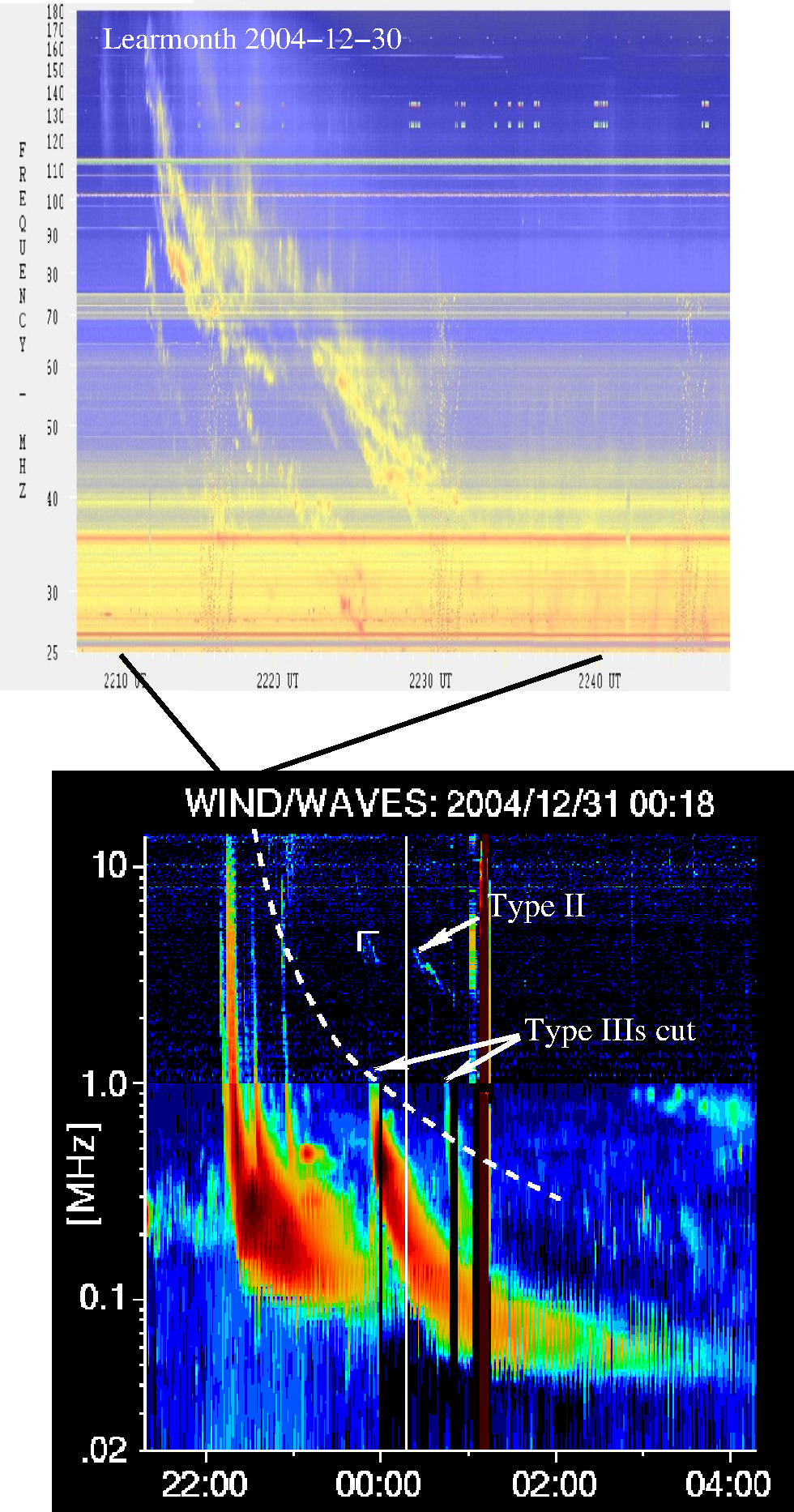} 
    \caption{On 22 December 2002, the burst source origin is at N23W42. The leading front of the CME (dashed white line, {\it bottom left}) is located 6.7 R$_{\odot}$ higher than the type II burst source. On 31 December 2004, the burst source origin is at N04E46. The leading front of the CME is located 7.2 R$_{\odot}$ higher than the type II burst source ({\it bottom right}). The DH type III bursts along the CME propagation paths were 'cut',{\it i.e.}, their emission did not cross the whole spectral range. Both events showed also metric type II burst activity at 140 - 25 MHz, observed by the RSTN Learmonth station.
     }
\label{lear-wind}      
\end{figure}

\subsection{Events on 22 December 2002 and 31 December 2004}
\label{events22122002and31122004}
  
The DH type II burst events on 22 December 2002 and 31~December 2004 look very similar (Figure \ref{lear-wind}). Both were on-the-disk events that showed EUV dimmings. The associated GOES flares were class M1.1 and M4.2, and the flare/CME source locations were N23W42 and N04E46, respectively. The CME speed at the type II start was 1100~km s$^{-1}$ in both events. The durations of the type II emission were 30 and 29 minutes, the type II formation heights were 4.5 and 4.8~R$_{\odot}$, and the respective CME heights were 11.2 and 12.0~R$_{\odot}$. The height differences were 6.7 and 7.2~R$_{\odot}$, respectively.

In both events there was strong type II burst activity at metric wavelengths (Figure \ref{lear-wind}, top panels), starting near 140 MHz and continuing down to 25 MHz (observational limit for ground-based observations, RSTN Learmonth). In both cases the metric type II emission was band-split: On 22 December 2002 at start the fundamental frequency was split to 34 and 39 MHz and the harmonic to 68 and 78 MHz, and on 30 December 2004 split lanes can be identified to start at 80 and 100 MHz. The 30 December 2004 spectrum is more complicated, for example at 22:25 UT a separate lane is observed to start at 30 MHz but due to the observational limit it cannot be followed below 25 MHz. We calculated the radio source heights using the lower split-lanes.

\begin{figure}[!h]
  \centering
    \includegraphics[width=0.45\textwidth]{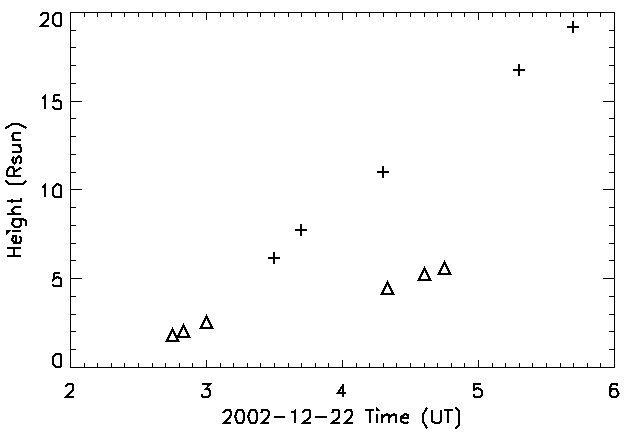}
    \includegraphics[width=0.45\textwidth]{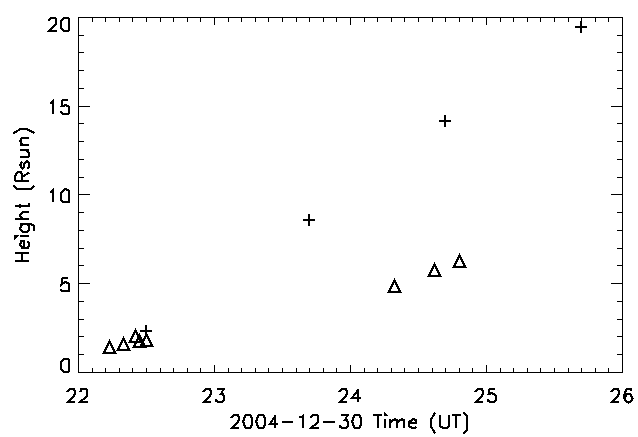} 
    \caption{Height-time plots for the 22 December 2002 and the 30-31 December 2004 events. The heights of the CME leading
      fronts (crosses) can be compared with the heights of the metric and DH type II bursts (triangles).}
\label{events2002and2004heights}      
\end{figure}

The height-time plots in Figure \ref{events2002and2004heights} show that the metric type II burst heights are in line with the CME heights, but the later DH type II burst sources are at much lower heights. The coronagraph images in Figures \ref{spectra20021222} and
\ref{spectra20041231} in the Appendix show streamers near the sides of the CMEs, and it is possible that the type II bursts were caused by shocks near the CME flanks, that were due to CME-streamer interaction at low heights.  

Type III bursts were observed within the full frequency range in both events near the CME launch times, but later on the type III bursts showed 'cuts', {\it i.e.}, the bursts were visible only at frequencies below the CME propagation heights, below $\sim$1 MHz. This suggests that either the type III electron beams were accelerated near the CME fronts or the CMEs blocked their visibility. Unfortunately these events occurred in the pre-STEREO era, so we cannot verify visibility toward other directions. 

%%%%%%%%%%%%%%%%%%%%%%%%%%%%%%%%%

\subsection{Event on 1 February 2005}
\label{event01022005}
  
\begin{figure}[!h]
  \centering
   \includegraphics[width=0.35\textwidth]{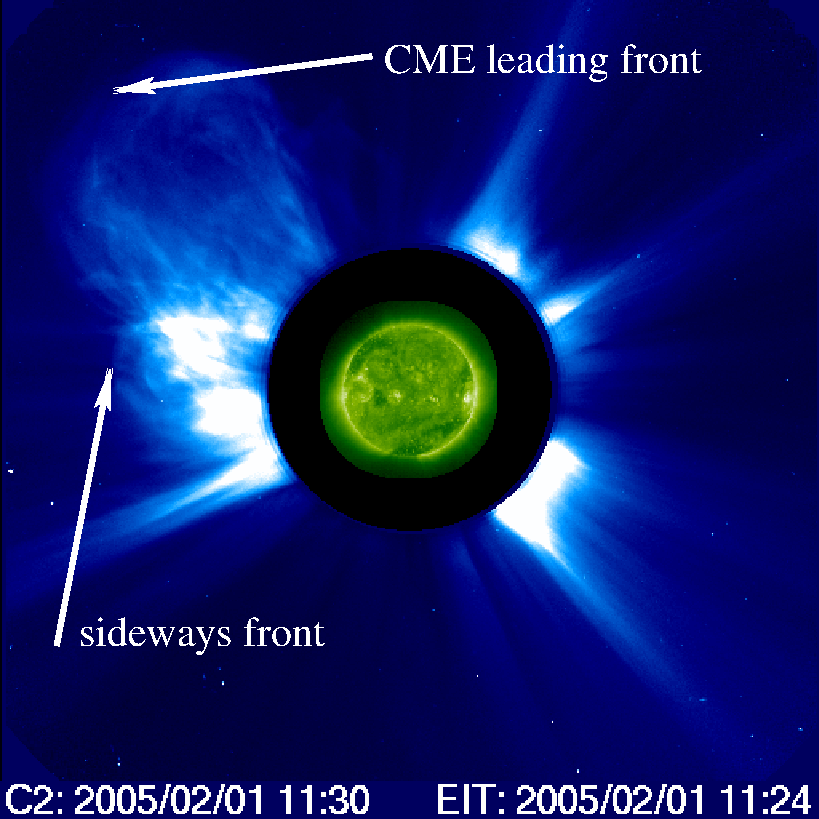}
  \includegraphics[width=0.35\textwidth]{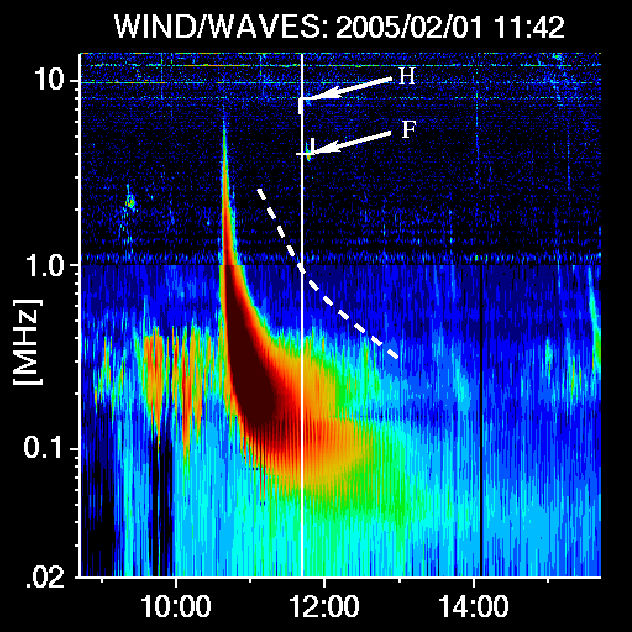}
    \caption{Event on 1 February 2005, source location at NE90b (on the far side). The leading front of the CME (white dashed line in the radio spectrum) is located 6.5 R$_{\odot}$ higher than the type II burst source. A lower, sideways propagating front is also visible near the south-east streamer in the LASCO C2 image ({\it left}). }
  \label{newspectra20050201}%
\end{figure}

In the 1 February 2005 event the CME originated from behind the east limb, and no X-ray flare was observed in the Earth view. The short, only five minute duration DH type II burst showed both fundamental and harmonic emission (Figure \ref{newspectra20050201}). The leading front of the CME was faint and diffuse, and denser structures were located below it, at lower heights.

The CORIMP map (Figure \ref{spectra20050201} in the Appendix) also shows a lower front in the eastern flank of the CME (yellow lines).
This lower front made a sudden south-westward movement in between the
LASCO images at 11:30 UT and 11:54~UT, which matches with the short 
lifetime of the type II burst at 11:45-11:50~UT. It is most probable that the DH type II burst was a short-duration shock,
created by the fast sideways movement within the CME.

%%%%%%%%%%%%%%%%

\section{Observations from Different Viewing Angles}
\label{stereocomp}

In 12 cases the solar event could be observed from at least two different viewing angles, with the instruments onboard {\it Wind}, STEREO-A, and STEREO-B satellites. These events, with the DH type II burst start times and frequencies, are listed in Table~\ref{table3} in the Appendix. In five events the radio emission start time and start frequency were different when observed from a different viewing angle. The start time delays were 5 min, 7 min, 8 min, 26 min, and 83 min, in the order of time delay. In one event the start time differed 2 min, but as the start frequency was the same in both, we concluded that there was no real delay in the type II start. We now describe the delayed events in more detail.

{\it 5 July 2012}. The 5 min delay in the DH type II start looks to be due to band-splitting of the fundamental emission lane. The lower band-split lane start is observed at the same time with both {\it Wind} and STEREO-A instruments at 22:57 UT at $\sim$1.4 MHz, but STEREO-A records emission from the upper band-split lane earlier, at 22:52 UT at 2.2 MHz. The STEREO-A/WAVES dynamic spectrum also shows emission patches that can be interpreted as harmonic emission. Therefore the difference between the observations is most probably due to sensitivity, that {\it Wind}/WAVES did not record the less intense emission from the upper split band. Moreover, a frequency drift from 2.2 MHz to 1.4 MHz within 5 min would require a burst driver speed of $~$3000 km s$^{-1}$, which also suggests that the emission structures did not belong to the same emission lane.

{\it 22 October 2012}. The type II burst became visible 26 minutes earlier in the STEREO-B spectrum, compared to {\it Wind} and STEREO-A. By STEREO-B the burst started as a narrow lane at 01:24 UT at 1.4 MHz, but became much wider after 01:45~UT. The type II burst emission is very faint in the STEREO-A spectrum, but the start time is the same as in the {\it Wind} spectrum, at 01:50 UT. The beginning of the DH type II burst in the {\it Wind} spectrum is visible only partially below 1 MHz (Figure \ref{spectra20121022} in the Appendix), and therefore we conclude that this could also be a sensitivity issue between the {\it Wind}/WAVES RAD1 and RAD2 receivers. 

\begin{figure}[!h]
  \centering
  \includegraphics[width=0.9\textwidth]{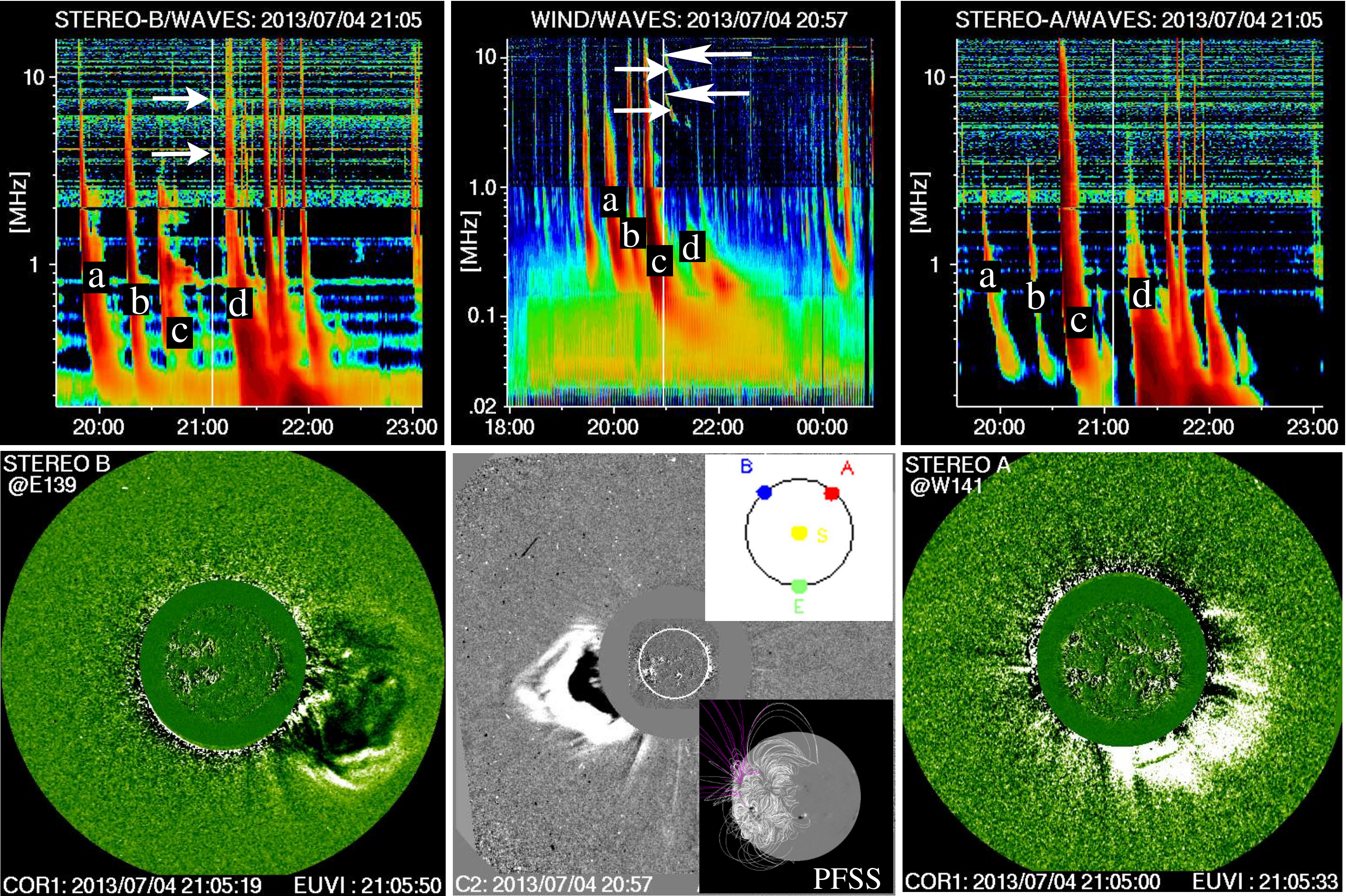}
 \caption{Comparison between STEREO-B, {\it Wind}, and STEREO-A observations on 4~July 2013. Longer arrows point to the first fundamental-harmonic structures observed by {\it Wind} and shorter arrows to the later emission structures observed also by STEREO-B. Type III bursts are labelled a-d, to compare their visibility with the different instruments. The burst source region was located at S14E62. Coronagraph and EUV difference images from the three different viewing angles are also shown, with the PFSS magnetic field extrapolation map (Earth view). Cartoon insert shows the positions of the three satellites, {\it Wind} in Earth view.
    }
  \label{comp20130704}%
\end{figure}

{\it 4 July 2013}. The time delay in the type II burst start was 8 min, as {\it Wind} observed the burst at 20:57 UT at 5 MHz and STEREO-B at 21:05 UT at 4~MHz (Figure \ref{comp20130704}). 
There were several type III bursts that preceded the type II burst, the closest-in-time are listed as a, b, c in the spectral plots. We note that the bursts b and c have a lower start frequency in the STEREO-B/WAVES spectrum, 8 MHz and 2.5 MHz, compared to the {\it Wind}/WAVES spectrum where the bursts are already visible at 14 MHz. This suggests that the emission of the type III bursts and the type II burst were occulted at low coronal heights in the STEREO-B view.  

\begin{figure}[!h]
\centering
\includegraphics[width=0.45\textwidth]{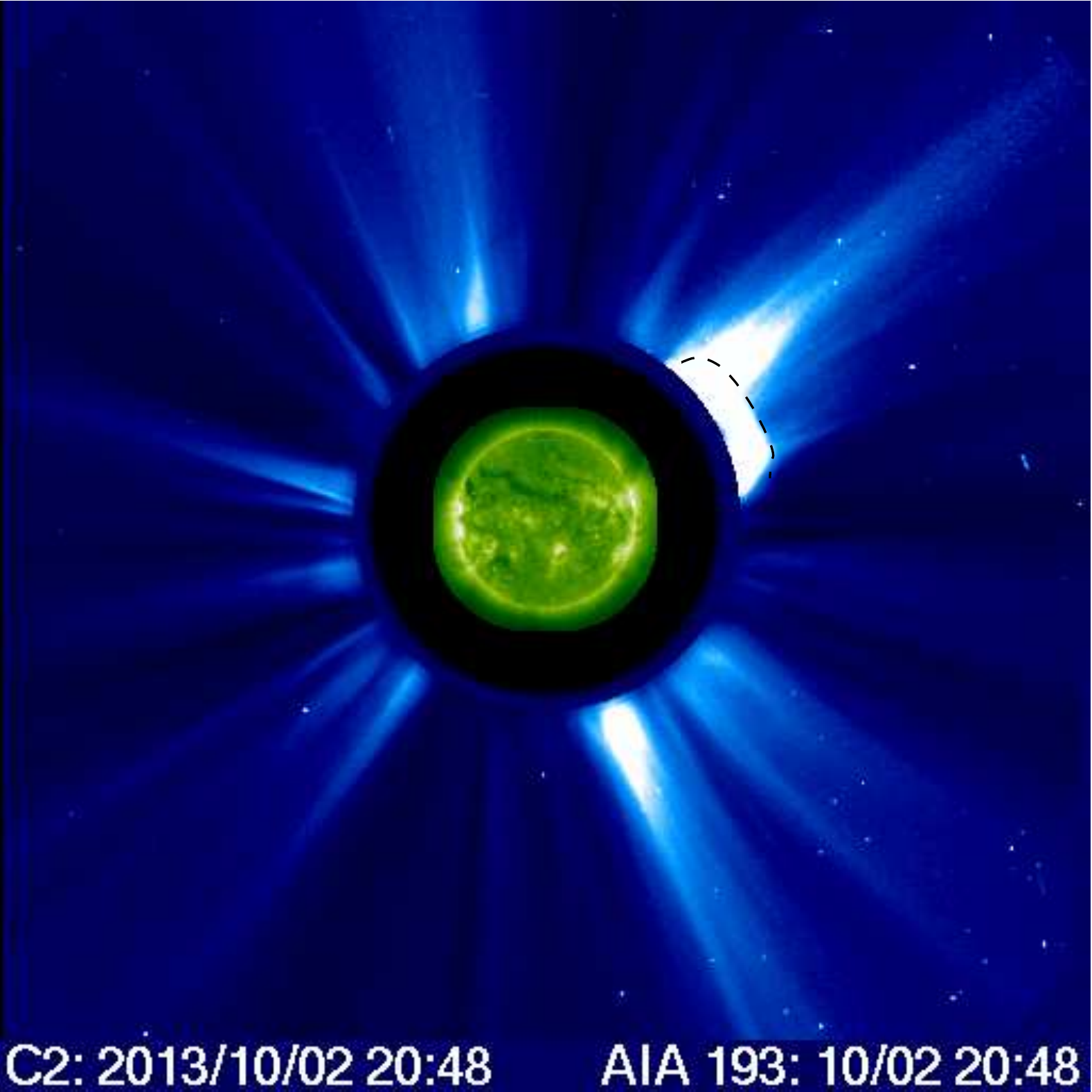}
\includegraphics[width=0.45\textwidth]{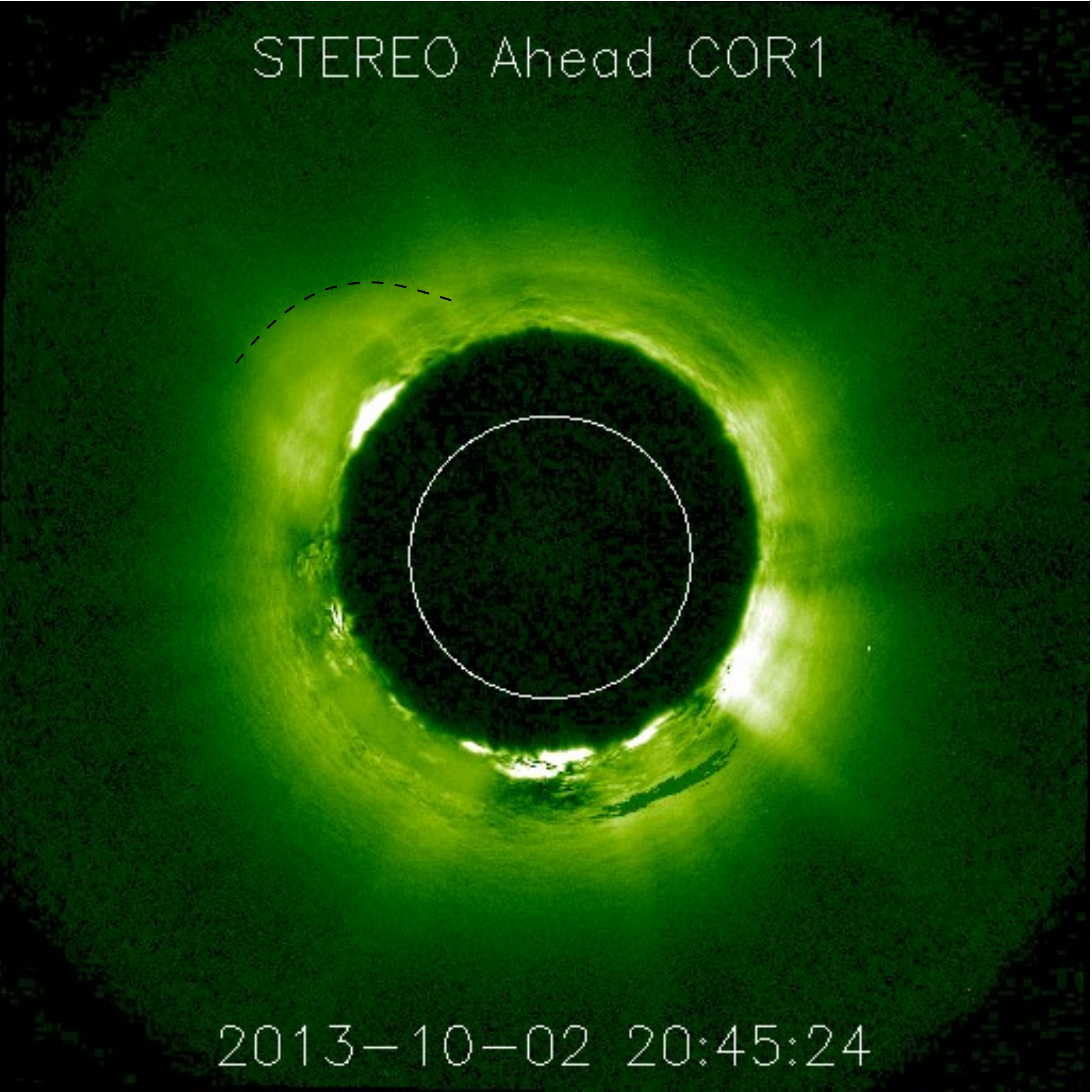}  
\caption{SOHO/LASCO C2-SDO/AIA composite image ({\it left}) of the CME and its source region at N20W85 on 2 October 2013. The STEREO-A/COR1 coronagraph image ({\it right}) shows the same structure. The CME leading front is marked with a dashed line in both images. The LASCO C2 image shows a streamer that crosses the CME (in projection), but it is uncertain weather the streamer is located in front or behind of the CME bubble.  
  }
 \label{comp20131002}
\end{figure}

{\it 2 October 2013}. STEREO-A observed the type II burst starting at 20:39~UT at 12~MHz and {\it Wind} later at 20:46 UT at 10 MHz. With the hybrid atmospheric density model a frequency drift from 12 to 10 MHz in 7 min would indicate a burst driver speed of $\sim$460 km s$^{-1}$. The CME speed at that time was $\sim$600~km~s$^{-1}$, so the type II burst speed would be quite realistic.

We find there is a streamer visible in the SOHO/LASCO image, located across the (projected) CME loop (Figure \ref{comp20131002}). If the streamer location was such that it lay in between {\it Wind} and the type II shock, then STEREO-A would have been able to observe the type II directly, and {\it Wind} only through the streamer. Modeled streamer densities \citep{Decraemer2019} are very near to the type II density values at these heights. 

It is therefore possible that the DH type II burst on 2 October 2013 became visible only when the type II burst emission frequency got above the plasma frequency of the streamer, when the streamer density and the critical plasma frequency decreased with height.

%%%%%%%%%%%%%
\begin{figure}[!h]
  \centering
  \includegraphics[width=0.9\textwidth]{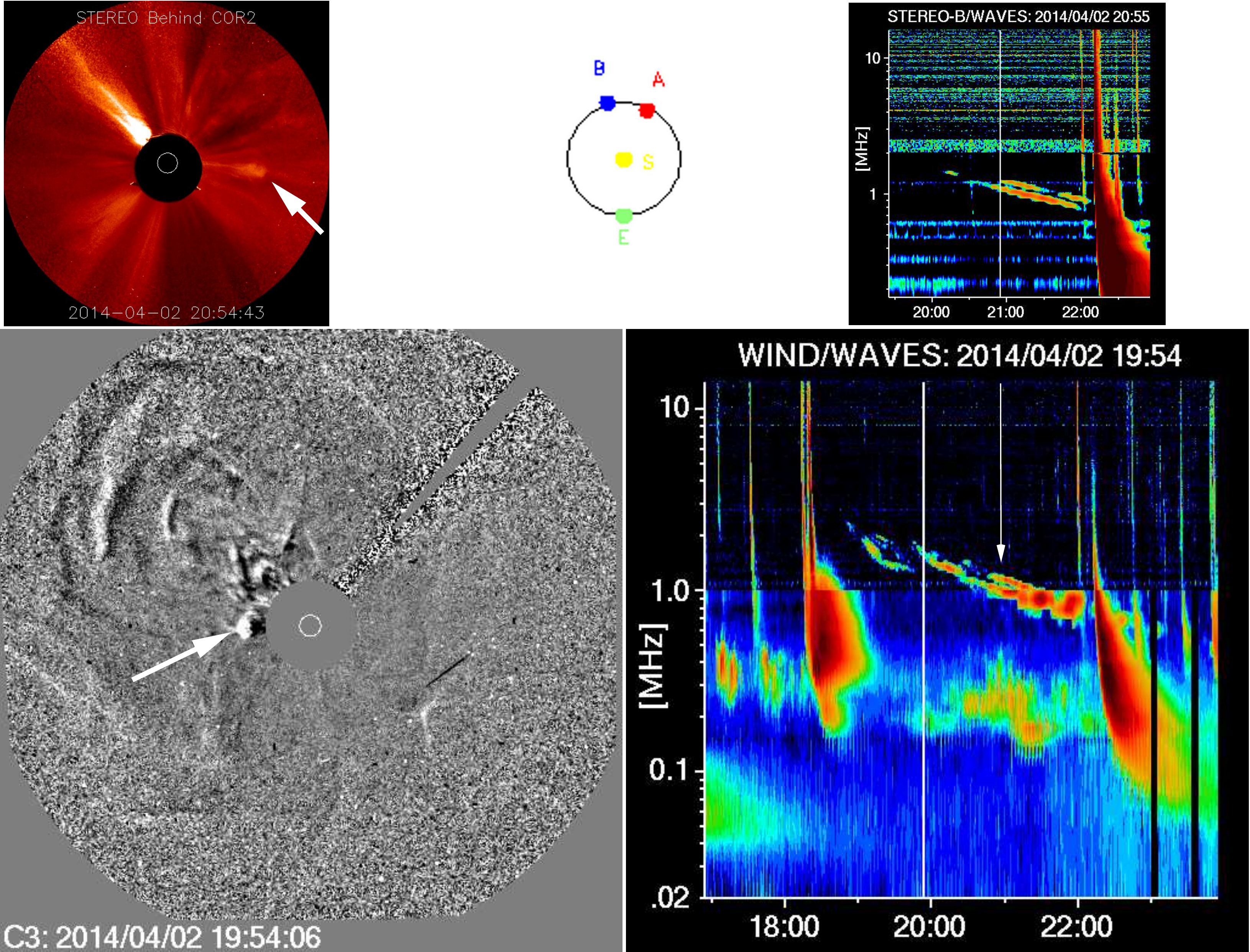}
 \caption{
       {\it Wind}-SOHO and STEREO-B observations on 2 April 2014. The active region associated with this event was located behind the north-eastern solar limb. Arrows point to the later CME structure (unlisted as a CME) that propagated inside the main CME. The fragments of the main CME are visible in the LASCO C3 difference image, north-east of the solar disk. STEREO-A observations are available only after 20:39 UT, and the recorded emission is similar to the STEREO-B observations. {\it Wind}/WAVES recorded the start of the type II burst 83 minutes before STEREO-B/WAVES.  
      }
  \label{comp20140402}%
\end{figure}

{\it 2 April 2014}. This event showed the longest gap in the starting time of the type II burst, 83 min, as {\it Wind} recorded the burst start at 18:49 UT at 2.4 MHz and STEREO-B at 20:12 UT at 1.4 MHz. The type II burst was formed near a time when a new bright front appeared inside the main CME. The height difference between this new CME front and the type II burst was large only near the beginning of the radio burst (Figure \ref{spectra20140402}, Appendix), as after 19:40 UT the heights were a good match.

The curved structures in the {\it Wind}/WAVES spectrum, near the burst start, resemble those reported in \cite{pohjolainen2008}. They found that the emission fragmentation was due to a shock passing through regions with highly variable densities, inside and outside high-density CME loops. Earlier launched CME loops on the shock propagation path would therefore be a good explanation for the emission curvature.

The type II burst was observed by STEREO-B only after 20:12 UT, {\it i.e.}, the curved structures were not observed (Figure \ref{comp20140402}). STEREO-B was located 164 degrees back from Earth orbit, almost directly on the opposite side of the Sun, so it should have had a good view to this new bright CME front. On the other hand, the earlier CME fragments were propagating toward north-east, {\it i.e.}, toward STEREO-B, so occultation was possible.

%%%%%%%%%%%%%%%%

\section{Coronal Conditions for Type II Burst Formation}
\label{conditions}

A large fraction of interplanetary shocks has been found to be radio quiet \citep{Gopalswamy2010}, and so even the appearance of a DH type II burst  requires some specific coronal conditions. The simplest scenario would be that a shock is formed when the speed of a coronal transient exceeds the local magnetosonic speed, {\it i.e.}, Alfv\'en speed in the solar corona, and this leads to type II radio emission at some location along the shock front.

\begin{figure}[h!]
   \centering
   \includegraphics[width=0.9\textwidth]{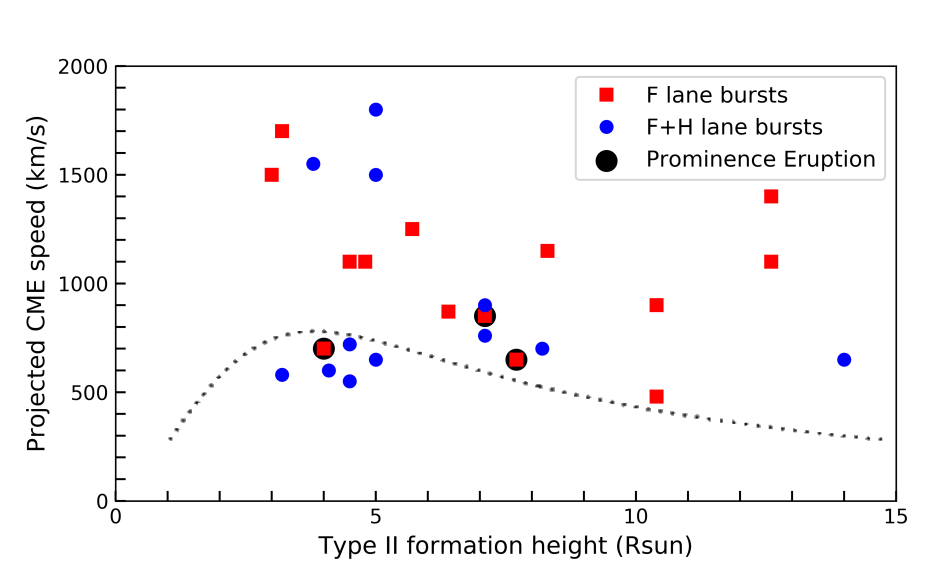}
   \caption{Type II burst formation height compared with the projected CME speed. The dotted line shows the local Alfv\'en speed, following the velocity curve presented in \cite{kim2012}. F and F+H lane bursts are plotted with different symbols and the three prominence eruption events are encircled.
   }
\label{alfvenspeed}%
\end{figure}

Figure \ref{alfvenspeed} shows the calculated DH type II burst formation heights compared with the projected CME speeds. The dotted line in the plot shows how the Alfv\'en speed rises with distance from the Sun above an active region, and reaches a local maximum of $\sim$800 km s$^{-1}$ near $\sim$4~R$_{\odot}$ as in \cite{kim2012}. Even a higher value for the Alfv\'en speed, 1100~km~s$^{-1}$ at 3.5 R$_{\odot}$ as in \cite{warmuth2005}, would not change the distribution of our events in most cases (below or above the Alfv\'en speed). 

In six events the CME speed looks to be less than the assumed Alfv\'en speed. All six type II bursts were estimated to be due to CME bow shocks, as the type II heights matched well with the CME leading front heights, within the reasonable height calculation error. The height difference was 0.1 - 0.5 R$_{\odot}$ in the four events that were well below the Alfv\'en speed peak value, and 1.0 and 1.2 R$_{\odot}$ in the two events that were closer to the Alfv\'en speed value (the CME speeds in these two events were 700 and 720 km s$^{-1}$). The only F lane event among the six was a prominence eruption. 

The four F+H lane type II bursts that had their CME speeds well below the modelled Alfv\'en speeds, had their source origin in similar locations on the solar disk, at longitudes E62, W63, W65, and W85. The CME speeds can be affected by projection effects in on-the-disk events, but a CME speed difference of 100 - 250 km s$^{-1}$ to the Alfv\'en speed cannot be explained with projection effects only. As these events were determined to be bow shock events, we conclude that the Alfv\'en speed had to be much lower in these coronal regions. This would be possible, for example, above some active regions and streamer regions \citep{evans2008}. But, for example \cite{Su2016} suggested that the generation of shocks, and type II bursts, may require larger values of compression ratio and Alfv\'en Mach number, rather than simply a higher speed of the disturbance.

Already \cite{roberts59} reported to have observed fundamental and harmonic emission in 60\% of all type II bursts. 
The intensity of the fundamental emission band has been observed to increase relative to the harmonic, as the burst evolves with heliocentric distance \citep{lengyel89}.
The fundamental and harmonic frequencies are assumed to be products of three-wave interaction processes of beam-excited Langmuir waves. Langmuir waves would first have to scatter off sound waves at a sufficient rate to create a suitable population of counter-propagating Langmuir waves, and if this process was too slow, harmonic emission would not be able to attain intensities comparable to the fundamental plasma emission process \citep{ganse2012a}.

Our correlation analysis showed that harmonic emission was present only in type II bursts that originated at longitudes larger than $\pm$60 degrees. Our small sample of events, where six events originated inside the longitude range of E60 - W60 and showed one emission band, may contain selection effects. The overall ratio, 14 events with F band and 12 events with F+H band looks to be in agreement with previous studies.

\section{Results and Discussion}

We have analysed 26 well-separated, isolated DH type II radio bursts and compared their characteristics with active region, flare and CME data. Association to metric type II emission and SEP (proton) events were also listed. First, we looked at the type II burst appearances and found bursts that showed a single emission lane (14 events, emission at the fundamental plasma frequency), and bursts with two lanes (12 events, fundamental and harmonic emission). 

From the 26 events 8 had their source origin on the far side of the Sun, and 7 were limb events (longitudes 85 - 90 degrees east or west). Of the remaining 11 events, 6 were located near the limb and 5 were located nearer to the disk center. No isolated DH type II bursts were observed inside the longitude range E30 - W40. At longitudes E30-50 and W40-50 all the isolated type II bursts were single lane bursts (no harmonic emission).  

The fastest CMEs associated with the DH type II bursts were observed in events that originated at or behind the solar limb. This is understandable, as height measurements of on-the-disk CMEs may suffer from projection effects. A similar result was obtained by \cite{vasanth2013}. The associated CMEs had variable widths, in the range of 40 - 230 degrees near the time of the DH type II burst start. The narrowest CMEs were in events that occurred at or behind the solar limb. The closer to the disk center the event originated, the larger the CME angular width. The largest angular widths were associated with low-frequency start of emission ({\it i.e.}, high formation height of the type II burst), although low-frequency onset of emission occurred also in events where the CME had very narrow width.

Of the 26 events, 16 were found to have a good match between the DH type II burst height and the CME leading front height, within the possible $\pm$2 R$_{\odot}$ error. In other five events we could also make a match to the CME leading front if the whole radio emission lane and later times along the type II lane were considered for the height calculation. These type II bursts could be due to CME bow shocks, located near the CME nose. We note that in several cases the radio burst height was closer to the bright front structure of the CME, and not to the diffuse region above it (if it existed) that is often taken as the CME height. Shock regions and CME structures are discussed, for example, in \cite{Kwon2018}, \cite{Song2019}, and \cite{Mei2020}, and references therein.  

Only five events showed a large height difference between the DH type II shock and the CME leading front, and these events were analysed in detail. For the DH type II burst that was located much higher than the CME leading front we found a possible explanation from two earlier CMEs that had comparable heights with the type II burst. This event was observed also with STEREO-A/WAVES, and the spectral features there suggest that the type II-like bursts could also have been radio enhancements along type III burst lanes, formed by electron beams passing through the earlier-launched CME material.

For the four DH type II bursts that showed heights much lower than the CME leading front we found the following explanations. The two DH type II bursts with source origins at longitudes W42 and E46 and emission only at the fundamental frequency, a CME flank shock looks to be the most probable explanation. In one event a later-launched CME structure, inside the main CME, was visible and the curvature of the type II lane matched with the height evolution of the later CME. And for one event, a fast sideways movement of a lower CME structure, creating a short-duration shock, may have been the cause for the type II burst. 

Our sample of isolated DH type bursts where most bursts look to be due to shocks near the CME leading front, contradicts earlier findings. DH type II bursts, using for example radio triangulation method, have in most studies been found to be located near the CME flank regions, and only in few cases have been located close to the CME nose, see \cite{jebaraj2020} and references therein. 
A statistical study of 153 interplanetary type II radio bursts observed by the two STEREO spacecraft in 2008-2014 \citep{krupar2019} 
also suggested that interplanetary type II bursts are more likely to have a source region situated closer to the CME flanks than the CME leading edge. The radio bursts located near the CME flanks have also been noted to be enhanced by shock-streamer interaction \citep{shen2013}.

The results of \cite{kahler2019} suggested that fast and narrow CMEs can create mostly bow shocks, and fast and wide CMEs predominately expansion shocks. Also \cite{vourlidas2009} have suggested that the bow shock morphology is associated almost exclusively with narrow CMEs. \cite{kahler2019} further concluded that no CMEs with widths less than 60$^{\circ}$ are associated with metric type II bursts. Our study confirms this finding, but we point out that CMEs with widths less than 60$^{\circ}$ can still be associated with DH type II bursts. 

Of our 26 DH type II burst events, 10 were associated with earlier metric type II bursts. All 10 metric bursts died out before reaching the DH instrumental limit of 16/14 MHz, and the later-appearing DH bursts were isolated. We note that the metric type II bursts were formed only with the fastest CMEs, with a CME velocity range of 1100~- 1800~km s$^{-1}$. It has been suggested earlier that metric type II radio bursts are formed by a distinct coronal shock and only produce radio emission in the low corona, see, {\it e.g.}, \cite{reiner2000} and \cite{magara2000}, but opposite arguments have also been made, see \cite{Gopalswamy1999} and references therein. 

In three of our events the start time of the DH type II burst matched well with the solar energetic particle (SEP) release time for protons in the 55-80 MeV range. In two SEP-associated events earlier metric type II emission was observed, but 21 of the isolated DH type II bursts were not associated with listed proton events. 

Comparison between the {\it Wind} and STEREO observations, available for 12 events, showed that the DH type II burst start was not always observed toward all directions. There was a delay in the starting times in five events. In two events only {\it Wind} recorded the DH type II burst, and both were eruptive prominences with source longitude at E30. Of the five events that showed time delays, two seemed to suffer from sensitivity issues in the radio spectral data. Less intense emission near the DH type II start could therefore go unnoticed. In the other three events, occultation by a dense streamer or earlier-launched CME fragments looks possible. Radio emission from the DH type II burst would become visible only when the type II burst frequency exceeds the plasma frequency of the occulting structure, {\it i.e.}, when the density of the occulting structure is less than in the type II shock region. 

In Section \ref{conditions} we discussed the required conditions in the corona for the development of a shock wave. It is well known that not all coronal shocks produce DH type II bursts. In most of our analysed events, the CME speed looked to be well above the local Alfv\'en speed, and therefore type II emission would be possible. Some of the DH type II bursts had very short duration, which indicates that either the emission ended or it became unobservable (invisible). In general, the DH type II burst duration was found to depend on the type II burst start frequency, {\it i.e.}, the lower the start frequency the longer the duration.

Type II bursts have a quite narrow emission band (narrow range of densities), and therefore the emission region cannot be spread out along a large area of the (curved) shock front. The small-scale type II emission region is thus expected to have special properties, setting it apart from the rest of the shock \citep{ganse2012b}.
A specific local region would also provide the conditions for the enhancement of emission near the plasma frequency harmonics \citep{annenkov2020}.

\section{Conclusions}

We conclude that isolated DH type II bursts can be narrow-band or wide-band, can have short or long duration, and can show only fundamental or both fundamental and harmonic emission. The bursts typically have their source origin some distance away from the central meridian, at longitudes 30-90 degrees east or west, but only some of the bursts originate from the far side of the Sun.

Almost all of the analysed 26 isolated DH type II bursts could be associated with a shock near the leading front of the main CME. In only two events a shock near the CME flanks was estimated to be the cause of the DH type II burst, and in three events the radio bursts could be explained by earlier or later CME structures. Our analysis suggests that for the  majority of isolated DH type II bursts a CME bow shock is more probable than a CME flank shock. 

We also analyzed the radio spectral data from different viewing angles in 12 events, provided by the STEREO and {\it Wind} spacecraft observations. We compared the spectral features of the DH type II bursts, and their timing. In five events a time delay in the DH type II start was evident, and only one of the five was a far side event. In three events occultation by dense matter between the source and the observer could have been the cause for the delay in the emission onset, and in two events spectral sensitivity could have affected the observed starting time.

Why do some DH type II bursts appear isolated, without other radio emission signatures, and why do many of them disappear from the spectrum as suddenly as they appear? Our findings suggest that isolated DH type II bursts could be a special subgroup within DH type II bursts, where the radio emission requires particular coronal conditions to form, and the bursts can die out rapidly if these conditions change.

\begin{acks}
The authors want to thank all the individuals who have contributed in creating and updating the various solar event catalogues. The authors also acknowledge the useful comments and suggestions from the anonymous referee, on how to improve the paper. The CME catalogue is generated and maintained at the CDAW Data Center by NASA and the Catholic University of America in cooperation with the Naval Research Laboratory. The {\it Wind}/WAVES radio type II burst catalog was initially prepared by Michael L. Kaiser and it is maintained at the Goddard Space Flight Center. SOHO is a project of international cooperation between ESA and NASA. STEREO is part of the NASA Solar Terrestrial Probes (STP) Program.
\end{acks}

\vspace{5mm}

{\bf Disclosure of Potential Conflicts of Interest}
The authors declare that they have no conflicts of interest.

%%%%%%%%%%%%%%%%%%%%%%%%%%%%%

%%%%%%%%%%

\pagebreak
\clearpage
\newpage

\appendix

\section{Event images}

For each event we show a coronagraph-EUV difference image (SOHO/LASCO - SOHO/EIT),
a {\it Wind}/WAVES radio dynamic spectrum that shows the DH type II burst, and
a CORIMP map or a coronagraph image (LASCO or STEREO) that shows the streamer
locations. In the radio dynamic spectra the white vertical lines indicate the time
of the coronagraph image, and white arrows point to the type II burst structures
(one arrow for a single F lane event and two arrows labelled F and H if a harmonic 
lane is visible).
The white dashed lines in the radio spectra indicate CME heights converted
to frequencies, with the hybrid plasma density model by \cite{vrsnak2004}.
The LASCO C2-C3 image from the CORIMP catalog (or the coronagraph image)
is from a time close to the type II start or during the type II burst. In the
CORIMP images the yellow lines detail the detected CME structure, and the red
points correspond to the outer features that are used for tracking the CME
heights (see details in the CORIMP catalog at
\url{http://alshamess.ifa.hawaii.edu/CORIMP/}).  
EP stands for eruptive prominence.

\begin{figure}[H]
  \centering
  \includegraphics[width=0.65\textwidth]{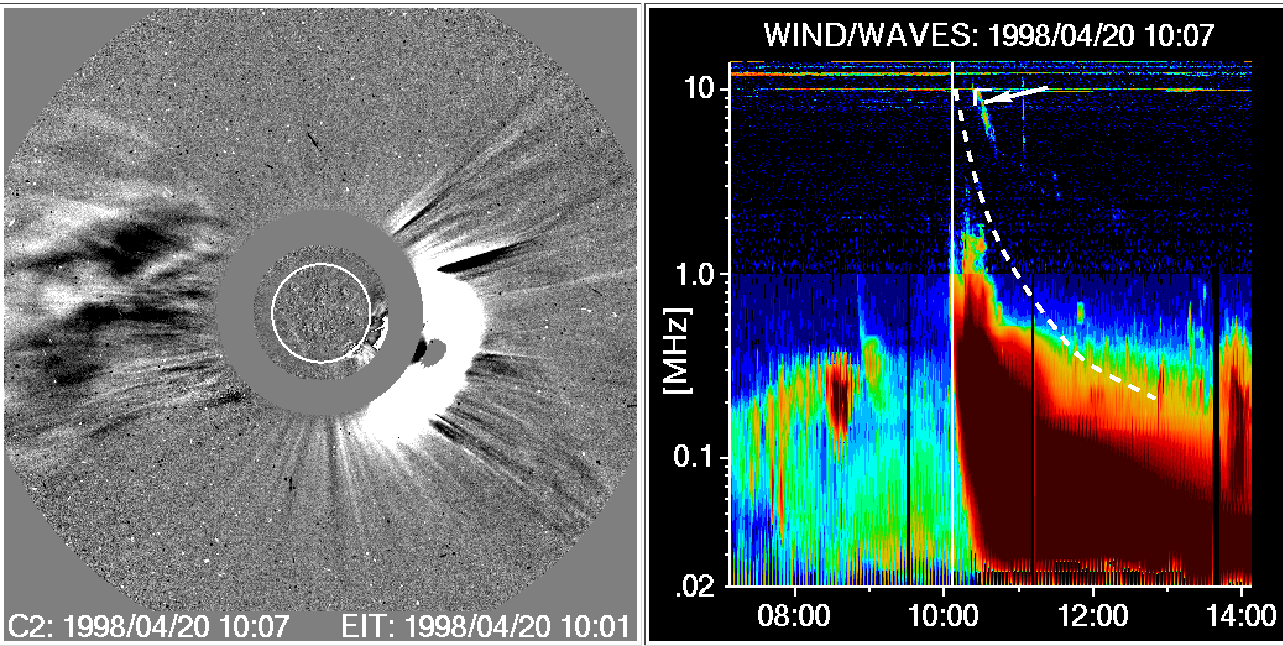}
  \includegraphics[width=0.32\textwidth]{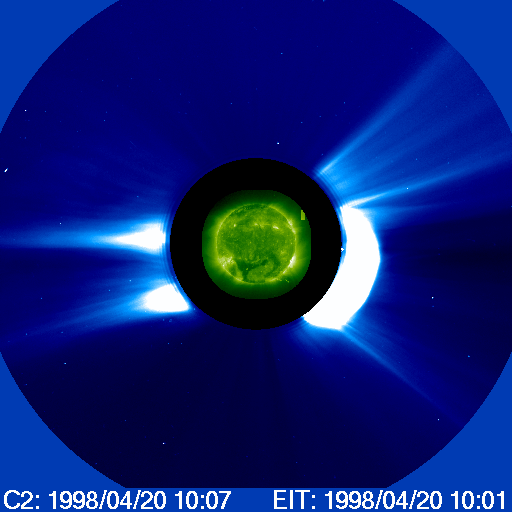}
    \caption{20 April 1998, source location S22W90.}
\label{spectra19980420}%
\end{figure}

\begin{figure}[H]
  \centering
  \includegraphics[width=0.65\textwidth]{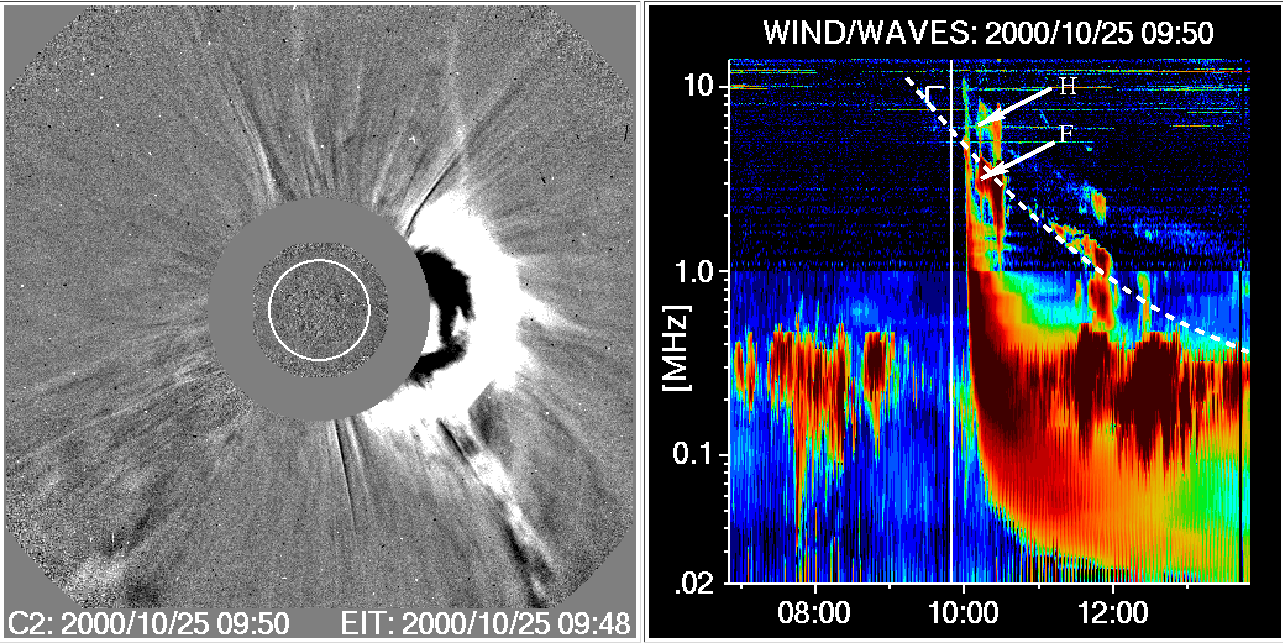}
  \includegraphics[width=0.32\textwidth]{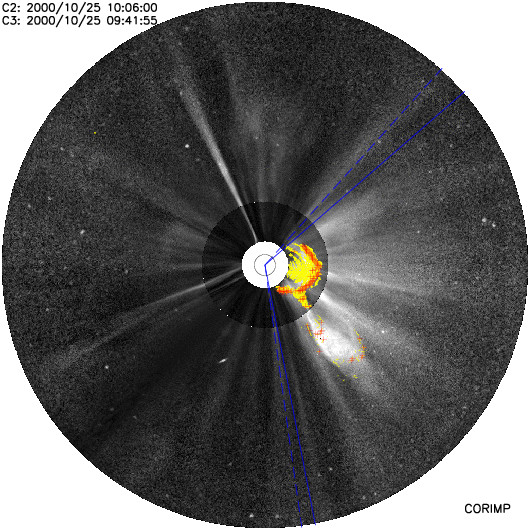}
\caption{25 October 2000, source location N09W63.}
\label{spectra20001025}%
\end{figure}

\begin{figure}[H]
  \centering
  \includegraphics[width=0.65\textwidth]{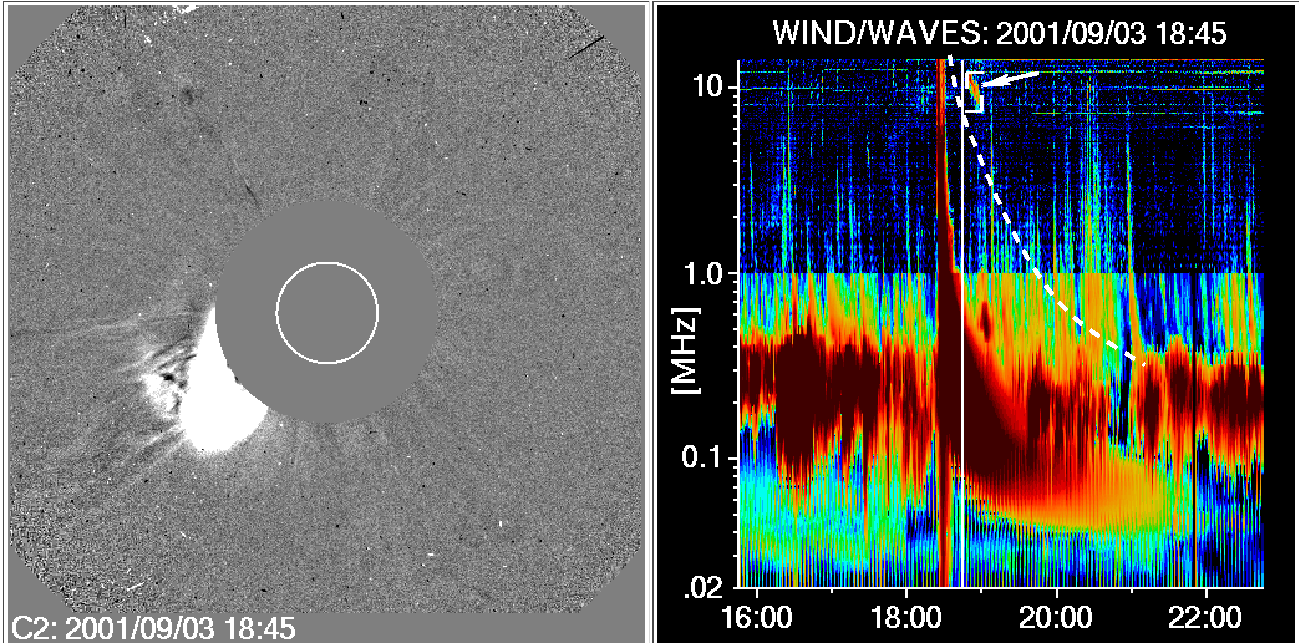}
  \includegraphics[width=0.32\textwidth]{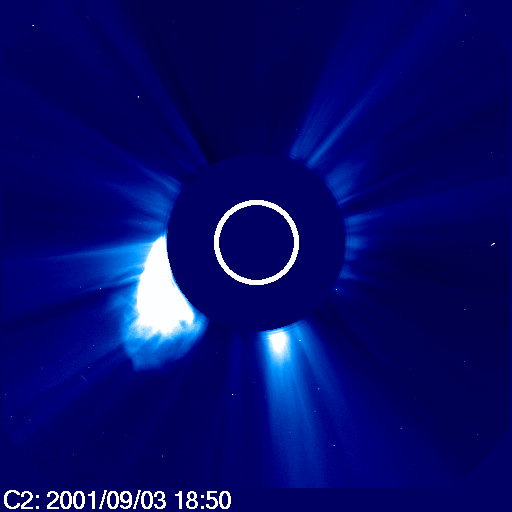}
  \caption{3 September 2001, source location S23E90.}
  \label{spectra20010903}%
\end{figure}

\begin{figure}[H]
  \centering
  \includegraphics[width=0.65\textwidth]{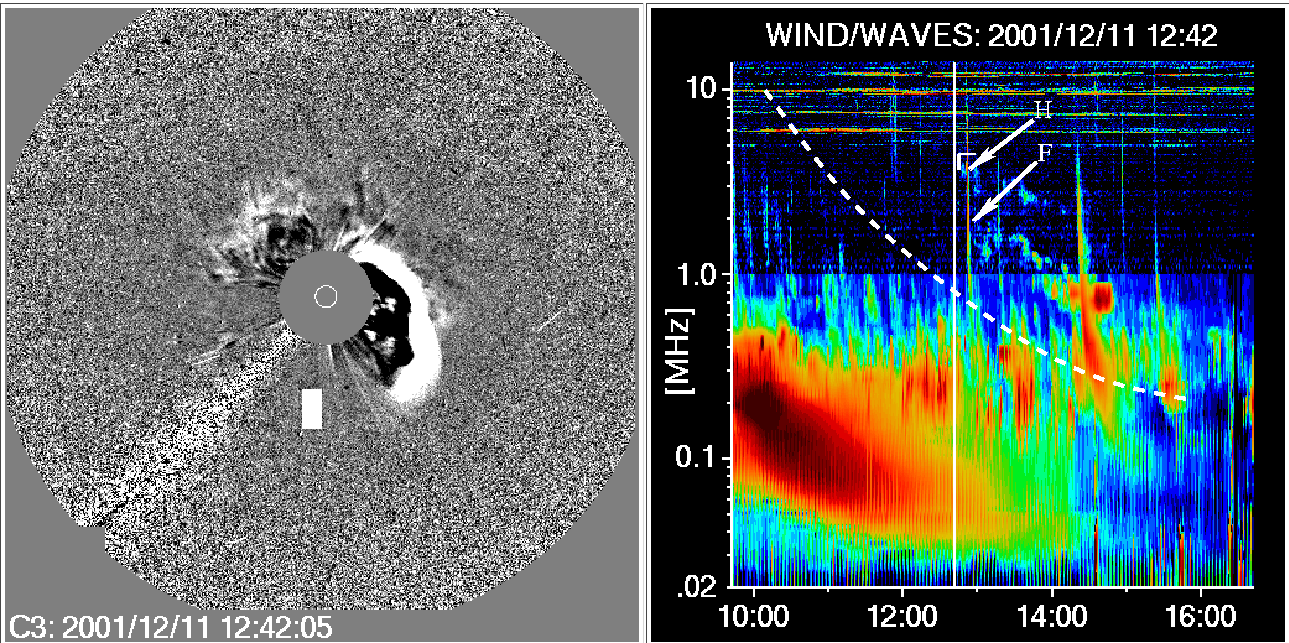}
  \includegraphics[width=0.32\textwidth]{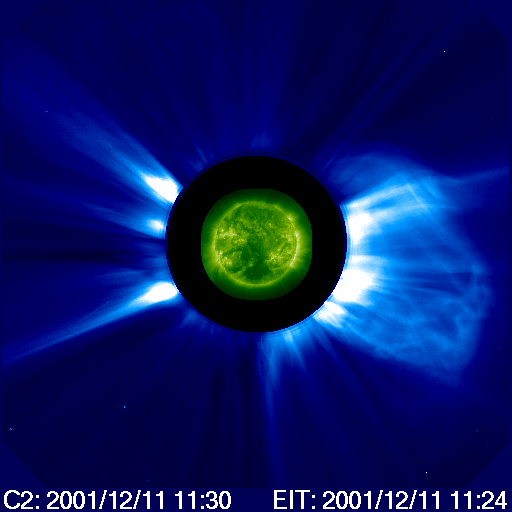}
  \caption{11 December 2001, source location SW90b.
  The DH type II burst was estimated to be located $\sim$6 R$_{\odot}$
  lower than the CME leading front, and 
  this event is analysed in more detail in Section \ref{event11122001}.}
  \label{spectra20011211}%
\end{figure}

\begin{figure}[!h]
  \centering
  \includegraphics[width=0.65\textwidth]{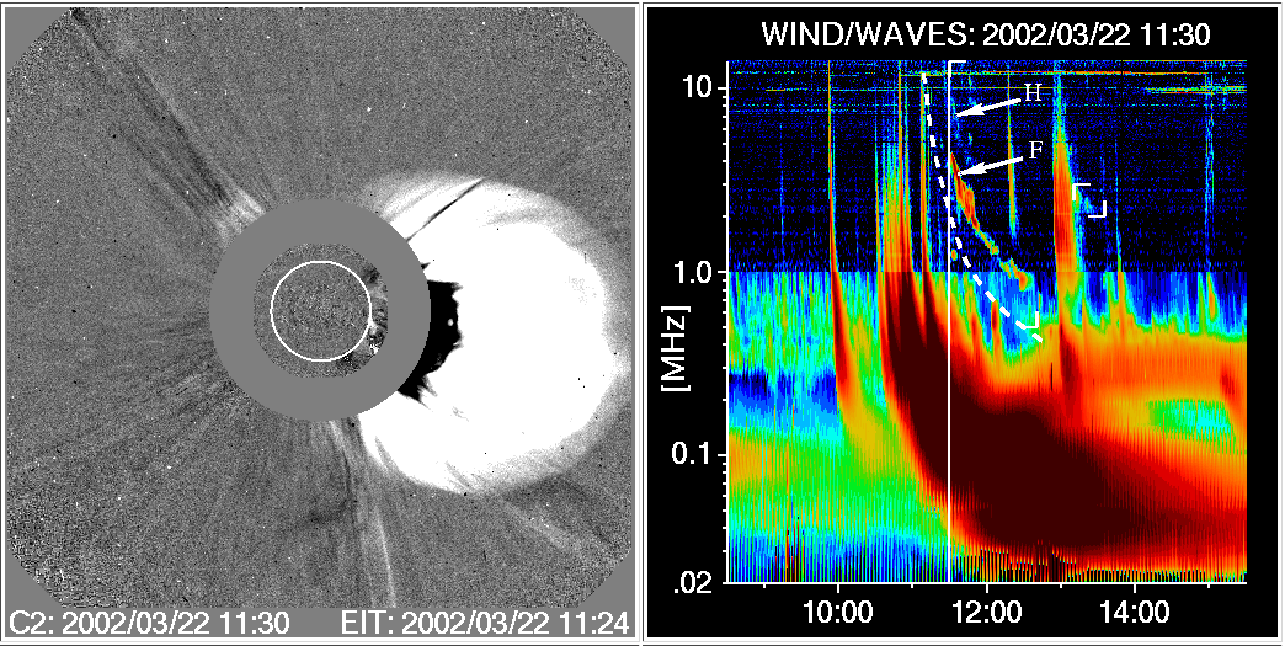}
  \includegraphics[width=0.32\textwidth]{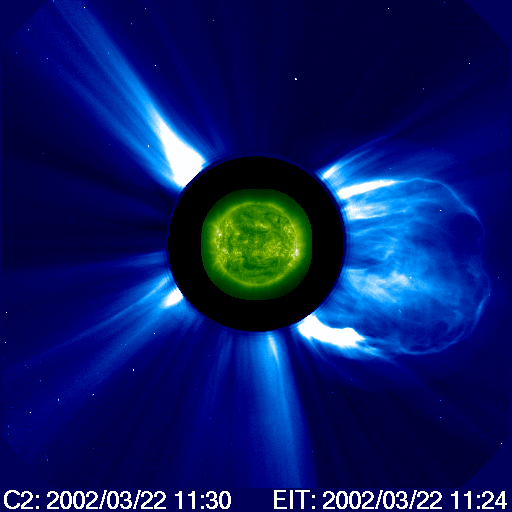} 
  \caption{22 March 2002, source location S09W90.}
  \label{spectra20020322}%
\end{figure}

\begin{figure}[H]
  \centering
\includegraphics[width=0.65\textwidth]{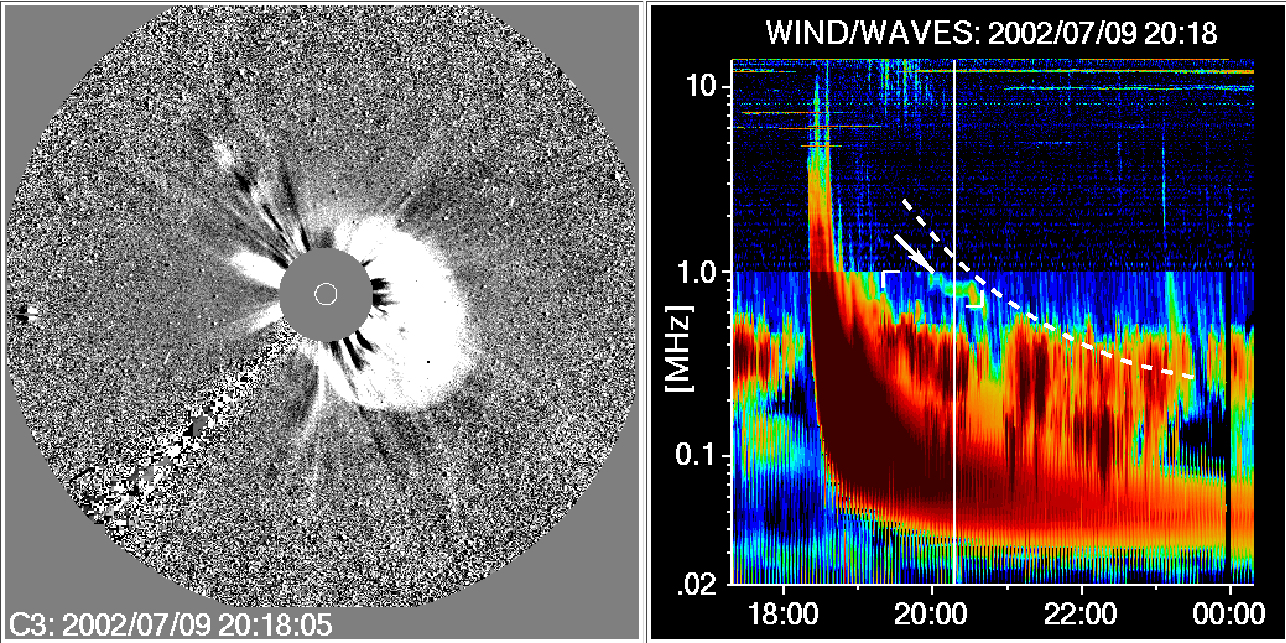}
\includegraphics[width=0.32\textwidth]{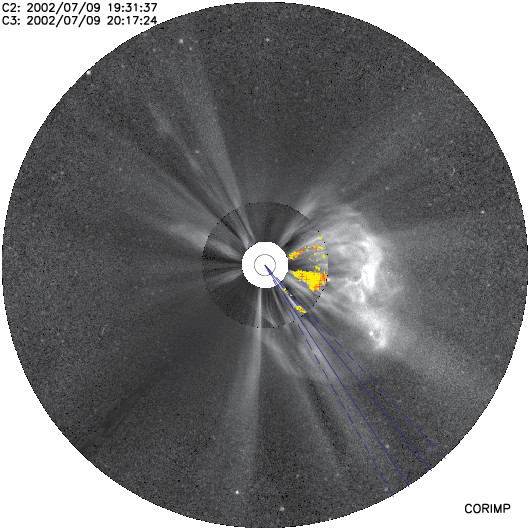}
    \caption{9 July 2002, source location W90b.}
\label{spectra20020709}%
\end{figure}

\begin{figure}[H]
  \centering
  \includegraphics[width=0.65\textwidth]{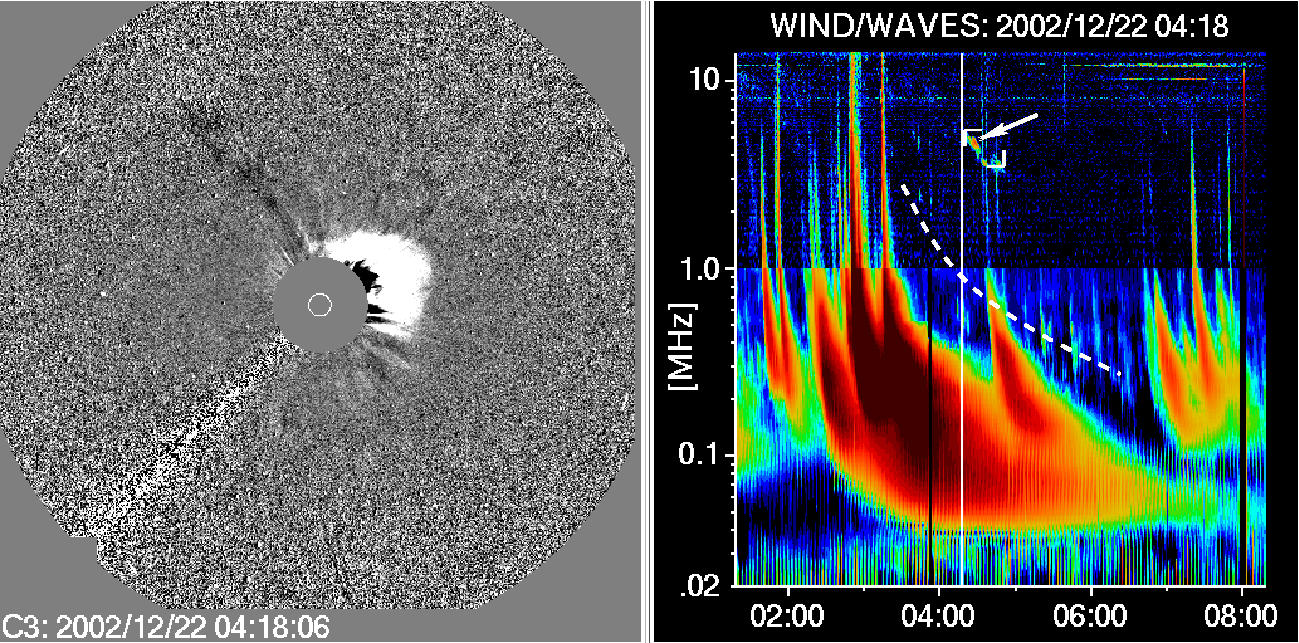}
  \includegraphics[width=0.32\textwidth]{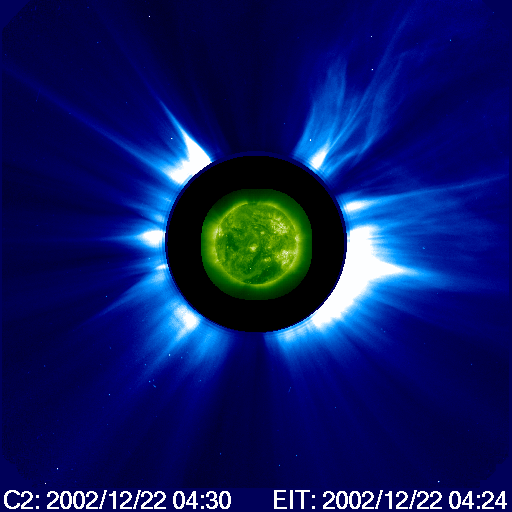}
  \caption{22 December 2002, source location N23W42. 
  The DH type II burst source was estimated to be located $\sim$7 R$_{\odot}$
  lower than the CME leading front, and  
  this event is analysed in more detail in Section \ref{events22122002and31122004}.}
  \label{spectra20021222}%
\end{figure}

\begin{figure}[H]
  \centering
\includegraphics[width=0.65\textwidth]{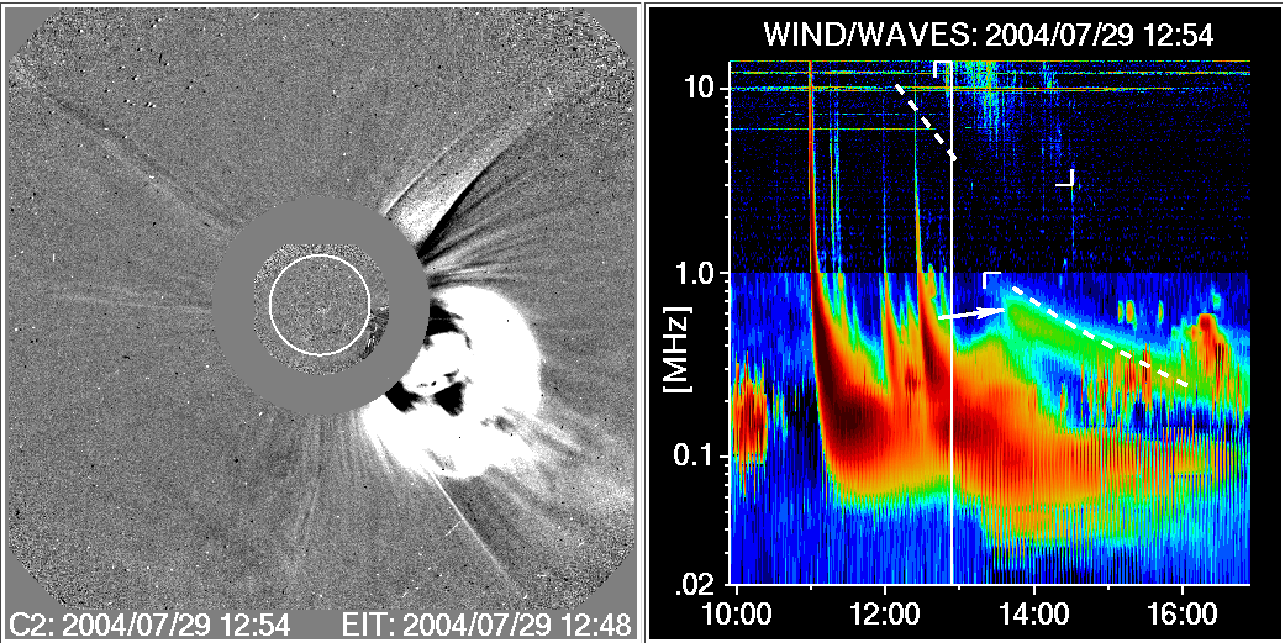}
\includegraphics[width=0.32\textwidth]{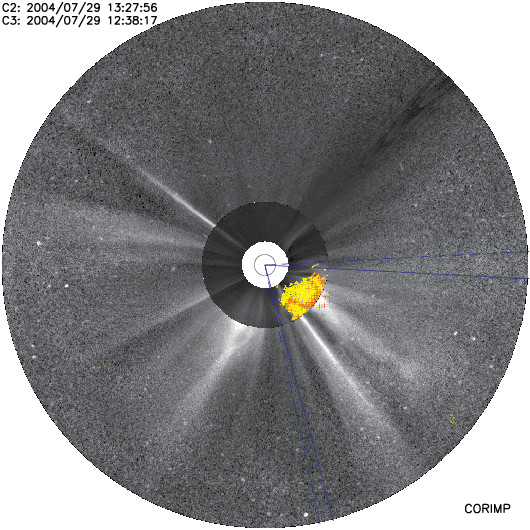}
\caption{29 July 2004, source location N00W90.}
\label{spectra20040729}%
\end{figure}

\begin{figure}[H]
  \centering
  \includegraphics[width=0.65\textwidth]{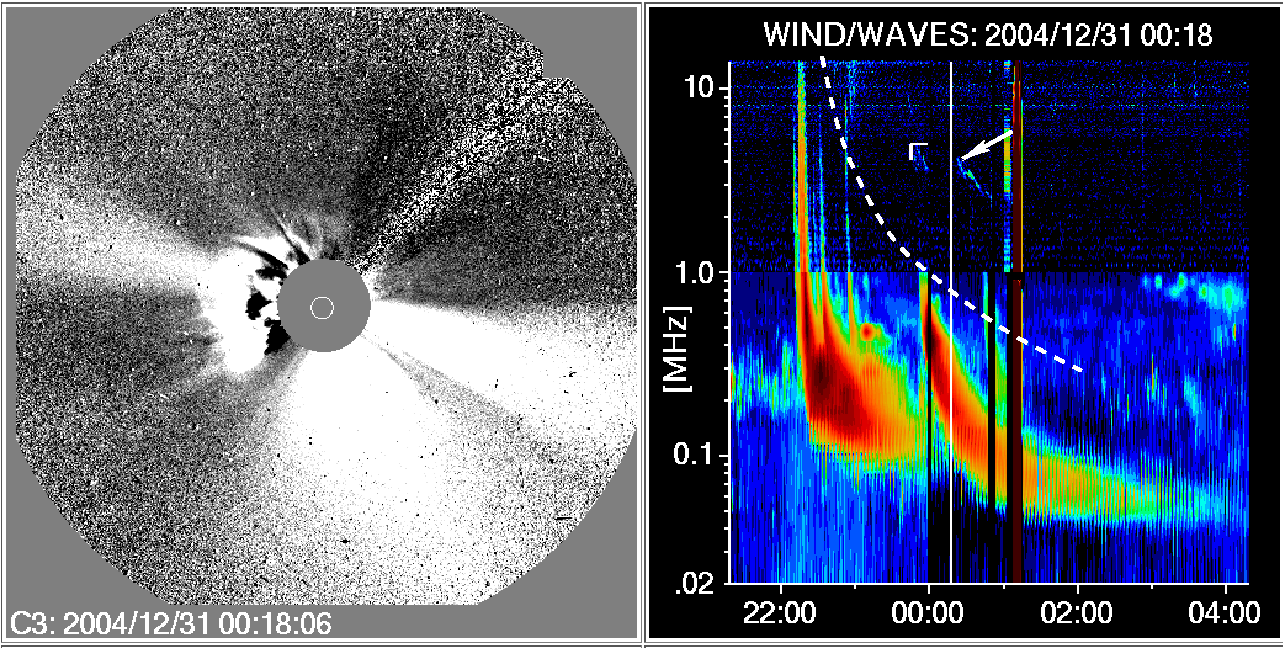}
  \includegraphics[width=0.32\textwidth]{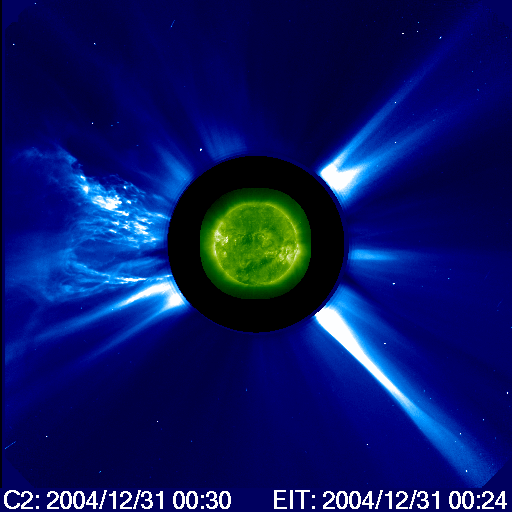}
  \caption{31 December 2004, source location N04E46.
  The DH type II burst source was estimated to be located $\sim$7 R$_{\odot}$
  lower then the CME leading front, and 
  this event is analysed in more detail in Section \ref{events22122002and31122004}.}
  \label{spectra20041231}%
\end{figure}

\begin{figure}[H]
  \centering
  \includegraphics[width=0.65\textwidth]{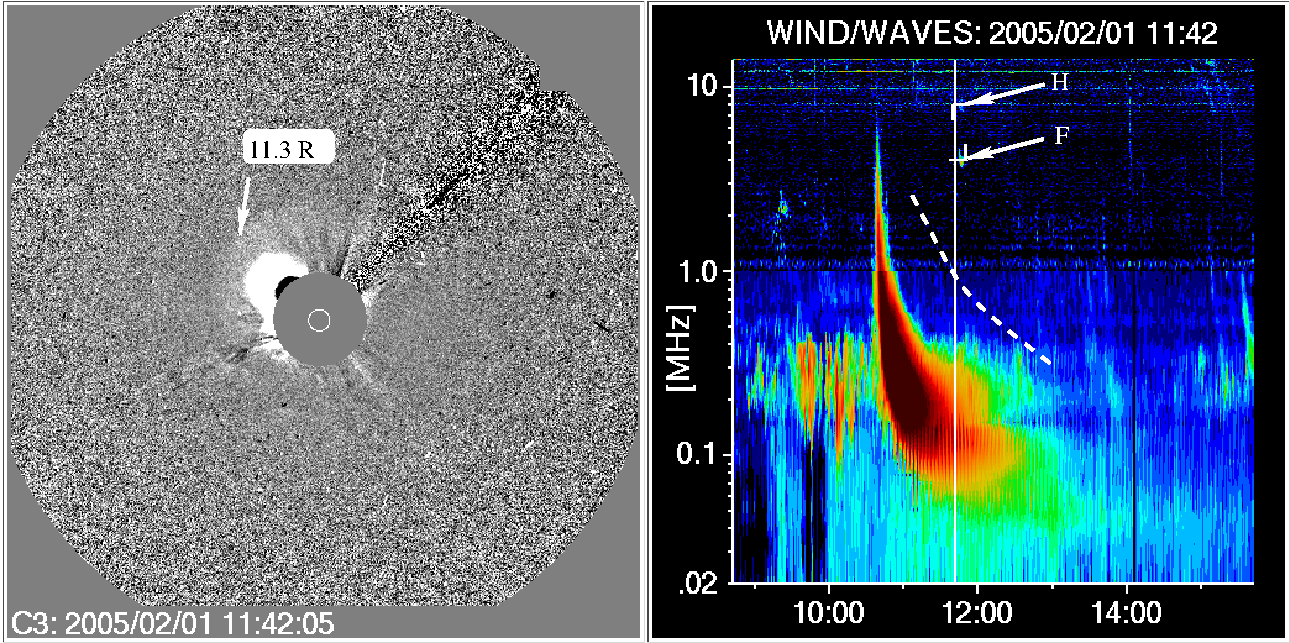}
  \includegraphics[width=0.32\textwidth]{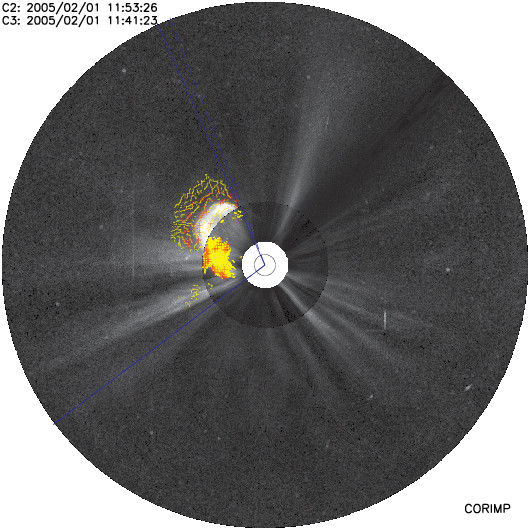}
  \caption{1 February 2005, source location NE90b. 
  The DH type II burst was estimated to be located $\sim$6 R$_{\odot}$ lower 
  than the CME leading front, 
  and this event is analysed in more detail in Section \ref{event01022005}.}
  \label{spectra20050201}%
\end{figure}

\begin{figure}[H]
  \centering
\includegraphics[width=0.65\textwidth]{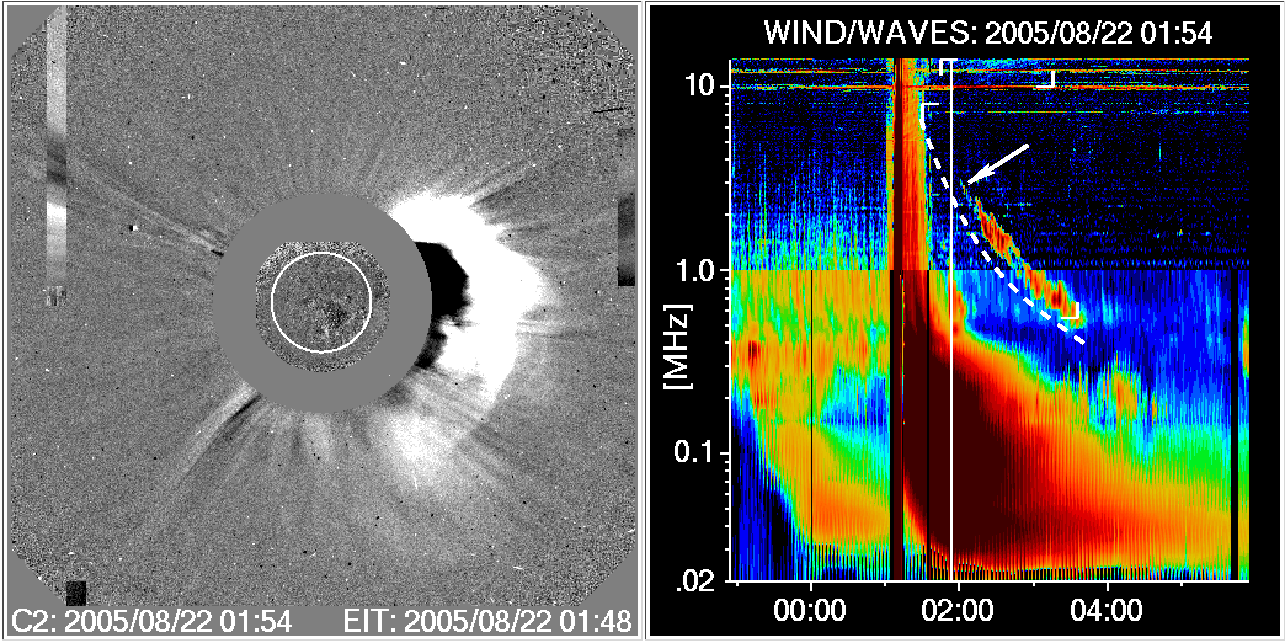}
\includegraphics[width=0.32\textwidth]{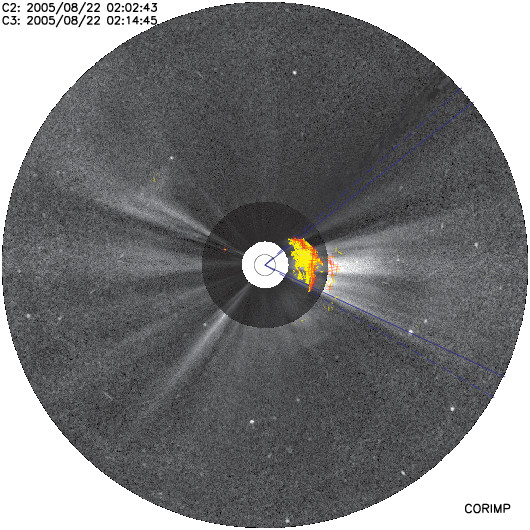}
    \caption{22 August 2005, source location S11W54.}
\label{spectra20050822}%
\end{figure}

\begin{figure}[H]
  \centering
  \includegraphics[width=0.65\textwidth]{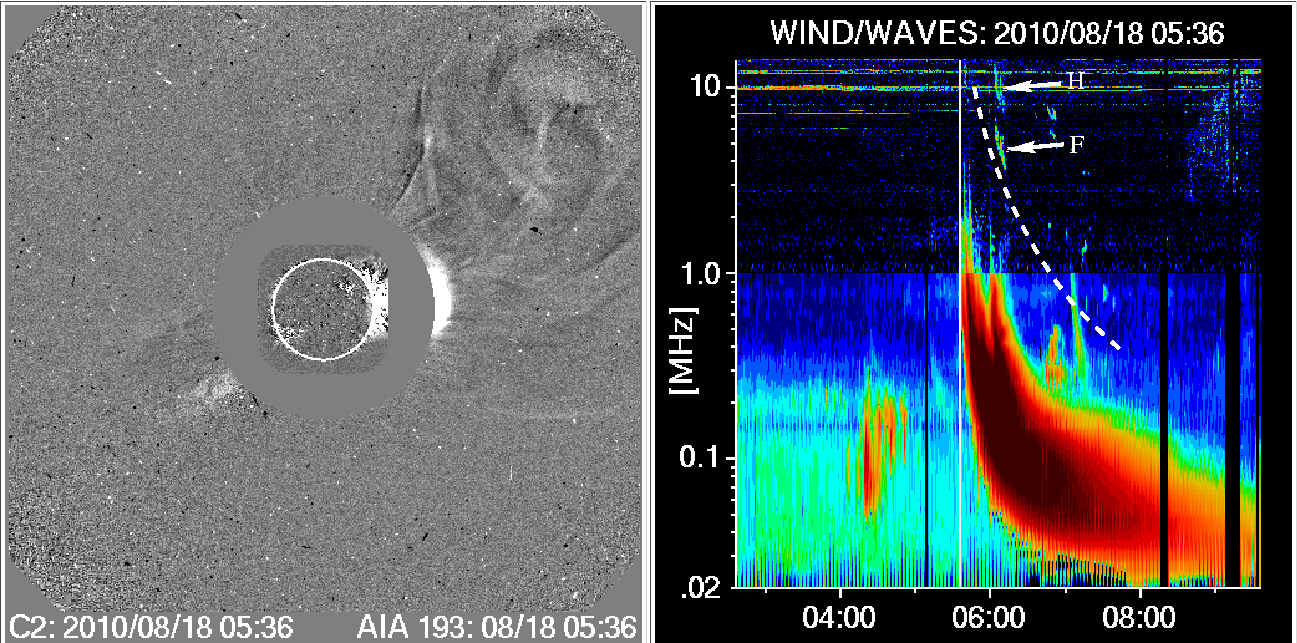}
  \includegraphics[width=0.32\textwidth]{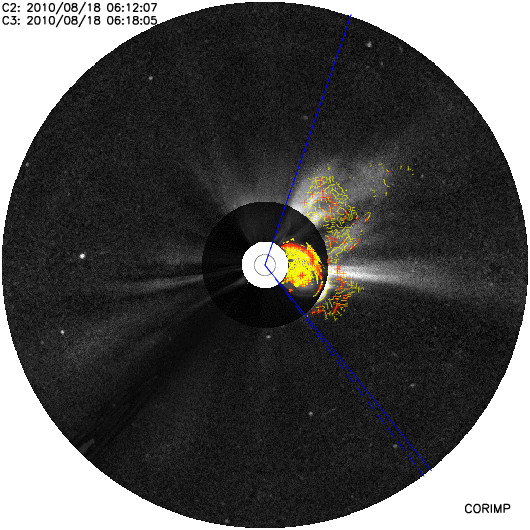}
  \caption{18 August 2010, source location N18W88.}
  \label{spectra20100818}%
\end{figure}

\begin{figure}[H]
  \centering
  \includegraphics[width=0.65\textwidth]{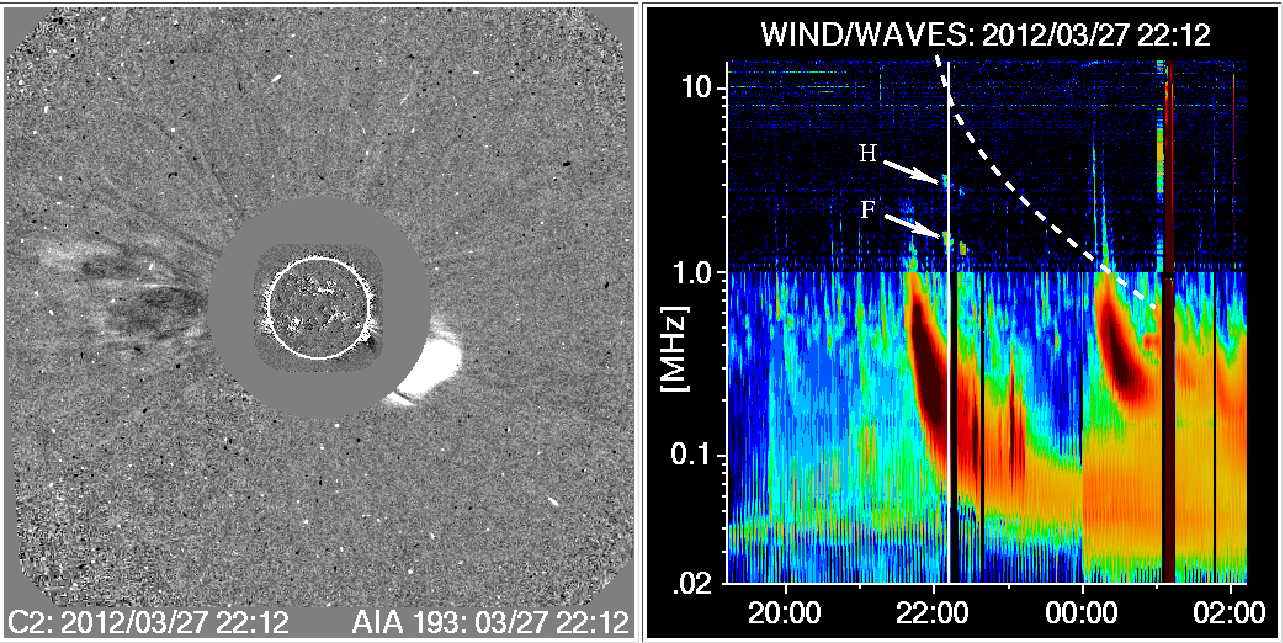}
  \includegraphics[width=0.32\textwidth]{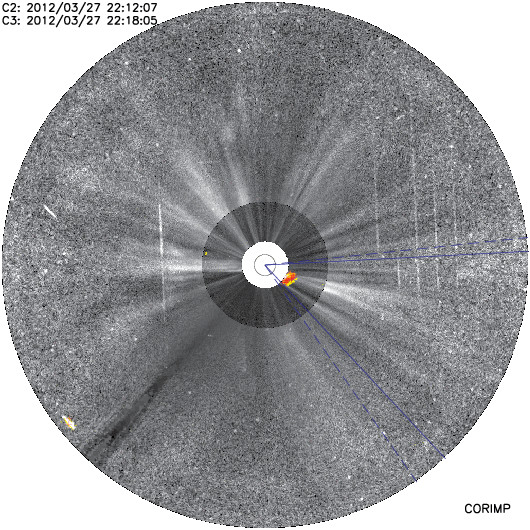}
  \caption{27 March 2012, source location SW90b.
  The leading front of the CME was located $\sim$5 R$_{\odot}$ lower than
  the DH type II burst source, and 
  this event is analysed in more detail in Section \ref{event27032012}.}
  \label{spectra20120327}%
\end{figure}

\begin{figure}[H]
  \centering
\includegraphics[width=0.65\textwidth]{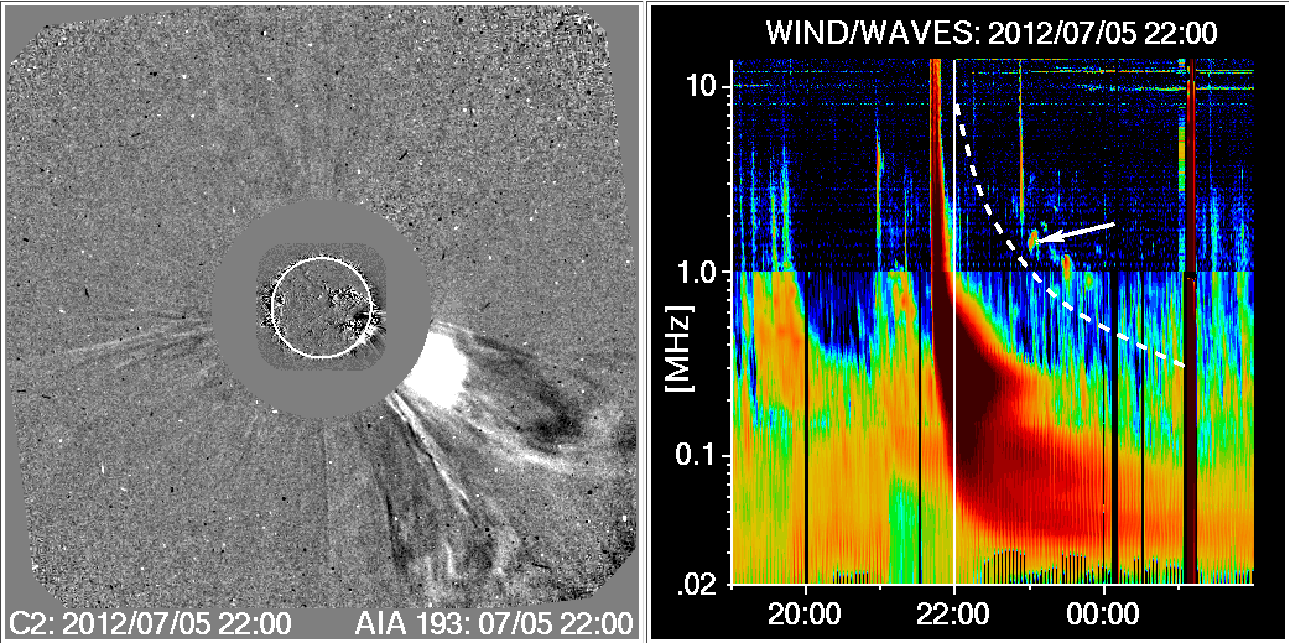}
\includegraphics[width=0.32\textwidth]{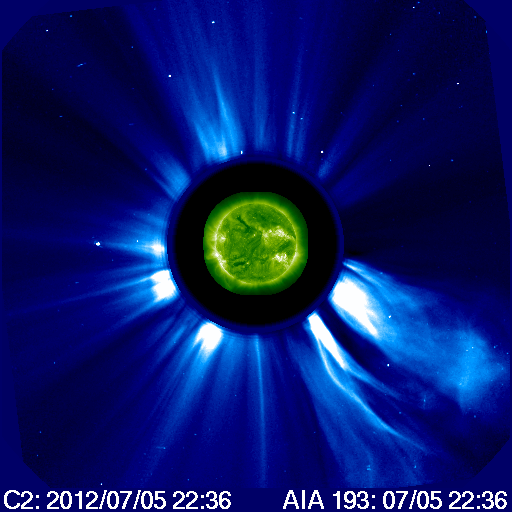}
    \caption{5 July 2012, source location S12W46.
    This event is discussed in more detail in Section \ref{stereocomp} (delay in
    DH type II burst start, STEREO observations).
    }
\label{spectra20120705}%
\end{figure}

\begin{figure}[H]
  \centering
  \includegraphics[width=0.65\textwidth]{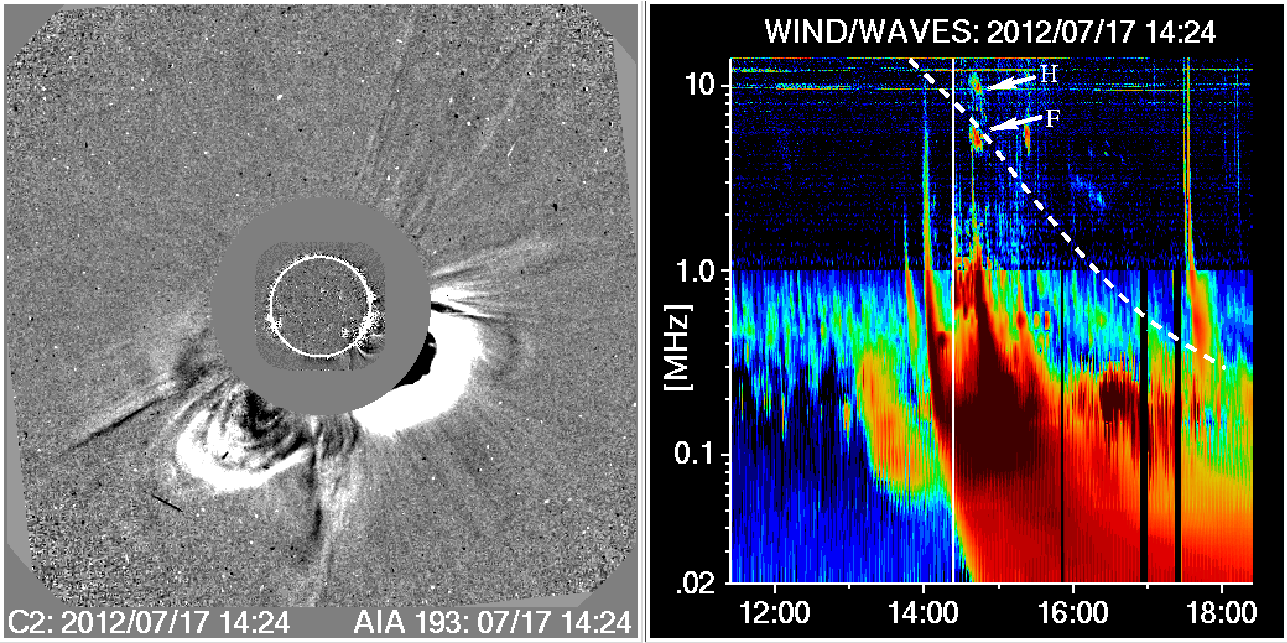}
  \includegraphics[width=0.32\textwidth]{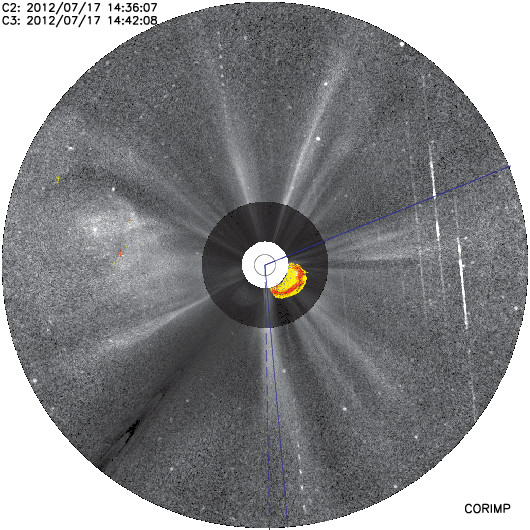}
  \caption{17 July 2012, source location S28W65.}
  \label{spectra20120717}%
\end{figure}

\begin{figure}[H]
  \centering
\includegraphics[width=0.65\textwidth]{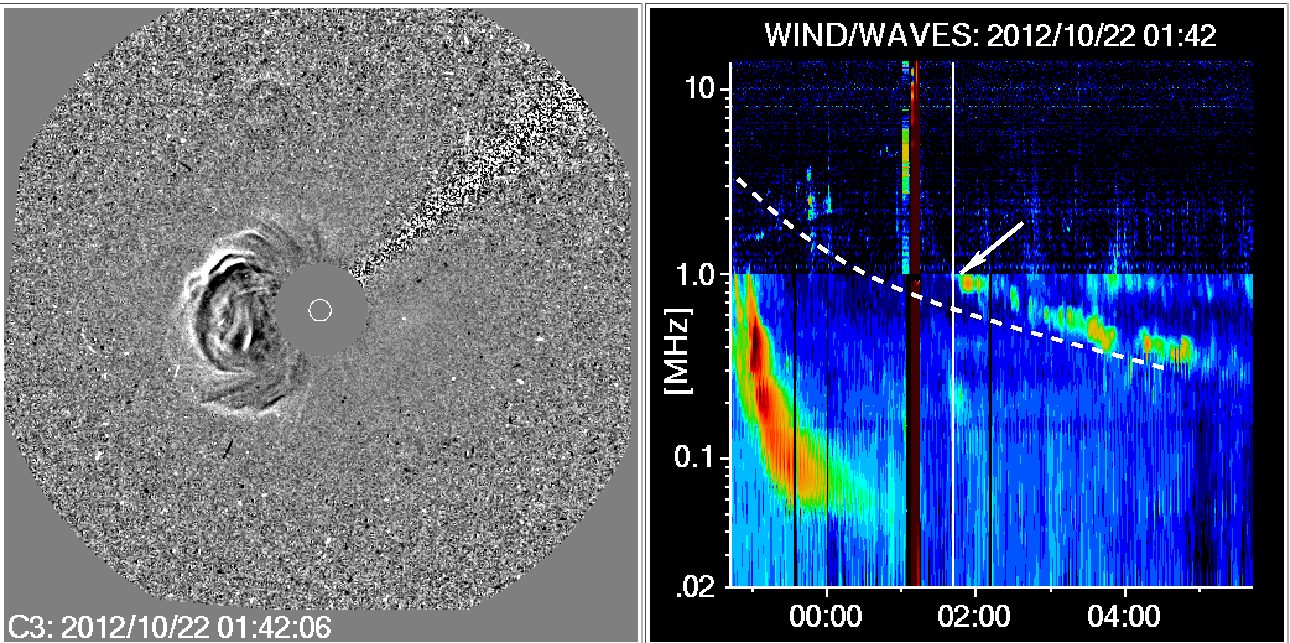}
\includegraphics[width=0.32\textwidth]{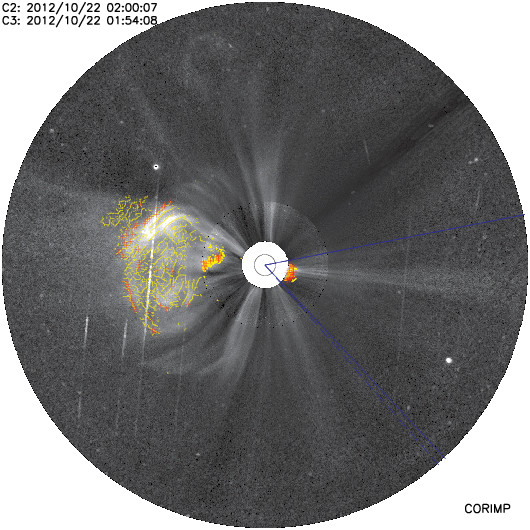}
\caption{22 October 2012, source location S10E76.
  This event is discussed in more detail in Section \ref{stereocomp}
  (delay in DH type II burst start, STEREO observations).
}
\label{spectra20121022}%
\end{figure}

\begin{figure}[H]
  \centering
  \includegraphics[width=0.65\textwidth]{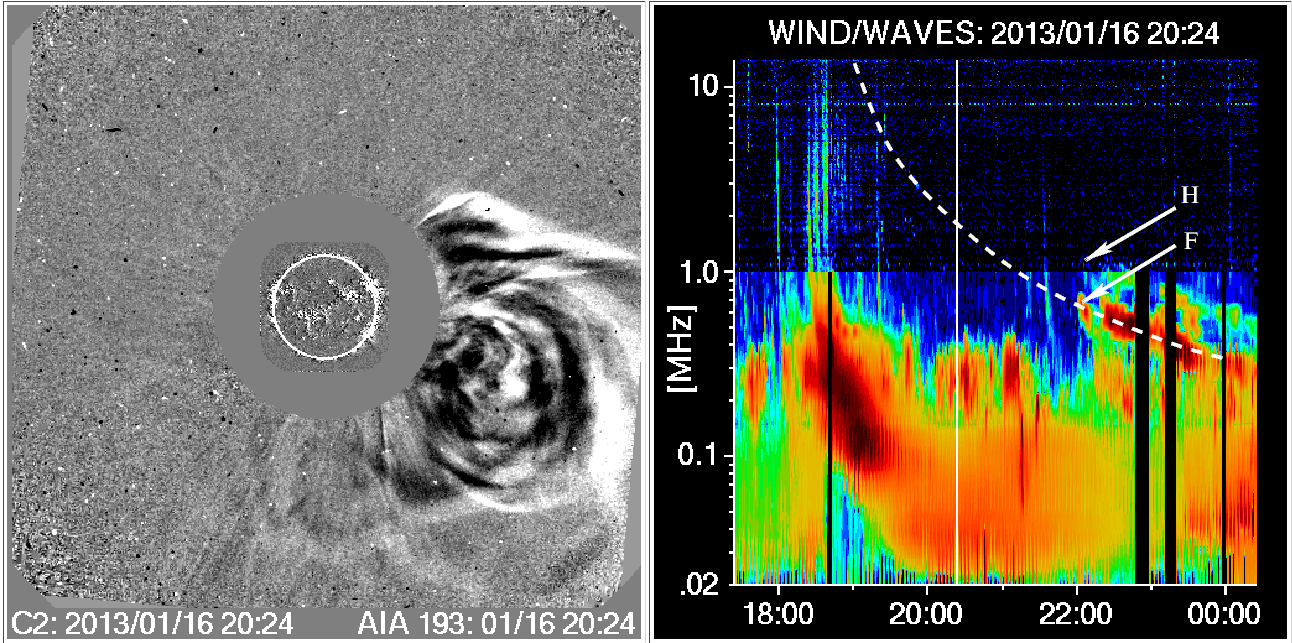}
  \includegraphics[width=0.32\textwidth]{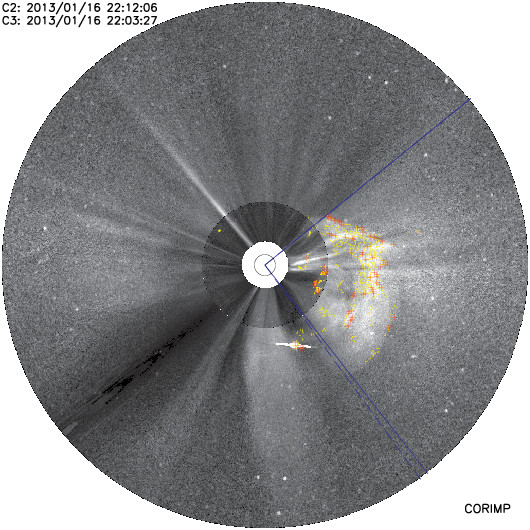}
  \caption{16 January 2013, source location S33W64.}
  \label{spectra20130116}%
\end{figure}

\begin{figure}[H]
  \centering
  \includegraphics[width=0.65\textwidth]{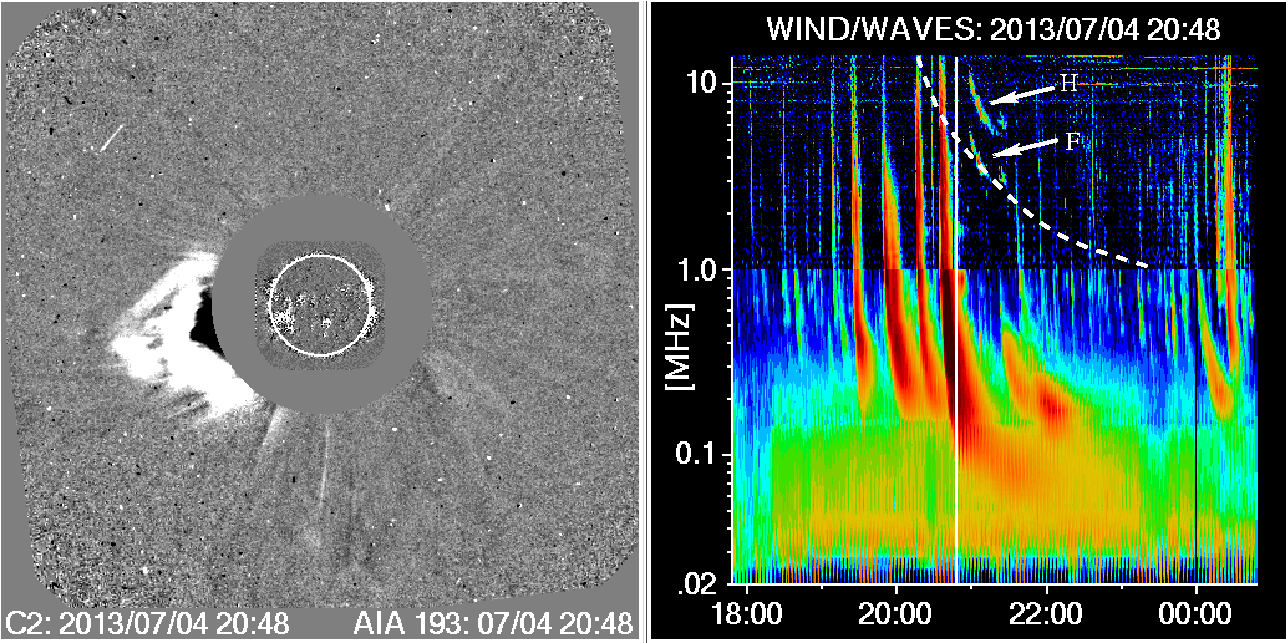}
  \includegraphics[width=0.32\textwidth]{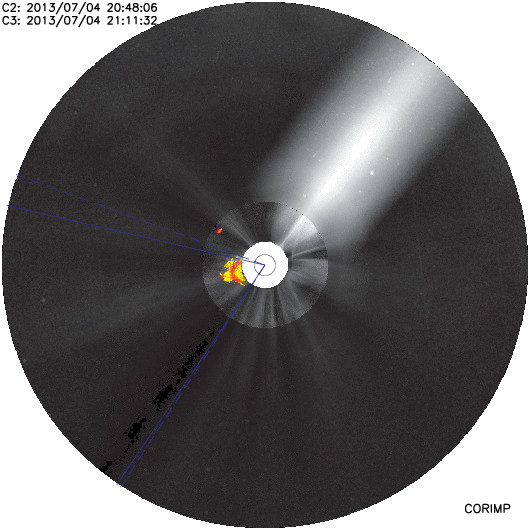}
\caption{4 July 2013, source location S14E62.
  This event is discussed in more detail in Section \ref{stereocomp}
  (delay in DH type II burst start, STEREO observations).
}
\label{spectra20130704}%
\end{figure}

\begin{figure}[H]
  \centering
  \includegraphics[width=0.65\textwidth]{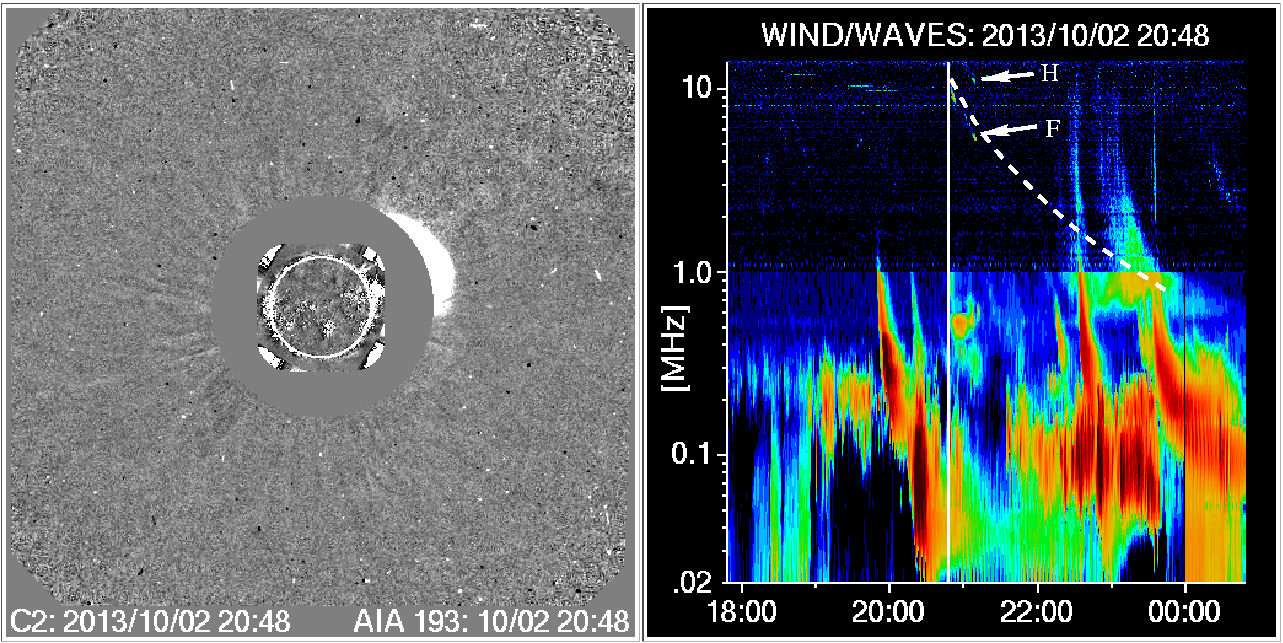}
  \includegraphics[width=0.32\textwidth]{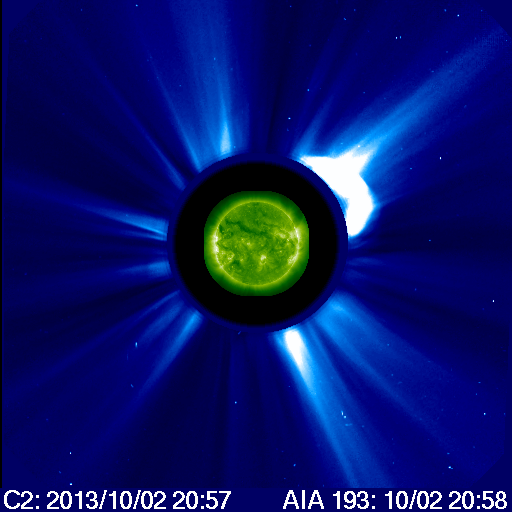}
  \caption{2 October 2013, source location N20W85.
    This event is discussed in more detail in Section \ref{stereocomp}
    (delay in DH type II burst start, STEREO observations).
  }
\label{spectra20131002}%
\end{figure}

\begin{figure}[H]
  \centering
  \includegraphics[width=0.65\textwidth]{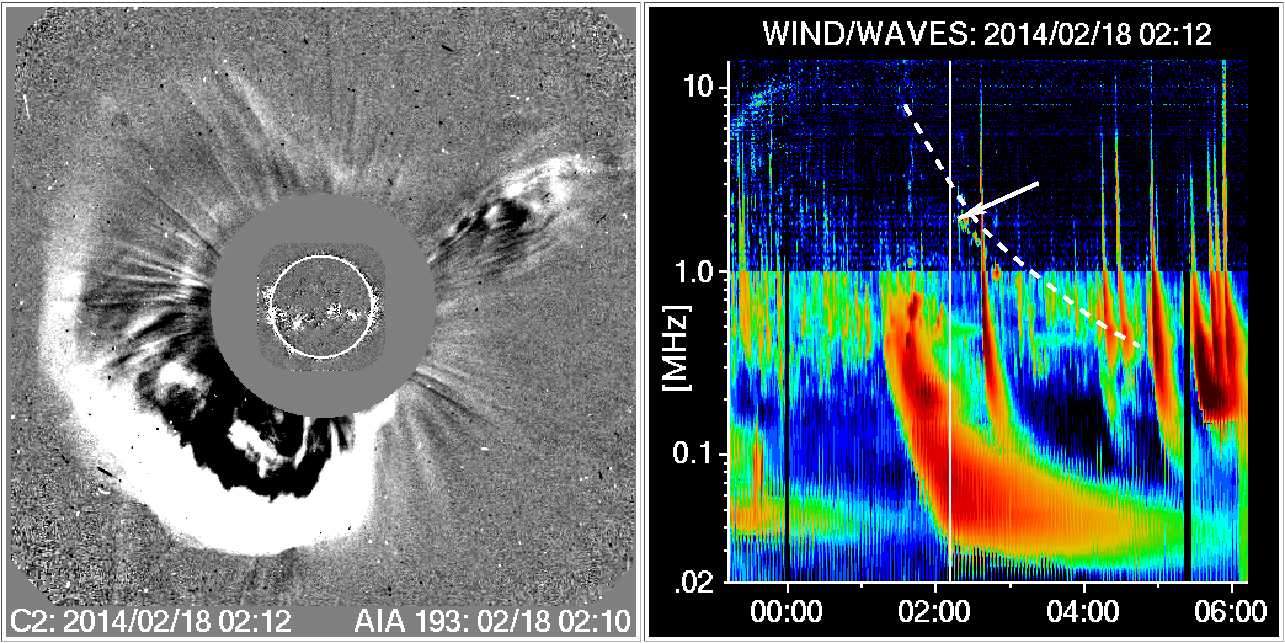}
  \includegraphics[width=0.32\textwidth]{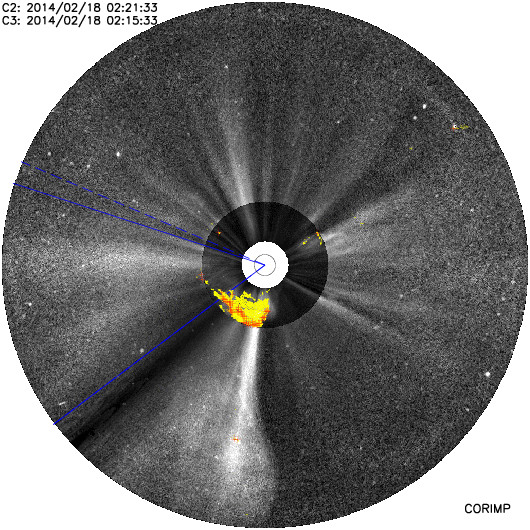}
  \caption{18 February 2014, source location SE30 (EP).}
  \label{spectra20140218}%
\end{figure}

\begin{figure}[H]
  \centering
\includegraphics[width=0.65\textwidth]{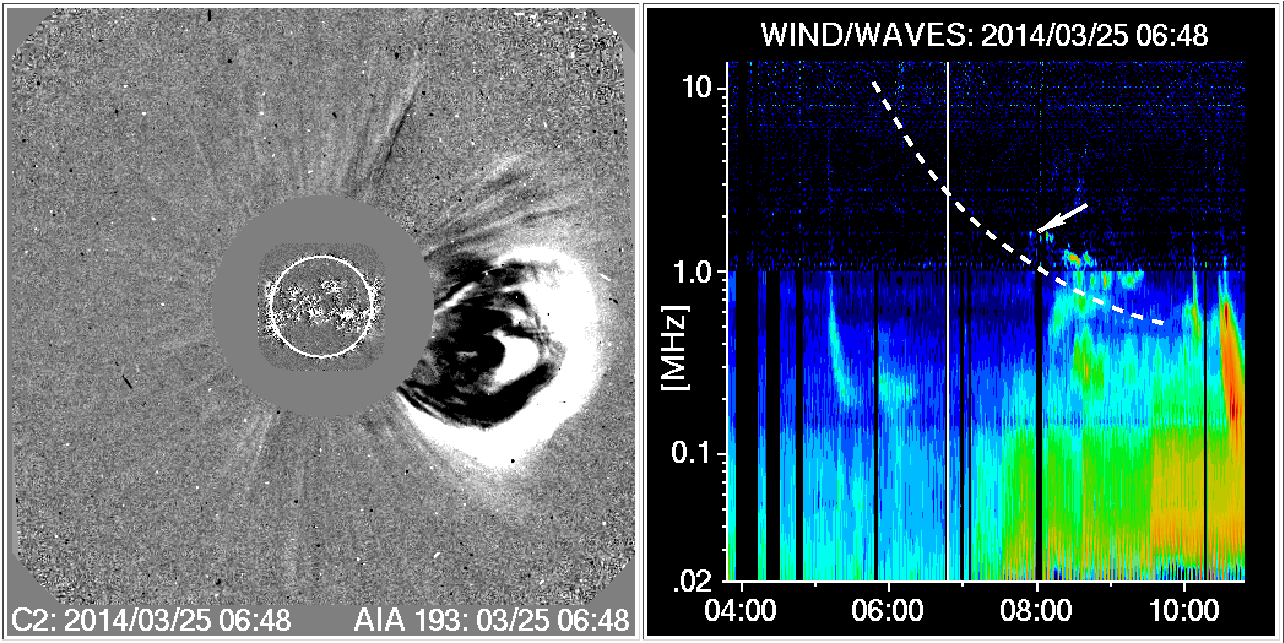}
\includegraphics[width=0.32\textwidth]{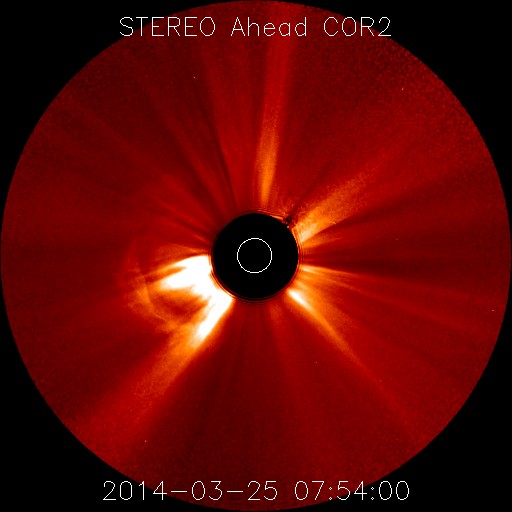}
\caption{25 March 2014, source location S23W90b (EP).}
\label{spectra20140325}%
\end{figure}

\begin{figure}[H]
  \centering
\includegraphics[width=0.65\textwidth]{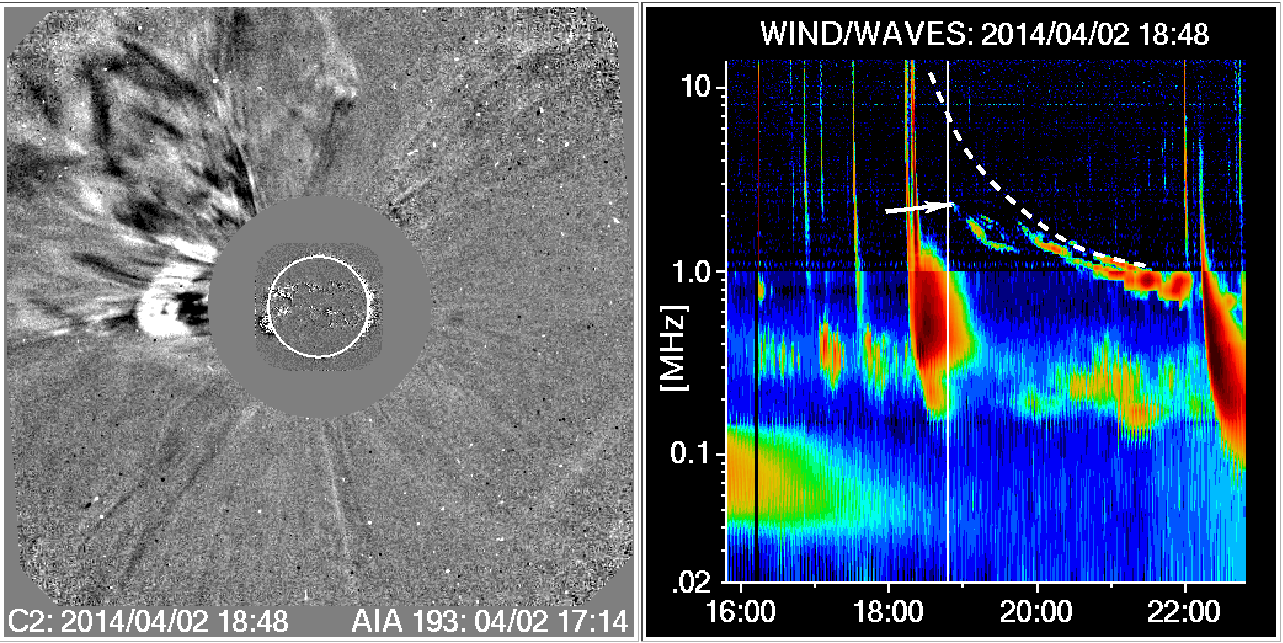}
\includegraphics[width=0.32\textwidth]{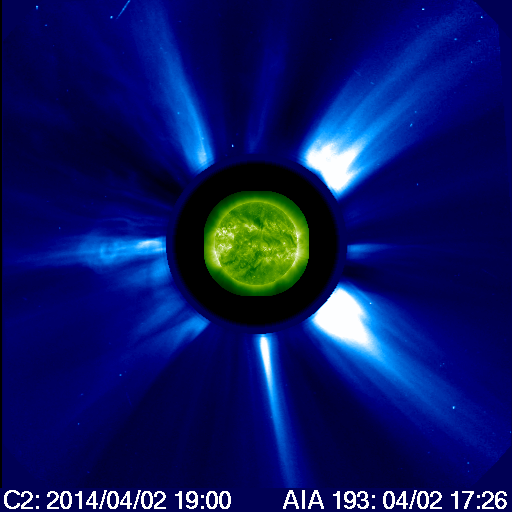}
\caption{2 April 2014, source location NE90b.
  The unlisted CME front was propagating inside an earlier CME.
  This later front propagation is indicated with the dashed line
  in the dynamic spectrum. The earlier, listed CME front was last observed at height 
  28~R$_{\odot}$ at 16:54 UT, which corresponds to $\sim$0.2 MHz frequency.
  This event is discussed in more detail in Section \ref{stereocomp}
  (delay in DH type II burst start, STEREO observations).
  }
\label{spectra20140402}%
\end{figure}

\begin{figure}[!ht]
  \centering
  \includegraphics[width=0.65\textwidth]{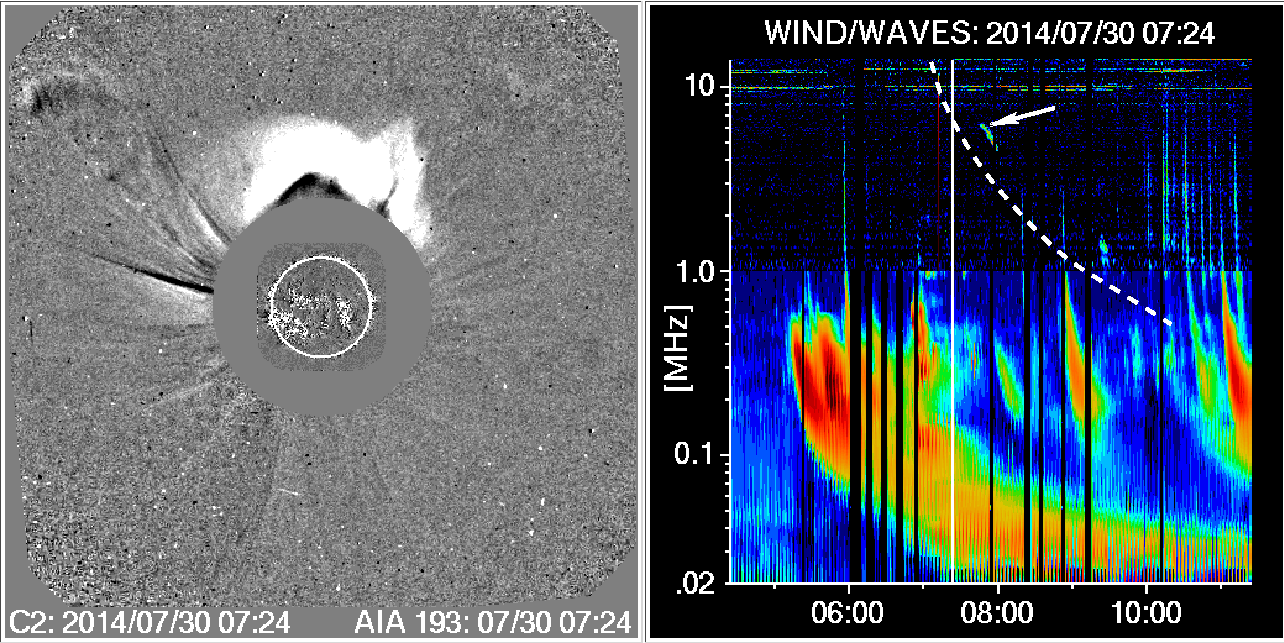}
  \includegraphics[width=0.32\textwidth]{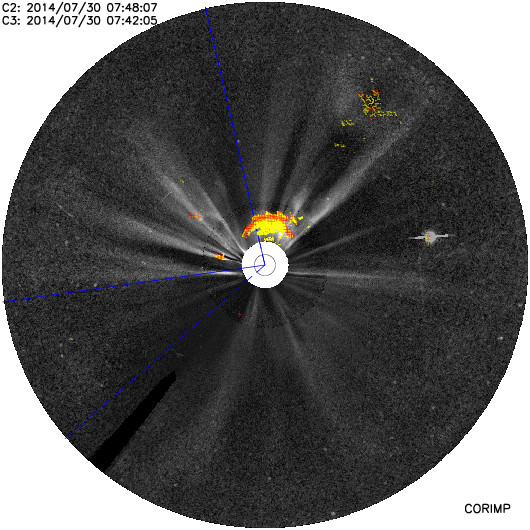}
  \caption{30 July 2014, source location N10E30 (EP).}
  \label{spectra20140730}%
\end{figure}

\begin{figure}[H]
  \centering
  \includegraphics[width=0.65\textwidth]{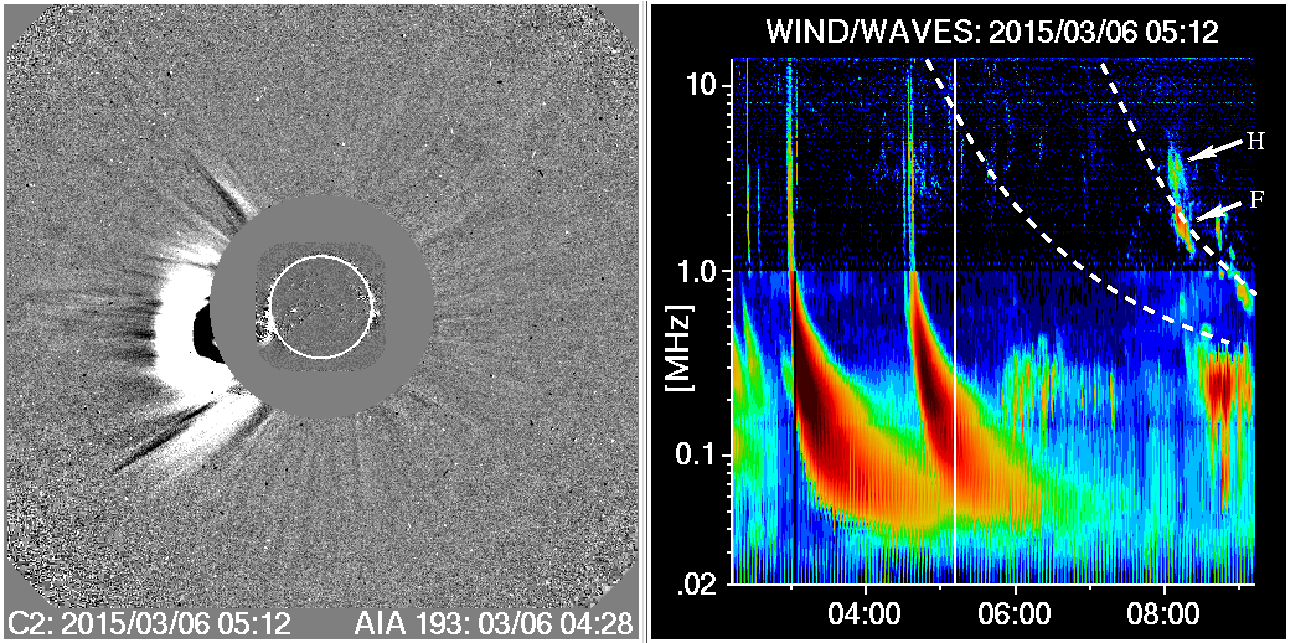}
  \includegraphics[width=0.32\textwidth]{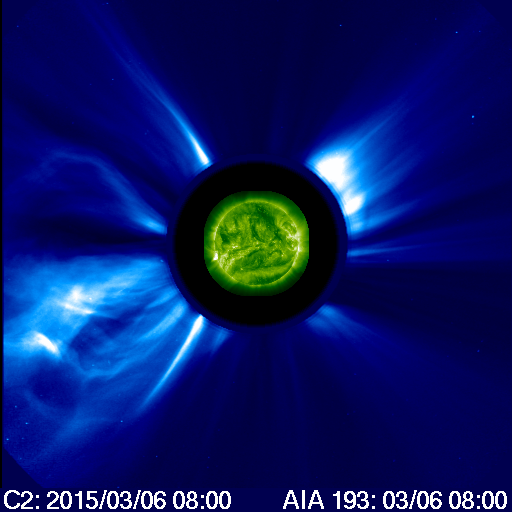}
  \caption{6 March 2015, source location S20E87.
    The CME was propagating in the wake of an earlier CME, and both CMEs are listed
    in the CME Catalog.}
  \label{spectra20150306}%
\end{figure}

\begin{figure}[H]
  \centering
\includegraphics[width=0.65\textwidth]{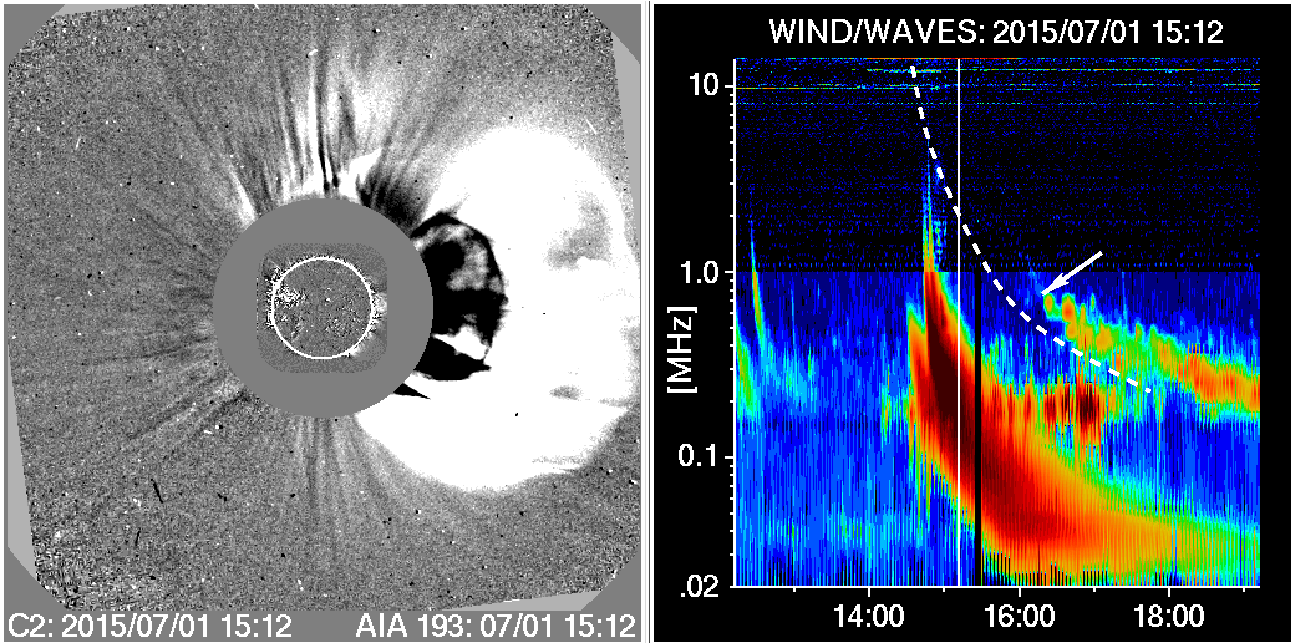}
\includegraphics[width=0.32\textwidth]{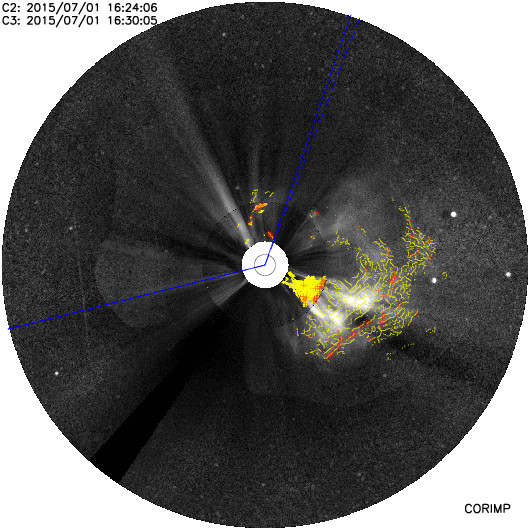}
    \caption{1~July 2015, source location W90b.}
\label{spectra20150701}%
\end{figure}

\begin{figure}[H]
  \centering
  \includegraphics[width=0.65\textwidth]{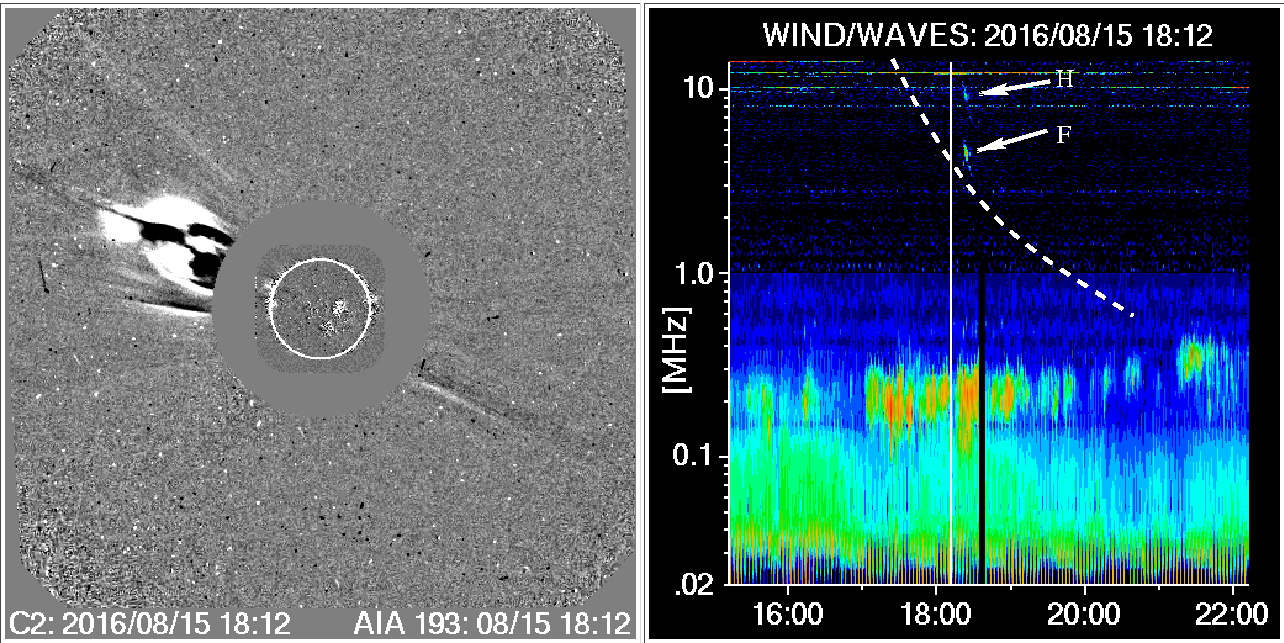}
  \includegraphics[width=0.32\textwidth]{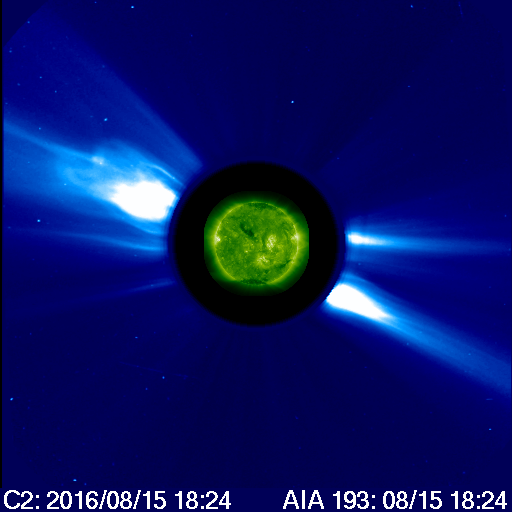} 
\caption{15 August 2016, source location E90b.}
\label{spectra20160815}%
\end{figure}

\begin{table}[!h]
\caption{Selected DH type II bursts, a total of 26 events. 
Flare/prominence location, CME type (H-halo, PH-partial halo, - not halo), 
and comparison between the type II height and the simultaneous CME height 
($\Delta$h) are listed.}
\label{table1}
\begin{tabular}{l l r r r r c c l}
\hline       
Date       & Type  & Dur. & F-    & F-     &  CME   & $\Delta$h & Flare/ & CME\\
           & II    &      & lane  & height &  height& CME-F     & prom.  & type\\
           & start &      &       &        &        &           & loc.   & \\
\tiny{yyyy-mm-dd} & UT    & min. & MHz  &R$_{\odot}$&R$_{\odot}$&R$_{\odot}$& & \\ 
\hline
1998-04-20 & 10:25 &  10  & 10  & 3.2  &    5.0  &   1.8  &  S22W90 & PH\\
2000-10-25 & 10:00 & 120+ &  4  & 5.0  &    4.5  &  -0.5  &  N09W63 & H\\
2001-09-03 & 18:48 &  12  & 12  & 3.0  &    4.1  &   1.1  &  S23E90 & PH\\
2001-12-11 & 12:45 & 120  &  2  & 7.1  &   13.0  &   5.9  &  SW90b  & PH\\
2002-03-22 & 11:30 &  70  &  4  & 5.0  &    7.0  &   2.0  &  S09W90 & H\\  
2002-07-09 & 19:46 &  54  &  1  & 10.4 &    7.3 &    -3.1 &   W90b  & H\\ 
2002-12-22 & 04:20 &  30  &  5  & 4.5  &   11.2  &   6.7  &  N23W42 & PH\\ 
2004-07-29 & 13:30 & 300+ & 0.7 & 12.6 &    10.0 &   -2.6 &  N00W90 & H\\
2004-12-31 & 00:19 &  29  & 4.2 & 4.8  &   12.0  &   7.2  &  N04E46 & H\\
2005-02-01 & 11:45 &   5  &  4  & 5.0  &   11.5  &   6.5  &  NE90b  & H\\
2005-08-22 & 02:00 &  95  &  3  & 5.7  &    7.0  &   1.3  &  S11W54 & H\\
2010-08-18 & 06:01 &  11  &  7  & 3.8  &    4.8  &   1.0 &   N18W88 & PH\\
2012-03-27 & 22:10 &  20  & 1.5 & 8.2 &     3.0 &   -5.2 &   SW90b  & - \\
2012-07-05 & 22:57 &  60  & 1.4 & 8.3 &     9.0 &    0.7 &   S12W46 & - \\
2012-07-17 & 14:40 &   8  &  6  & 4.1 &     4.0 &   -0.1 &   S28W65 & PH\\
2012-10-22 & 01:50 & 500+ &  1  & 10.4  &  13.3 &    2.9 &   S10E76 & PH\\
2013-01-16 & 22:00 & 180  & 0.6 & 14.0 &   13.5 &   -0.5 &   S33W64 & PH\\
2013-07-04 & 20:57 &  18  &  5  & 4.5  &    4.9  &   0.4 &   S14E62 & PH\\
2013-10-02 & 20:46 &  24  & 10  & 3.2  &    2.8  &  -0.4 &   N20W85 & PH\\
2014-02-18 & 02:16 &  35  &  2  & 7.1  &    6.2  &  -0.9 &  SE30    & H\\
2014-03-25 & 07:52 &  83  & 1.7 & 7.7  &    9.6 &    1.9 &  S23W90b & PH\\
2014-04-02 & 18:49 & 230+ & 2.4 & 6.4  &    3.8 &   -2.6 &  NE90b   & H\\ 
2014-07-30 & 07:44 &  16  & 6.3 & 4.0  &    5.2 &    1.2 &  N10E30  & PH\\
2015-03-06 & 08:10 & 140  & 2.0 & 7.1  &    7.2 &    0.1 &  S20E87  & PH\\
2015-07-01 & 16:22 & 260  & 0.7 & 12.6 &   16.0 &    3.4 &  W90b    & H\\
2016-08-15 & 18:21 &   7  & 5.0 & 4.5  &    5.5 &    1.0 &  E90b    & - \\
\hline                  
\end{tabular}\\
\end{table}

\begin{table}[!h]
\caption{Selected DH type II bursts and associated events.}
\label{table2}
\begin{tabular}{l r r l r r r l l}
\hline       
Date       & AR    & GOES  & CME   & CME         & CME         & Metric & DH     & DH \\
           &       & class & first & speed$^{1}$ & width$^{2}$ & type   & band   & lane\\
           &       &       & obs.  &             &             & II     &        &    \\
\tiny{yyyy-mm-dd} &    &   & UT    & km s$^{-1}$ & deg         &     & $\Delta$f/f & \\ 
\hline
1998-04-20 & 8194 & M1.4 & 10:07 &  1700 ac.  & 120    & yes    & 0.04    & F\\
2000-10-25 & 9199 & C4.0 & 08:26 &   650 ac.  & 120    & -      & 0.30    & F+H\\
2001-09-03 & 9607 & M2.5 & 18:35 &  1500 de.  & 70     & yes    & 0.06    & F\\
2001-12-11 & -    & -    & 09:54$^3$ & 900 ac. & 120   & -      & 0.25    & F+H\\
2002-03-22 & 9866 & M1.6 & 11:06 &  1800 de.  & 110    & yes    & 0.12    & F+H\\  
2002-07-09 &  -   & -    & 19:31 &   900 de.  & 200    & -      & 0.20    & F\\ 
2002-12-22 &10223 & M1.1 & 03:30 &  1100 de.  & 120    & yes    & 0.12    & F\\ 
2004-07-29 & 10652& C2.1 & 12:06 &  1100 ac.  & 100    & -      & 0.36    & F\\
2004-12-31 &10715 & M4.2 & 22:30 &  1100 co.  & 130    & yes    & 0.07    & F\\
2005-02-01 & -    & -    & 11:06 &  1500 de.  & 100    & yes    & 0.10    & F+H\\
2005-08-22 &10798 & M2.6 & 01:31 &  1250 de.  & 180    & yes    & 0.10    & F\\
2010-08-18 &11099 & C4.5 & 05:48 &  1550 de.  & 90     & yes    & 0.14    & F+H\\
2012-03-27 &  -   &  -   & 22:00  &   700 co.  & 45    & -      & 0.20    & F+H\\
2012-07-05 & 11515& M1.6 & 22:00  &  1150 de.  & 60    & yes    & 0.21    & F\\
2012-07-17 & 11520& C9.9 & 13:48  &   600 ac.  & 80    & -      & 0.18    & F+H\\
2012-10-22 & 11598& M1.3 & 20:57  &   480 ac.  & 180   & -      & 0.20    & F\\
2013-01-16 & 11650& C2.2 & 19:00  &   650 de.  & 180   & -      & 0.33    & F+H\\
2013-07-04 & 11787& C6.8 & 20:12 &   550 de.  & 85     & -      & 0.10    & F+H\\
2013-10-02 & 11850& C1.2 & 20:36 &   580 ac.  & 70     & -      & 0.05    & F+H\\
2014-02-18 & EP   & -    & 01:25 &   850 de.  & 230    & -      & 0.10    & F\\
2014-03-25 & EP   & -    & 05:36 &   650 ac.  & 80     & -      & 0.09    & F\\
2014-04-02 & -    & -    & 13:36$^4$&870 de.  & 50     & -      & 0.12    & F\\ 
2014-07-30 & EP?  & -    & 07:00 &   700 co.  & 100    & -      & 0.05    & F\\
2015-03-06 & 12297& M1.5 & 07:12$^5$&760 ac.  & 120    & -      & 0.30    & F+H\\
2015-07-01 & -    & -    & 14:36 &  1400 de.  & 200    & yes    & 0.21    & F\\
2016-08-15 & -    & -    & 17:24 &   720 de.  & 40     & -      & 0.10    & F+H\\
\hline                  
\end{tabular}\\
$^1$CME speed at the time of DH type II burst start, from second order fit, with acceleration/deceleration/constant speed information (CDAW CME Catalog).\\
$^2$CME angular width at the time of type II burst start.\\ 
$^3$Later CME front observed at 11:30 UT, inside the listed CME, speed 500 km s$^{-1}$ and accelerating.\\
$^4$Long duration X-ray flare, bright propagating CME front appears at 18:36 within the listed CME, height and speed are for the new feature.\\
$^5$Propagating in the wake of an earlier CME.\\
\end{table}

%%%%%%%%%

\begin{table}[!h]
\caption{DH type II bursts observed with  {\it Wind} and STEREO (total of 12 events).}
\label{table3}
\begin{tabular}{l l l l l l}
\hline       
Date       & STEREO-B      & {\it Wind}     & STEREO-A      & Source &  Comment\\
           & UT / MHz      & UT / MHz       & UT / MHz      &        & \\
\hline
2010-08-18 & -             & 06:01 / 7.0    & 06:01 / 7.0   & N18W88 & No delay\\
2012-03-27 & -             & 22:06 / 1.5    & 22:06 / 1.5   & SW90b  & No delay \\
2012-07-05 &   -           & 22:57 / 1.4    & 22:52 / 2.2   & S12W46 & A first, 5 m\\
2012-07-17 & -             & 14:40 / 6.0    & 14:40 / 6.0   & S28W65 & No delay\\
2012-10-22 & 01:24 / 1.4   & 01:50 / 1.0    & 01:50 / 0.9   & S10E76 & B first, 26 m\\
2013-07-04 & 21:05 / 4.0   & 20:57 / 5.0    & -             & S14E62 & {\it Wind} first, 8 m\\
2013-10-02 & -             & 20:46 / 10.0   & 20:39 / 12.0  & N20W85 & A first, 7 m\\
2014-02-18 & -             & 02:16 / 2.0    & -             & SE30   & {\it Wind} only, EP\\
2014-03-25 & 07:54 / 1.7   & 07:52 / 1.7    & 07:54 / 1.7   & S23W90b& No delay, EP \\
2014-04-02 & 20:12 / 1.4   & 18:49 / 2.4    & no observ.    & NE90b  & {\it Wind} first, 83 m \\ 
2014-07-30 & -             & 07:44 / 6.3    & -             & N10E30 & {\it Wind} only, EP?\\
2016-08-15 & no observ.    & 18:21 / 5.0    & 18:21 / 5.0   & E90b   & No delay\\
\hline                  
\end{tabular}\\
\end{table}

%%%%%%%%%

\end{article}
\end{document}